\documentclass[usegraphicx,useAMS]{mn2e}
\usepackage{rotating}
\usepackage{subfigure}
\usepackage{lscape}

\newcommand{\be}{\begin{equation}}
\newcommand{\ee}{\end{equation}}
\newcommand{\bd}{\begin{displaymath}}
\newcommand{\ed}{\end{displaymath}}

\title[XMM-Newton observations of NGC 2547]
  {An {\it XMM-Newton} observation of the young open cluster NGC 2547:
  coronal activity at 30 Myr}
\author[R.D. Jeffries et al.]
  {R.D.~Jeffries$^1$, P.A.~Evans$^1$, J.P. Pye$^2$ and K.R. Briggs$^{3}$ \\
  $^1$ Astrophysics Group, School of Chemistry and Physics, Keele University, Keele, 
      Staffordshire ST5 5BG, United Kingdom\\
  $^2$ Department of Physics and Astronomy,
  University of Leicester, Leicester LE1 7RH\\
  $^3$ Paul Scherrer Institut, 5232 Villigen PSI, Switzerland\\
}
\date{Submitted October 2005}

\pagerange{\pageref{firstpage}--\pageref{lastpage}} \pubyear{2004}

\def\LaTeX{L\kern-.36em\raise.3ex\hbox{a}\kern-.15em
    T\kern-.1667em\lower.7ex\hbox{E}\kern-.125emX}

\begin{document}

\label{firstpage}

\maketitle

\begin{abstract}
We report on {\em XMM-Newton} observations of the young open cluster
NGC~2547 which allow us to characterise coronal activity in solar-type
stars, and stars of lower mass, at an age of 30\,Myr.  X-ray emission
is seen from stars at all spectral types, peaking among G-stars at
luminosities (0.3--3\,keV) of $L_{\rm x}\simeq 10^{30.5}$~erg\,s$^{-1}$
and declining to $L_{\rm x} \leq 10^{29.0}$~eg\,s$^{-1}$ among M-stars
with masses $\geq 0.2\,M_{\odot}$. Coronal spectra show evidence for
multi-temperature differential emission measures and low coronal metal
abundances of $Z\simeq 0.3$. The G- and K-type stars of NGC~2547 follow
the same relationship between X-ray activity and Rossby number
established in older clusters and field stars, although most of the
solar-type stars in NGC~2547 exhibit saturated or even super-saturated
X-ray activity levels. The median levels of $L_{\rm x}$ and $L_{\rm
x}/L_{\rm bol}$ in the solar-type stars of NGC~2547 are very similar to
those in T-Tauri stars of the Orion Nebula cluster (ONC), but an order
of magnitude higher than in the older Pleiades. The spread in X-ray
activity levels among solar-type stars in NGC~2547 is much smaller than
in older or younger clusters.

Coronal temperatures increase with $L_{\rm x}$, $L_{\rm x}/L_{\rm bol}$
and surface X-ray flux. The most active solar-type stars in NGC~2547
have coronal temperatures intermediate between those in the ONC and the
most active older ZAMS stars. We show that simple scaling arguments
predict higher coronal temperature in coronally saturated stars with
lower gravities.  A number of candidate flares were identified among
the low-mass members and a flaring rate (for total flare energies
[0.3--3\,keV] $>10^{34}$~erg) of 1 every $350^{+350}_{-120}$\,ks 
was found for solar-type stars, which is similar to rates found in the ONC and
Pleiades. Comparison with {\it ROSAT} HRI data taken 7 years previously
reveals that only 10--15 percent of solar-type stars or stars with
$L_{\rm x}>3\times10^{29}$~erg\,s$^{-1}$ exhibit X-ray variability by
more than a factor of two. This is comparable with clusters of similar
age but less than in both older and younger clusters. The similar
median levels of X-ray activity and rate of occurrence for large flares
in NGC~2547 and the ONC demonstrate that the X-ray radiation
environment around young solar-type stars remains relatively constant
over their first 30\,Myr.

\end{abstract}


\section{Introduction}

X-ray emission from the hot coronae of cool stars is now a well
established phenomenon (e.g. see the review by G\"udel 2004). The
emission arises from magnetically confined and heated structures with
temperatures in excess of $10^{6}$\,K. In stars that have reached the
zero age main sequence (ZAMS) or older, the driving mechanism for this
magnetic activity is thought to be a stellar dynamo: stars with
convective envelopes and rapid rotation are relatively luminous X-ray
sources compared with slower rotating stars of similar spectral
type. There is now a well-founded age-rotation-activity paradigm (ARAP
-- see Jeffries 1999; Randich 2000), established via observations of
many open clusters with ages from 50\,Myr to several Gyr (e.g. Stauffer
et al. 1994; Stern, Schmitt \& Kahabka 1995; Jeffries, Thurston \& Pye 1997),
whereby younger stars tend to be more rapidly rotating and hence
exhibit strong X-ray emission up to a saturated level, where the ratio
of X-ray to bolometric flux, $L_{\rm x}/L_{\rm bol} \simeq 10^{-3}$. As
stars get older, they lose angular momentum and eventually spin down to
rotation rates where $L_{\rm x}/L_{\rm bol}<10^{-3}$ and decreases
further thereafter.

For very young stars in star forming regions with ages $<10$\,Myr, a
direct connection between rotation and X-ray activity is much less
clear and the presence of an (accretion) disc may play a role in either
stimulating or inhibiting the observed levels of X-ray activity (see
Feigelson et al. 2003; Flaccomio et al. 2003a; Flaccomio, Micela \&
Sciortino 2003b; Stassun et al. 2004; Preibisch et al. 2005).
Feigelson et al. (2003) find no correlation between rotation and X-ray
activity and a ``saturation level'' of only $L_{\rm x}/L_{\rm
bol}\simeq 10^{-3.8}$ for pre-main-sequence (PMS) stars, both with and
without discs, in the Orion Nebula Cluster (ONC). They suggest that a
less efficient, turbulent ``distributed'' dynamo may act throughout the
convective zones of these stars. On the other hand Flaccomio et
al. (2003a,b) and Stassun et al. (2004) suggest that ONC stars with
accretion discs bias the average $L_{\rm x}/L_{\rm bol}$ downwards,
perhaps as a result of intrinsic absorption or changes in the magnetic
field geometry.  Preibisch et al. (2005) show that active accretion,
rather than the mere presence of a disc is possibly responsible for the
wider spread and lower median level of X-ray activity among the very
young ONC stars.

NGC 2547 is an interesting open cluster in the context of studying the
transition between the early behaviour of stellar coronae in star
forming regions and the development of the well tested ARAP at older
ages.  It has a precisely determined age of either $30\pm5$\,Myr
determined from fitting isochrones to its 0.3--1.2$M_{\odot}$ stars as
they descend their PMS tracks (Naylor et al. 2002),
or $35\pm 3$\,Myr determined from the re-appearance of lithium in the
atmospheres of even lower mass stars (Jeffries \& Oliveira 2005). It is
old enough that inner circumstellar discs have dispersed -- no
accretion-related H$\alpha$ emission or $L$-band near infrared excesses
are seen from its solar-type members (e.g. Jeffries, Totten \& James
2000; Young et al. 2004).  However, cluster members with
$M<1.4\,M_{\odot}$ are still in the PMS phase, stars with
$M<0.4\,M_{\odot}$ are fully convective (Siess, Dufour \& Forestini
2000; D'Antona \& Mazzitelli 1997) and it is significantly younger than
other well-studied open clusters like IC\,2391 ($50\pm 5$\,Myr) and the
Alpha Per cluster ($90\pm10$\,Myr).

NGC 2547 was observed at X-ray wavelengths by the {\it ROSAT} High
Resolution Imager (HRI). Jeffries \& Tolley (1998) found a rich
population of low mass, X-ray active cluster candidates with
$10^{29}<L_{\rm x}<10^{31}$\,erg\,s$^{-1}$. Puzzlingly, the solar-type
stars of NGC 2547 seemed slightly {\it less} X-ray active than their
counterparts in older clusters. Their activity peaked at $L_{\rm
x}/L_{\rm bol}=10^{-3.3}$, whereas the most active stars at lower
masses had $L_{\rm x}/L_{\rm bol}=10^{-3}$ as expected.  Jeffries,
Totten \& James (2000) ruled out anomalously slow rotation rates in the
cluster as an explanation; they found both very fast and slow rotators
among the X-ray selected members.  A further possibility is that the
energy distribution of the X-ray emitting plasma is rather different
for the solar-type stars of NGC 2547 than for those in older clusters
or cooler stars -- the HRI observations had no spectral resolution and
so a uniform conversion factor was used to estimate X-ray fluxes from
count rates.

In this paper we present the results of an {\it XMM-Newton} observation
of NGC 2547 using the European Photon Imaging Camera (EPIC), which seeks
to characterise the coronal emission of solar-type (and lower mass)
stars at $\simeq 30$\,Myr.  The sensitivity of these X-ray images is
better than the {\it ROSAT} HRI data, enabling us to identify cluster
members with lower activity and measure X-ray emission from cluster
members with lower mass.  There is also some spectral resolution
available with the EPIC data that allows us to test whether the X-ray
spectra of the solar-type members of NGC 2547 are significantly
different to active stars in other open clusters. Finally, we are able
to look for possible variability in the level of X-ray emission of
these young stars on a timescale of 7 years, which is comparable with
the solar magnetic activity cycle.

Section~2 describes the observations, data analysis and identification
of X-ray sources with members of NGC~2547. Section~3 deals with
spectral analysis of the X-ray data, whilst section~4 uses the spectral
information to calculate intrinsic luminosities and search for evidence
of dynamo-related activity. Section~5 looks at the X-ray variability of
NGC~2547 members, both within the observation (flares, rotational
modulation) and on the longer 7 year timescale. Section 6 places
NGC~2547 in context with younger and older clusters and discusses the
evolution of X-ray activity, coronal temperatures and coronal
variability. Our conclusions appear in section~7.

\section{Observations and Data Analysis}

NGC 2547 was observed by {\it XMM-Newton} between UT 23:16:33 on 2
April 2002 and UT 13:17:53 on 3 April 2002 using the EPIC instrument,
for a nominal exposure time of 49.4\,ks.  The two EPIC-MOS cameras and
the EPIC-pn camera were operated in full frame mode (Turner et
al. 2001; Str\"uder et al. 2001), using the medium filter to reject
optical light. The nominal pointing position of the observation was
RA\,$=08$h\,10m\,12.0s, Dec\,$=-49$d\,13m\,0.0s (J2000.0).  As we shall
show in subsequent sections (2.1, 2.2 and 4.1), these data yield an X-ray
luminosity threshold (0.3--3.0\,keV) for the weakest detected sources
of $\ga 8\times10^{28}$\,erg\,s$^{-1}$ for NGC 2547 cluster members
near the centre of the EPIC field of view.

\subsection{Source Detection}

Version 6.0 of the {\it XMM-Newton} Science Analysis System was used
for the initial data reduction and source detection. Unfortunately, the
data were affected by several periods of high background. Data from the
three cameras were individually screened for high background periods
and these time intervals were excluded from all subsequent
analysis. Observation intervals were excluded where the total count
rate (for single events of energy above 10\,keV) in the instruments
exceed 0.35\,s$^{-1}$ and 1.0\,s$^{-1}$ for the MOS and pn detectors
respectively.  The remaining useful exposure times were 29.0\,ks and
29.4\,ks for the MOS1 and MOS2 cameras, but only 13.7\,ks for the pn
camera, which is more sensitive to high background intervals.

Images were created using the {\em evselect} task and a spatial
sampling of 2 arcseconds per pixel. The event lists were filtered to
exclude anomalous pixel patterns and edge effects by including only
those events with ``pattern''\,$\leq 12$.  The contrast between
background and source events was also increased by retaining only those
events with energies between 0.3 and 3\,keV. The {\sc edetect\_chain}
task was used to find sources with a combined maximum likelihood value
in all three instruments greater than 10 for the 0.3-3.0\,keV energy
range.  We expect 1-2 spurious X-ray detections at this level of
significance, though they would be highly unlikely to correlate with an
NGC~2547 member, so will not hamper any analysis in this paper.

Before performing the combined search we executed searches on
the individual images to confirm that there were no systematic
differences in the astrometry of the brightest sources.  Count rates in
each detector were determined for sources using exposure maps created
within the same task. In addition, count rates were determined for each
source in the 0.3--1.0\,keV and 1.0--3.0\,keV bands separately.  A total
of 163 significant X-ray sources were found. 
Some only have count rates measured
in a subset of the three instruments because they fell in gaps between
detectors, on hot pixels or lay outside the field of view.

\subsection{Source Identification}

Jeffries \& Tolley (1998) showed that the
majority of bright X-ray sources in this region are associated with
stars in the NGC 2547 cluster. As the purpose of this paper is to
examine the X-ray properties of cluster stars, we have restricted our
analysis to those X-ray sources with counterparts among
photometrically selected cluster members taken from the catalogues of
Naylor et al. (2002) and Jeffries et al. (2004).

The EPIC X-ray source list was correlated with: (1) the
photometrically selected members based on the D'Antona \& Mazzitelli
(1997) isochrones and $BVI$ photometry in Naylor et al. (2002 --
their Table~6), which incorporates $BV$ photometry of bright
cluster members from Clari\'a (1982); (2) photometrically selected
members based on {\em either} the D'Antona \& Mazzitelli or Baraffe et
al. (2002) isochrones and $RIZ$ photometry in Jeffries et al. (2004 --
their Tables A.2 and A.3)

The correlation took place in two stages. In stage 1
correlations were sought with bright (maximum log likelihood $>100$) X-ray sources
that have small formal position uncertainties. The purpose was to
establish how much additional systematic error there might be in the
{\em XMM-Newton} astrometry and whether there was any systematic offset in
the X-ray source positions\footnote{The absolute astrometric accuracy
of the optical catalogues is of order 0.2 arcseconds and they have
astrometry that is internally consistent to $<0.1$ arcseconds.}.
It was found that an offset of 1.48 and 0.05 arcseconds needed subtracting
from the RA and Dec of the X-ray positions and that an additional
systematic error of 1.2 arcseconds (1-$\sigma$) 
needed adding in quadrature to the
X-ray astrometric uncertainties in order to yield a reduced $\chi^{2}$
of unity in the fit to the mean offset of 38 bright X-ray/cluster-object
correlations. This additional error corresponds perfectly with the expected
precision of the current {\it XMM-Newton} attitude reconstruction
(Kirsch et al. 2005).
In stage 2 this offset and additional error were applied and
the full X-ray source list correlated with the membership catalogues using an error
circle of radius 3 times the total positional error. 103
correlations were found between $BVI$ selected members and X-ray sources and 67
correlations between $RIZ$ selected members and X-ray sources. There
are a total of 108 X-ray sources with a cluster counterpart, 5 of which
were not considered members by Naylor et al. (2002) using $BVI$
photometry (they look just too blue in $B-V$) and 8 of which are not
considered members by Jeffries et al. (2004) using $RIZ$ photometry (3
look just too blue in the $R-I$ vs $I-Z$ diagram and 5 lie a little too
far above the cluster isochrone in $I$ vs $R-I$).  A further 33 objects
that are members in the $BVI$ catalogue have no good photometry in the
$RIZ$ catalogue, mostly because these stars are too bright and were
saturated in the deeper $RIZ$ images. We choose to include all the
above objects as likely cluster members.

\begin{figure}
\includegraphics[width=84mm]{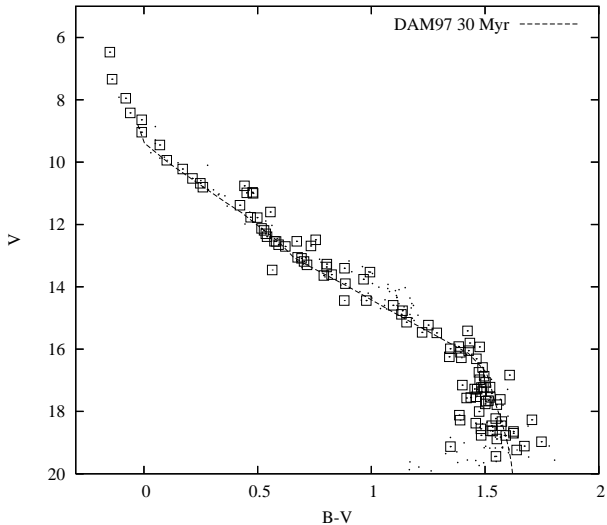}
\caption{V,B-V diagram for X-ray selected members of
  NGC 2547 (open squares). 
  The isochrone is derived from the models of D'Antona \&
  Mazzitelli (1997) using an age of 30\,Myr, a distance modulus of 8.1,
  reddening and extinction appropriate for $E(B-V)=0.06$ and a
  colour-$T_{\rm eff}$ relation calibrated using the Pleiades (see
  Naylor et al. 2002). Dots represent all {\em photometric} cluster
  candidates within 15 arcminutes of the X-ray pointing.
}
\label{vbvcmd}
\end{figure}

\begin{figure}
\includegraphics[width=84mm]{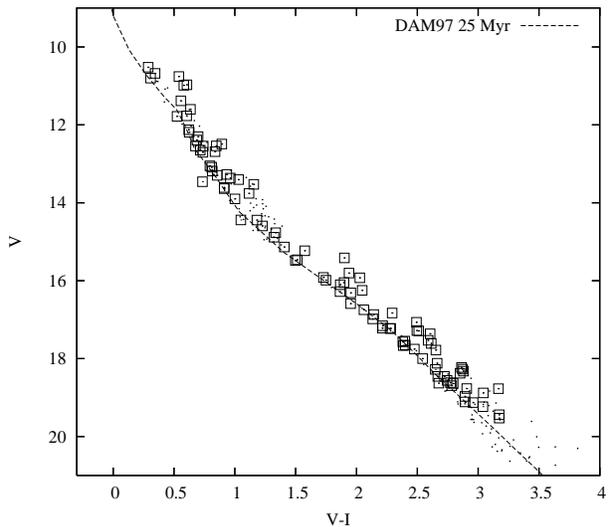}
\caption{V,V-I diagram for X-ray selected members of
  NGC 2547. Symbols as for Fig.~1. 
The isochrone is derived as for Fig.~\ref{vbvcmd}, but this
  time an age of 25\,Myr gives a better fit (see Jeffries \& Oliveira 2005).
}
\label{vvicmd}
\end{figure}

\begin{figure}
\includegraphics[width=84mm]{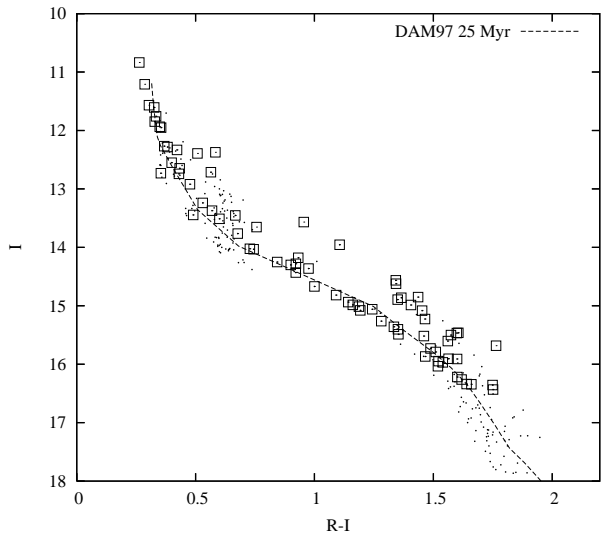}
\caption{I,R-I diagram for X-ray selected members of
  NGC 2547. Symbols as for Fig.~1. The isochrone is derived as for Fig.~\ref{vbvcmd}, but
  using an age of 25\,Myr.
}
\label{iricmd}
\end{figure}

Colour magnitude diagrams for the X-ray sources with cluster
counterparts are shown in Figs.~\ref{vbvcmd}, \ref{vvicmd} and
\ref{iricmd}. A distance of 417~pc, $E(B-V)=0.06$, $E(V-I)=0.077$ and
$E(R-I)=0.043$ were assumed (see Jeffries \& Oliveira 2005 and
references therein). Cluster candidates that lie within 15 arcminutes
of the {\it XMM-Newton} pointing and which were {\it not} detected as
X-ray sources are shown for comparison.  Details of the X-ray sources
with cluster counterparts are given in Tables~\ref{xraymembers}
and~\ref{counterparts}.  The number of spurious correlations was
estimated by applying random 30 arcsecond offsets to the X-ray
sources. From these tests fewer than 2 of the 108 cluster X-ray sources
are expected to be spurious correlations. X-ray sources are detected
from across the mass range covered by the cluster members. For low-mass
objects our census is limited by the sensitivity of the X-ray
observations (see section 4.1). The faintest detected cluster members
have $V\simeq 19.5$ and $I\simeq 16.5$, corresponding to about
0.25\,$M_{\odot}$ from the isochrones used in Figs.~\ref{vvicmd}
and~\ref{iricmd}. The observations are not sensitive enough (by a
factor of a few) to detect brown dwarfs if they have an X-ray to
bolometric flux ratio of $10^{-3}$.

The remaining 55 X-ray sources with no cluster counterpart are listed
in Table~\ref{nocounterparts}. In an exposure of this length it is
quite probable that many of these sources are extragalactic and there
may also be a number of magnetically active field stars unassociated
with the cluster. However, we have not listed possible correlations with
the full optical catalogues of Naylor et al. (2002) and Jeffries et
al. (2004) because the expected number of spurious correlations down to
the limits of these catalogues (which would still have plausible
X-ray-to-optical flux ratios for extragalactic sources) is of order 50
and many have several counterparts.  These X-ray sources will not be
discussed further in this paper.

\begin{landscape}
\begin{table}
\caption{X-ray sources that are correlated with photometric candidate
  members of NGC 2547. The Table is only available electronically and
  contains 108 rows. The first two rows are shown here as a guide to form
  and content. Columns are as follows: (1) running source identification number (produced
  by the SAS analysis system), (2) the conventional {\em XMM-Newton}
  source name, (3), (4) RA and Dec (J2000.0), (5) 
  1-sigma positional uncertainty 
  of the X-ray source  (corrected for a mean astrometry shift and including an additional
  systematic error -- see section 2.2), (6) the maximum likelihood
  statistic returned by the SAS analysis; then for the pn (7)--(10),
  MOS1 (11)--(14) and
  MOS2 (15)--(18) detectors we list the count-rates in the 0.3-3\,keV,
  0.3-1.0\,keV, 1.0-3.0\,keV ranges and their uncertainties, followed
  by a hardness ratio (e.g. defined as [col.9$-$col.8]/[col.9$+$col.8]
  for the pn) and its uncertainty.}
\begin{flushleft}
\begin{tabular}{ccccccc@{\hspace*{1mm}}c@{\hspace*{1mm}}c@{\hspace*{1mm}}c}
\hline
No.&Name  & RA        &  Dec        & $\Delta$ & ML &  \multicolumn{3}{c}{pn
  count rates (s$^{-1}$)} & HR (pn)\\
 &    &\multicolumn{2}{c}{J2000}& arcsec   &    &
(0.3-3.0)\,keV & (0.3-1.0)\,keV&
(1.0-3.0)\,keV& \\
(1)  & (2)       & (3)         & (4)      & (5) & (6) & (7) & (8) & (9) &(10)\\
\hline
   3&XMMU J081012.9-491408 & 8 10 12.90& -49 14 08.6 &  1.22 &2896.14& 
5.62E-02$\pm$2.33E-03& 3.46E-02$\pm$1.84E-03& 2.16E-02$\pm$1.43E-03& -0.231$\pm$0.040\\
   4&XMMU J080947.2-491305 & 8 09 47.25& -49 13 05.1 &  1.23 &2607.62& 
5.40E-02$\pm$4.29E-03& 4.15E-02$\pm$3.73E-03& 1.25E-02$\pm$2.12E-03& -0.536$\pm$0.068\\
\hline
\end{tabular}
\vspace*{5mm}
\begin{tabular}{c@{\hspace*{1mm}}c@{\hspace*{1mm}}c@{\hspace*{1mm}}cc@{\hspace*{1mm}}c@{\hspace*{1mm}}c@{\hspace*{1mm}}c}
\hline
 \multicolumn{3}{c}{MOS1 count rates (s$^{-1}$)} & HR (M1) &\multicolumn{3}{c}{MOS2
 count rates (s$^{-1}$)} & HR (M2)\\
(0.3-3.0)\,keV & (0.3-1.0)\,keV&
(1.0-3.0)\,keV& & (0.3-3.0)\,keV & (0.3-1.0)\,keV&
(1.0-3.0)\,keV&  \\
(10) & (11) & (12) & (13) & (14) & (15) & (16) & (17) \\
\hline
1.62E-02$\pm$8.15E-04& 8.77E-03$\pm$5.89E-04& 7.46E-03$\pm$5.62E-04&
 -0.081$\pm$0.050& 1.74E-02$\pm$1.03E-03& 9.38E-03$\pm$7.52E-04&
 8.01E-03$\pm$7.02E-04& -0.079$\pm$0.059\\
2.03E-02$\pm$9.37E-04& 1.24E-02$\pm$7.26E-04&
7.81E-03$\pm$5.93E-04&
 -0.229$\pm$0.045& 2.13E-02$\pm$9.96E-04& 1.46E-02$\pm$8.20E-04&
 6.77E-03$\pm$5.66E-04& -0.365$\pm$0.044\\
\hline
\end{tabular}
\end{flushleft}
\label{xraymembers}
\end{table}

\begin{table}
\caption{\ : The optical and derived X-ray properties of the cluster candidates
  that are found within 3-sigma of the X-ray sources in
  Table~\ref{xraymembers}. The Table is only available electronically and
  contains 108 rows. The first two rows are shown here as a guide to form
  and content. We list (1) the running source
  identification number, (2)--(6) the catalogue identifier, separation
  between optical and X-ray position, $V$, $B-V$ and $V-I$ photometry
  from Naylor et al. (2002), (7)--(10) the identifier,
  separation between optical and X-ray position, $I$ and
  $R-I$ from Jeffries et al. (2004), (11)--(13) the
  $L_{\rm x}/L_{\rm bol}$ (for the 0.3--3\,keV energy band) calculated
  using bolometric corrections derived from the $B-V$, $V-I$ and $R-I$
  indices respectively, (14) $L_{\rm x}$ (0.3--3\,keV, assuming a distance of
  417\,pc) and its 1-sigma uncertainty. }

\begin{flushleft}
\begin{tabular}{cccccccccccccc}
\hline
No. & ID (N02) & Sep & $V$ & $B-V$& $V-I$ & ID (J04) & Sep & $I$ & $R-I$ &
\multicolumn{3}{c}{$L_{\rm x}/L_{\rm bol}$} & $L_{\rm x}$ (0.3-3\,keV)\\
    &          & arcsec&   &      &       &          & arcsec&   &       &
bc($B-V$) & bc($V-I$) & bc($R-I$) & erg\,s$^{-1}$ \\                   
(1)    & (2)& (3) & (4) & (5)  & (6) & (7)& (8)  &
  (9) & (10) & (11) & (12) & (13) & (14) \\
\hline
3 &13    516& 1.08& 13.637&   0.791&   0.912&  13     98& 1.11&  12.741&   0.430&
9.17E-04& 8.76E-04& 7.46E-04& 2.26E+30$\pm$1.46E+29\\
4 &14     32& 0.69& 12.301&   0.536&   0.698&  13     34& 0.75&  11.605&   0.326&
3.45E-04& 3.31E-04& 2.80E-04& 2.54E+30$\pm$1.71E+29\\
\hline
\end{tabular}
\end{flushleft}
\label{counterparts}
\end{table}
\end{landscape}

\begin{table}
\caption{X-ray sources that were not identified with candidate members
  of NGC~2547. This table contains 55 rows and is only available in
  the electronic edition. The format is identical to that of
  Table~\ref{xraymembers}.}
\label{nocounterparts}
\end{table}

\section{Spectral Analysis}

\subsection{X-ray spectra}
\label{spectra}

\begin{figure}
\includegraphics[width=84mm]{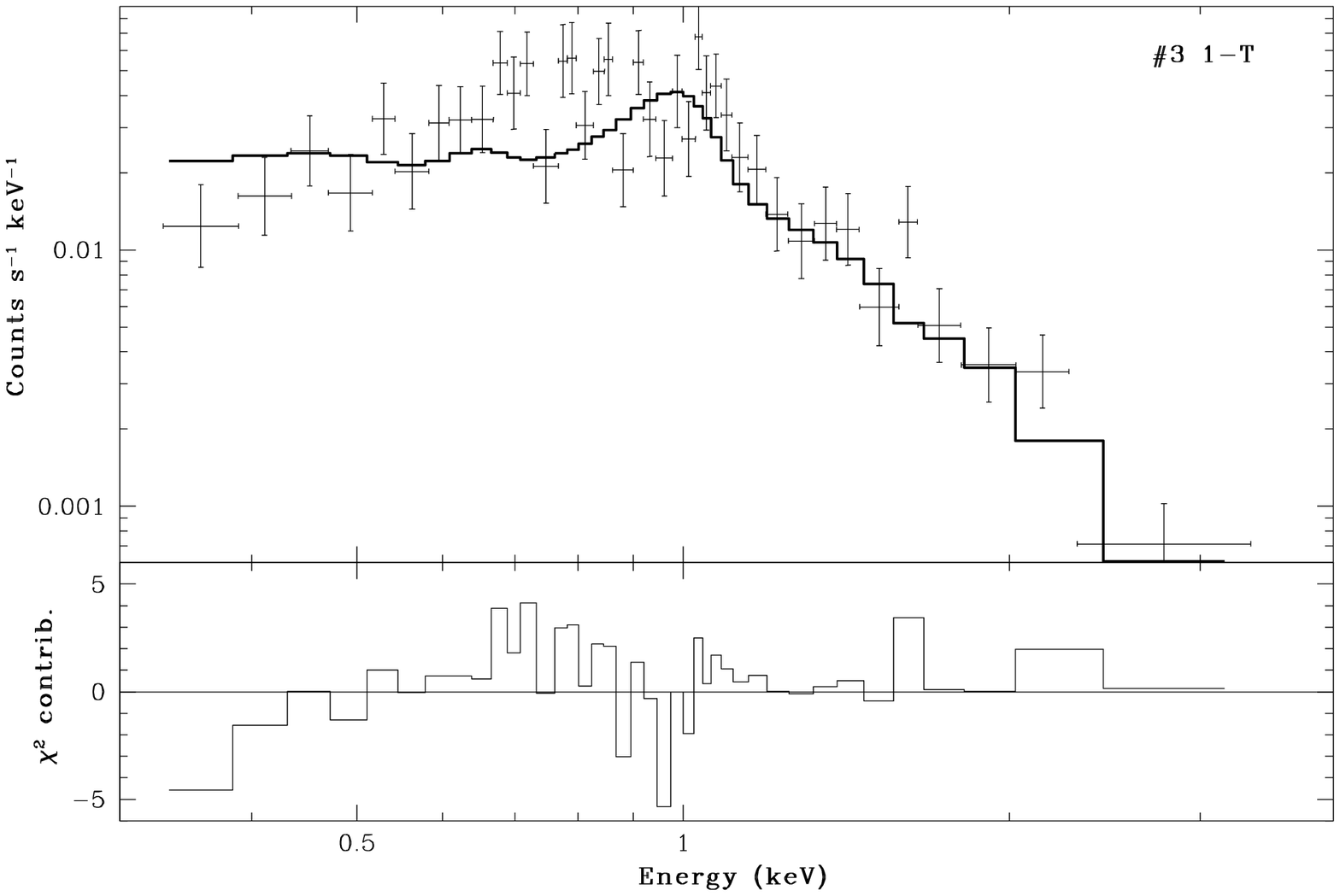}
\includegraphics[width=84mm]{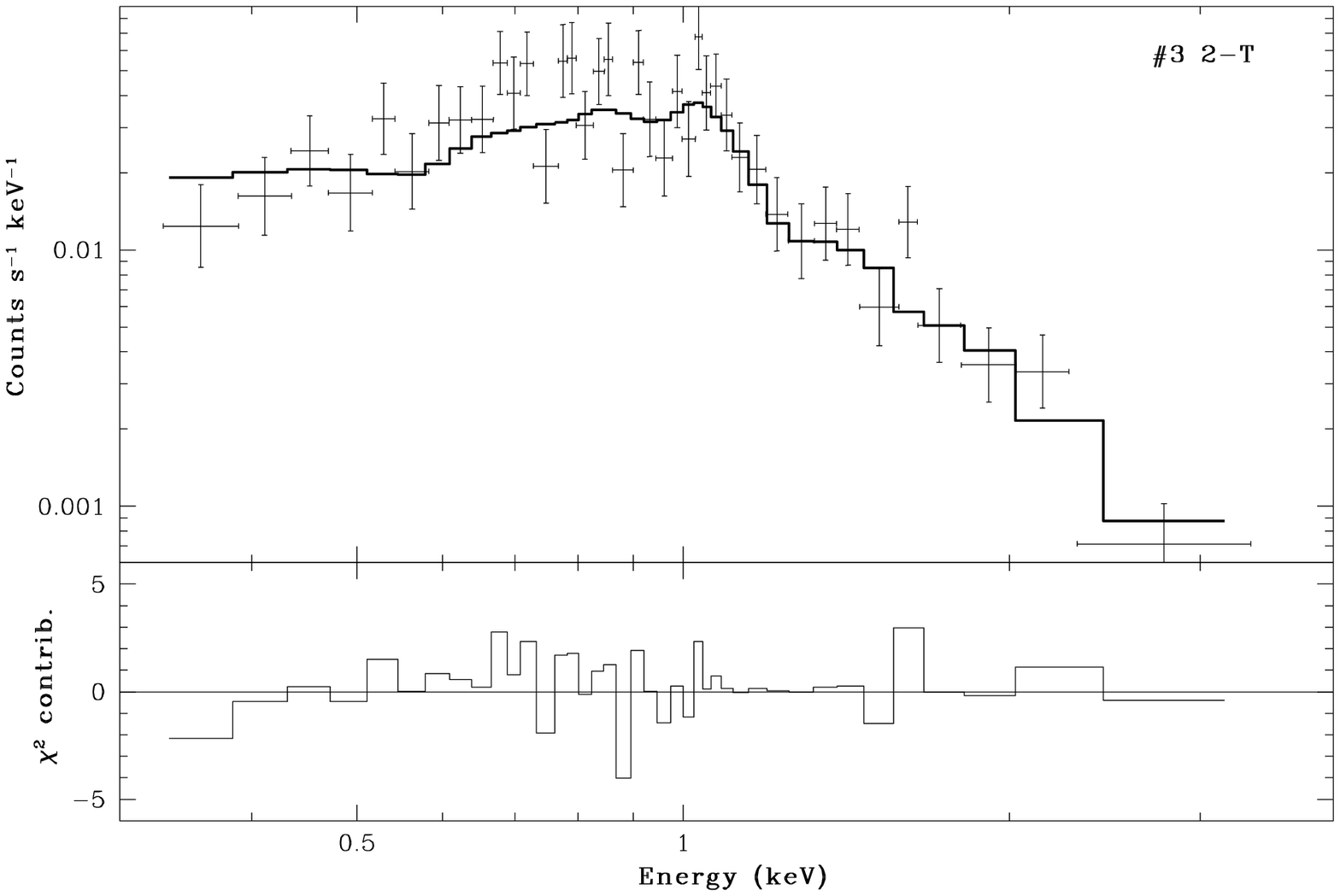}
\caption{Examples of one and two component thermal models (as described
  in the text) fitted to the pn data of star number 3.
}
\label{specfig}
\end{figure}

Ten cluster candidates were chosen for a detailed spectroscopic
examination. These ten sources were those with the largest number of
detected X-ray photons, with 300-670 counts in the pn detector and
230-530 counts in the MOS1/MOS2 detectors respectively.  Source spectra
of these stars were extracted from circular or elliptical regions with
radii $\simeq 12$ arc sec. This relatively small extraction region was
used to minimise the significant subtracted background.  
A larger (45 arcsec) extraction radius for the brightest 
source was used to check that the smaller extraction radius did not change 
the derived spectral parameters. The
same filtering expression used to generate the images was used for
source extraction. For 9 of the stars spectra were obtained from all
three EPIC instruments. Star 11 lay in a region of the pn detector
which was excluded by the selection expression described earlier, and
thus only MOS data were used for this star. Annuli around each source
were used to estimate the background.

Redistribution and ancillary response matrices were generated using the
{\sc rmfgen} and {\sc arfgen} tasks. One matrix was generated per
instrument (based on star 3) and then used for each of the
stars. Energy ranges above 0.3\,keV were considered in the analysis,
but the spectra were binned such that there were at least ten source
counts per bin and then modelled using {\sc xspec}.

A single optically thin thermal plasma ({\sc mekal} -- Mewe, Kaastra \&
Liedahl 1995) component modified by photoelectric absorption (a 1-T
model) was used as an initial model.  The column density of the
absorption was fixed at $3\times10^{20}$\,cm$^{-2}$. This corresponds
to the reddening estimated for bright cluster members and is unlikely
to be uncertain by more than a factor of two (see Jeffries \& Tolley
1998)\footnote{Experiments which allowed the column density to be a free
parameter showed that the X-ray spectra could only constrain the column
density to be $\la 10^{21}$\,cm$^{-2}$ because of the lack of
sensitivity to this parameter at energies $>0.3$\,keV and the
possibility to compensate for changes in column density with changes in
the emission measure of a cool coronal component}.  The
metal abundance was a free parameter (in the form of a multiple of the
solar metal abundances of Anders \& Grevesse 1989) while the
normalisation of the {\sc mekal} component was allowed to optimise
independently for the three EPIC instruments to counter the effects of
any cross-calibration uncertainties. In general, good agreement was
found between the three normalisations.  The best-fit parameters and
$\chi^{2}$ values of these model fits are given in
Table~\ref{tab:fits1}. An example spectral fit is shown in the top
panel of Fig.~\ref{specfig}.

The 1-T model fits are on the whole statistically acceptable. In two
cases (stars 3 and 10), the 1-T model is rejected at 99 per cent
confidence.  A second {\sc mekal} component was added to the models (a
2-T model), which of course improved the fit in all stars
(Table~\ref{tab:fits2} and see the lower panel of Fig.~\ref{specfig}). 
For four objects the upper bound to the temperature of
the second component could not be constrained; these stars are
indicated by an asterisk in Table~\ref{tab:fits2}. All of the 2-T
models are statistically acceptable, and there are significant
reductions chi-squared values in 8 cases (statistically justified at 
$>95$ per cent according to a likelihood ratio test).

Little weight should be attributed to this. It has commonly been found
that multi-temperature fits are required to fit coronal X-ray
spectra once a sufficiently precise spectrum is obtained. Even then, a
2-T model is probably a crude approximation to the true
differential emission measure (DEM). The pattern here appears to be
that the DEM could be approximated with a 2-T model with a lower
temperature $T_{1}\simeq 0.6$\,keV and an upper temperature
$T_{2}\simeq 1.5$\,keV. In spectra with insufficient counts or where
one component is significantly larger (in terms of the number of X-ray
photons produced) than the other, then a 1-T fit is adequate with
$T_{1}<T<T_{2}$. The relative emission measure of the two
components reveals that those 2-T models for which $T_{2}$ is
unconstrained are where the cooler component is
dominant. 

Perhaps the biggest difference between the 1-T and 2-T models is that
adding the extra {\sc mekal} component relaxes somewhat the requirement
for a very low metallicity in the 1-T fits. Even so, significantly
subsolar metallicities are implied by all the 2-T fits, with an average
$Z \simeq 0.3$. This value is likely dominated by the effects of a
group of strong (unresolved) iron lines around 1\,keV. The deduction of
low coronal metallicity is a common feature of spectral fits to X-ray
data from low-mass stars with high levels of magnetic activity
(e.g. Briggs \& Pye 2003; G\"udel 2004).  Telleschi et al. (2005)
studied solar analogues at a range of ages, finding that coronal iron
abundances decrease from solar values for an average coronal
temperature of 4\,MK to half-solar at temperatures of 10\,MK. The
NGC~2547 stars have average (emission measure weighted) coronal
temperatures $>10$\,MK, so a coronal metallicity of $Z \simeq 0.3$ is
not surprising.

\begin{table*}
\caption{Parameters and  $\chi^{2}$ (and reduced $\chi^{2}$) 
  values for 1-T fits to 
  sources 3--12 in Table~1. The tabulated emission measures assume a
  cluster distance of 417\,pc and are the average of the emission
  measures derived from the pn and MOS instruments. 
  Uncertainties quoted are 90 per cent
  confidence limits for one parameter. The final column gives the
  probability of attaining a
  $\chi^{2}$ at least as high as observed if the 1-T model were valid.}
\begin{center}
\begin{tabular}{cllllll}
\hline
Star  & $B-V$ & $kT$ &  $\log$ EM  & $Z$ & $\chi^2$ ($\chi^{2}_{\nu}$) &
$P(\geq \chi^{2})$  \\

 &      & (keV) & (cm$^{-3}$) &   & &\\
\hline
3   & 0.791&  1.05 (0.98 1.11)&53.39 (53.33 53.45)& 0.16 (0.14 0.21) &130.7
(1.49)& 0.01 \\
4   & 0.536&  0.68 (0.65 0.74)&53.55 (53.49 53.62)& 0.13 (0.10 0.18) & 57.5
(0.80)& 0.89\\
5   & 1.485&  1.38 (1.26 1.66)&53.43 (53.34 53.48)& 0.12 (0.07 0.21) &105.7
(1.22)& 0.08 \\
6   & 0.994&  0.80 (0.74 0.87)&53.52 (53.46 53.58)& 0.05 (0.03 0.07) & 71.7
(0.96)& 0.59\\
7   & -0.06&  0.69 (0.65 0.73)&53.16 (53.05 53.27)& 0.23 (0.16 0.35) & 59.7
(1.07)& 0.34 \\
8   & -0.08&  0.70 (0.66 0.77)&53.13 (53.00 53.24)& 0.25 (0.18 0.37) & 59.2
(0.99)& 0.50 \\
9   & 0.622&  0.87 (0.80 1.03)&53.15 (53.06 53.24)& 0.12 (0.08 0.17) & 40.5
(0.96)& 0.54  \\
10  & 0.693&  0.82 (0.76 0.88)&53.23 (53.14 53.30)& 0.12 (0.09 0.18) & 82.1
(1.55)& 0.01  \\
11  & 0.755&  1.02 (0.91 1.11)&53.47 (53.40 53.54)& 0.10 (0.06 0.15) & 54.5
(1.19)& 0.18  \\
12  & 0.704&  1.05 (0.95 1.35)&53.06 (52.96 53.15)& 0.13 (0.08 0.20) & 50.4
(1.33)& 0.09 \\
\hline
\end{tabular}
\label{tab:fits1}
\end{center}
\end{table*}

\begin{table*}
\caption{Parameters and  $\chi^{2}$ (and reduced $\chi^{2}$) 
  values for the 2-T model fits to
  sources 3--12 in Table~1. The tabulated emission measures assume a
  cluster distance of 417\,pc and are the average of the emission
  measures derived from the pn and MOS instruments. 
  Uncertainties quoted are 90 per cent
  confidence limits for one parameter. In
those fits marked with an asterisk (*), the upper bound to the temperature of
  the hotter component could not be constrained.}
\begin{center}
\begin{tabular}{cllllll}
\hline
Star   &  $kT_{1}$ & $kT_{2}$ & EM$_1$
& EM$_{2}$ & $Z$ & $\chi^{2}$  ($\chi^{2}_{\nu}$) \\
       & (keV) & (keV) &
(cm$^{-3}$)&(cm$^{-3}$) & &\\
\hline
3   & 0.61 (0.49 0.68) & 1.62 (1.37 1.96) & 52.70 (52.47 52.89) & 53.07
(52.94 53.18) & 0.45 (0.30 0.68)  & 93.1  (1.11) \\
4   & 0.64 (0.56 0.69) & 1.51 (0.97 5.93 )  & 53.34 (53.20 53.54) & 52.84
(52.65  53.07)   & 0.19 (0.13 0.30)  & 47.4  (0.70)\\
5   & 1.07 (0.96 1.30) & 9.99 (6.22 --) & 53.24 (53.19 53.45) & 52.74
(52.54 52.92) & 0.09 (0.05 0.13)  & 92.5  (1.10)* \\
6   & 0.68 (0.64 0.79) & 8.22 (1.69 --)  & 53.43 (53.23 53.53) & 52.58
(53.34 53.64) & 0.06 (0.04 0.10)  & 57.2  (0.79)*\\
7   & 0.67 (0.61 0.71) & 1.14 (0.61 --) & 53.04 (52.79 53.22) & 52.00
(-- 52.73) & 0.31 (0.19 0.56)  & 54.5  (1.05)* \\
8   & 0.57 (0.40 0.69) & 0.79 (0.71 2.81)  & 52.62 (52.35 53.18) & 52.92
(-- 53.18) & 0.31 (0.21 0.81)   & 53.7  (0.96)\\
9   & 0.57 (0.43 0.69) & 1.14 (0.99 1.35) & 52.62 (-- 52.89)    & 52.87
(52.80 53.07) & 0.22 (0.13 0.37)  & 25.6  (0.67) \\
10  & 0.63 (0.42 0.70) & 1.52 (1.18 2.06) & 52.66 (52.38 52.95) & 52.81
(52.60 53.00) & 0.36 (0.21 0.54)  & 60.5  (1.23)  \\
11  & 0.81 (0.72 0.93) & 9.99 (2.02 --) & 53.39 (53.26 53.51) & 52.64
(52.29 52.87) & 0.10 (0.06 0.19)  & 43.8  (1.00)* \\
12  & 0.64 (0.51 0.82) & 3.53 (2.28 7.75 )  & 52.40 (51.91 52.79) & 52.73
(52.54 52.90) & 0.43 (0.18 0.70)  & 25.6  (0.75)\\
\hline
\end{tabular}
\label{tab:fits2}
\end{center}
\end{table*}

\subsection{Hardness ratios}
\label{hardness}

\begin{figure}
\includegraphics[width=84mm]{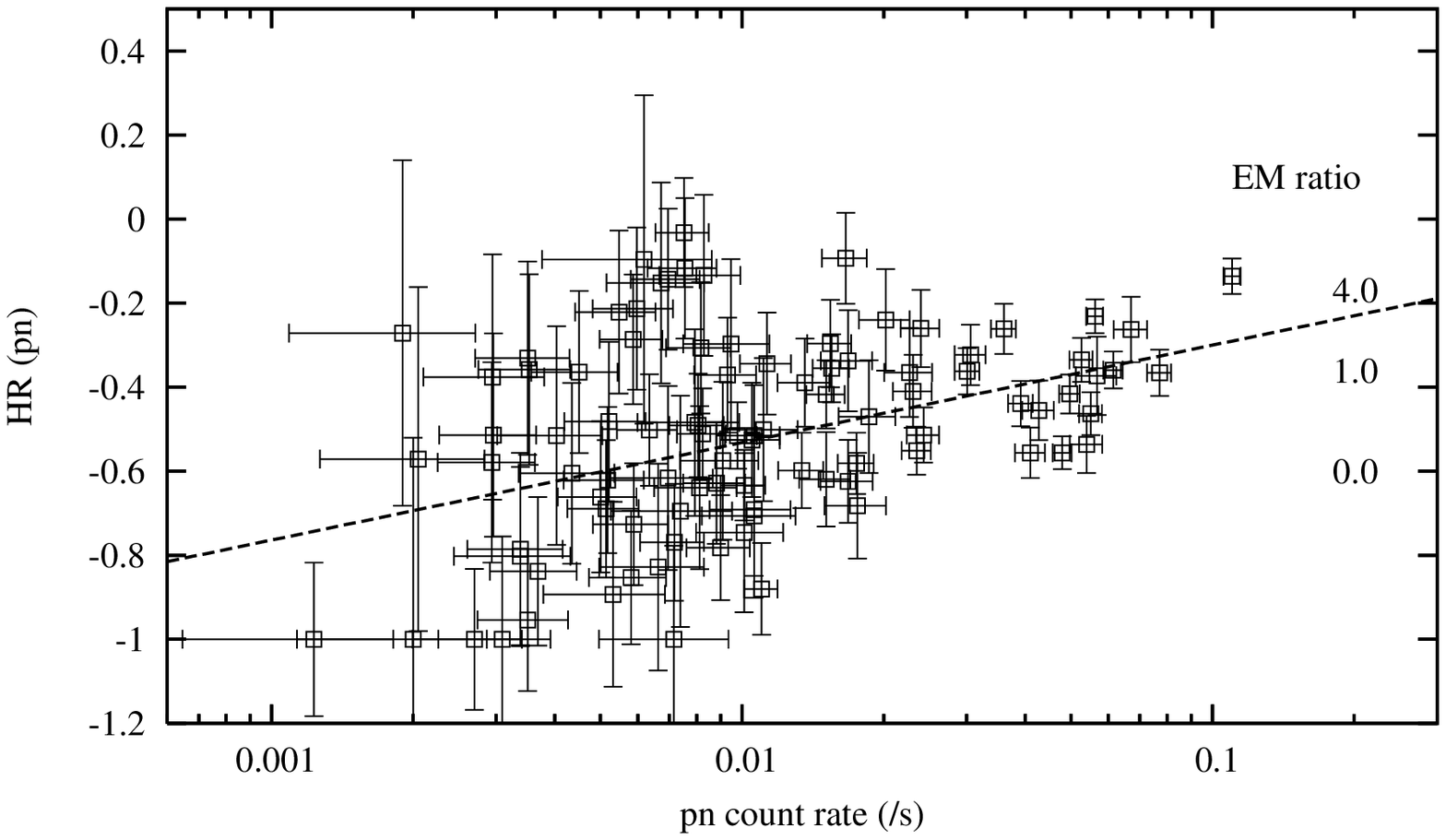}
\includegraphics[width=84mm]{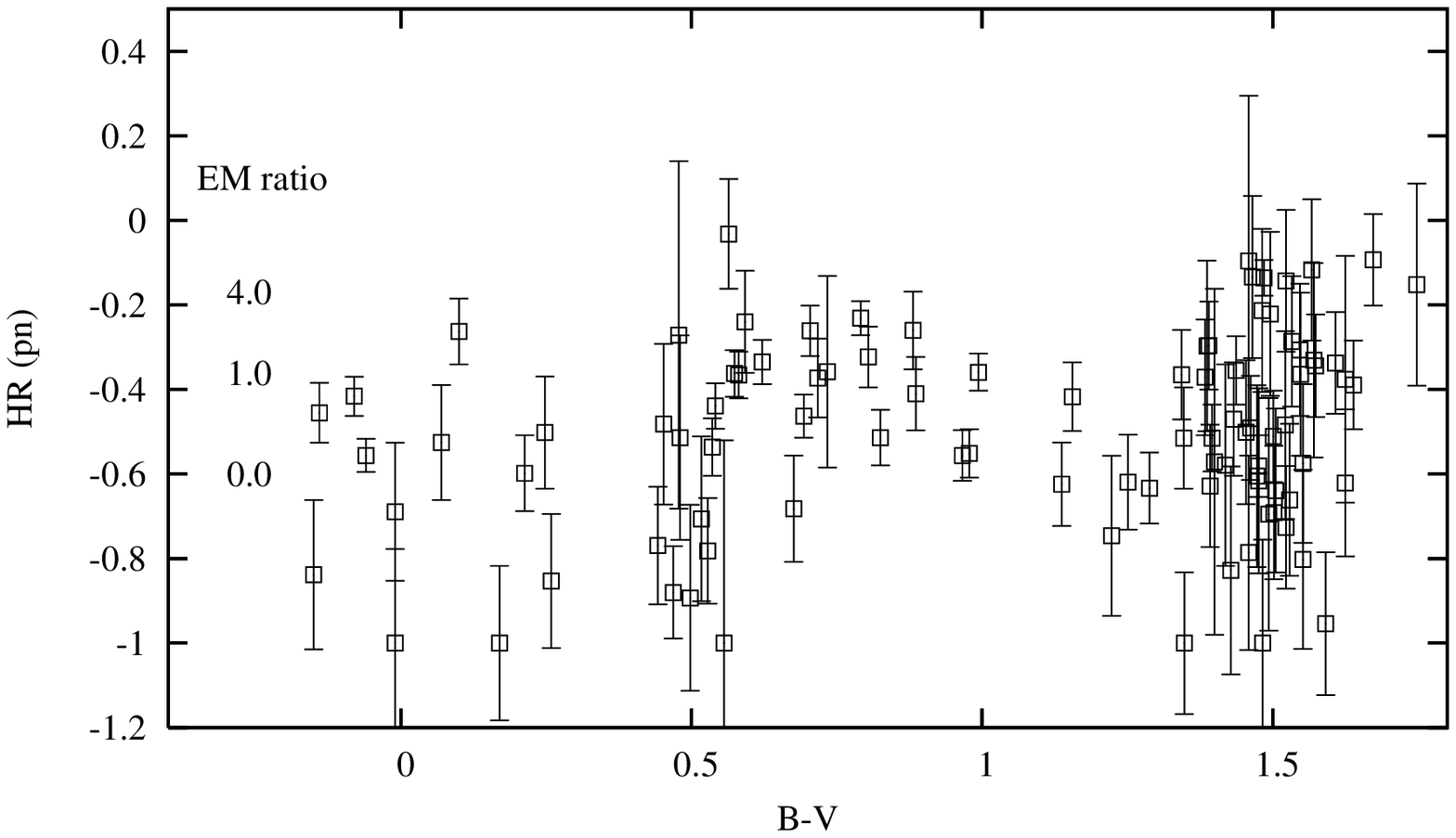}
\includegraphics[width=84mm]{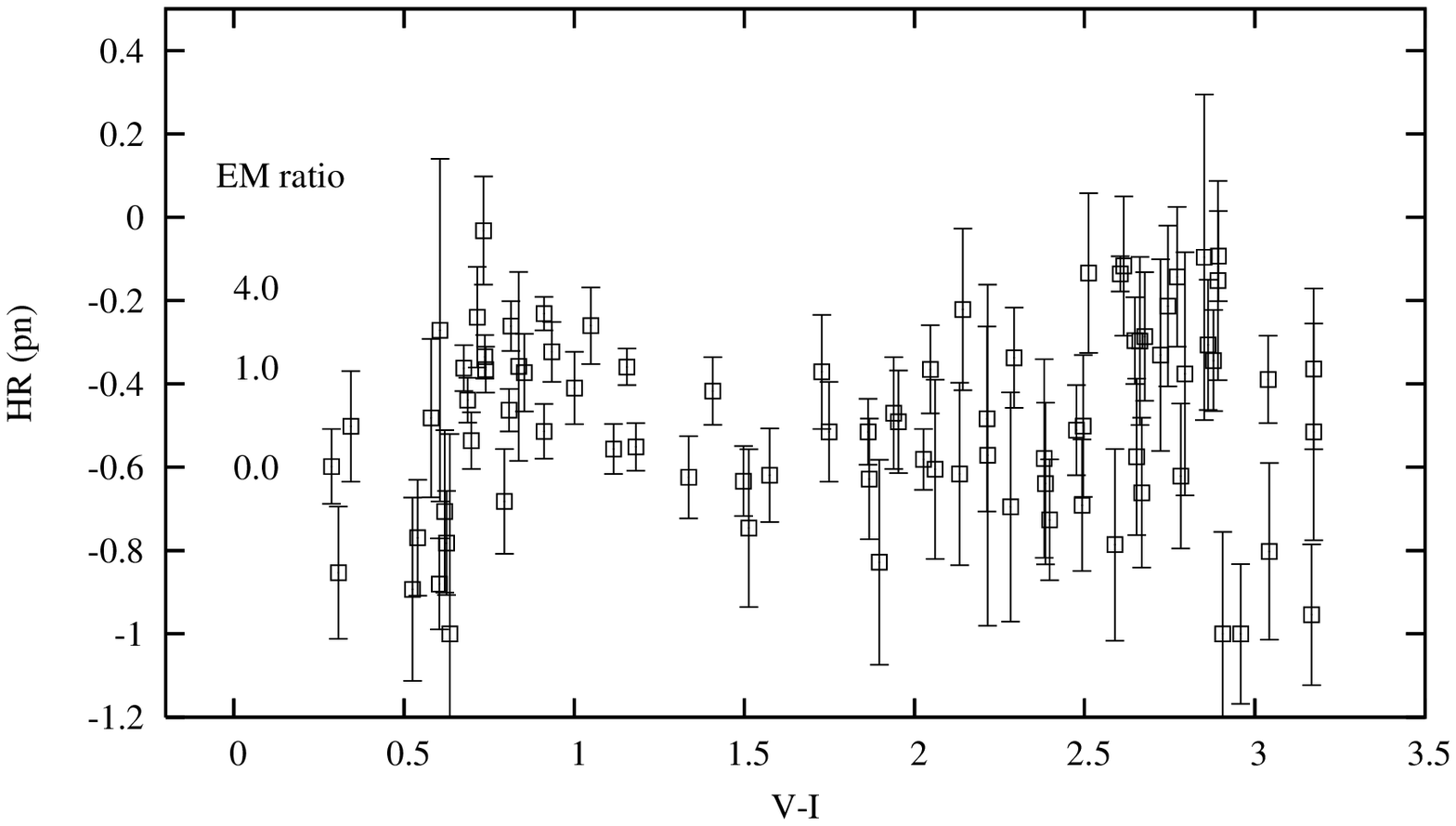}
\caption{The hardness ratio from the pn detector (see
  section~\ref{hardness}) for NGC 2547 members versus overall count
  rate in the pn detector, versus $B-V$ and versus $V-I$. Indicated against
  the y-axis of each plot are the emission measure ratios of the hot to cold
  components that would provide that observed hardness ratio in a
  simple 2-T coronal model with $T_{1}=0.6$\,keV and
  $T_{2}=1.5$\,keV. The dashed line in the upper plot is a simple
  straight line fit to the data.
}
\label{hardnessfig}
\end{figure}

Due to low numbers of counts, spectral fitting barely provides
constraints on the temperature distribution of the coronal plasma even
in the brightest NGC~2547 sources.  Nevertheless, there is sufficient
evidence here that these coronae follow the often observed pattern that,
given sufficient statistics, a multi-component thermal model fits the
data better than a single temperature. To extend our
analysis to fainter sources, the hardness ratio, defined as
$(H-S)/(H+S)$, where $S$ is the count rate in the 0.3--1.0\,keV band
and $H$ is the count rate in the 1.0--3.0\,keV band, was modelled in terms of a 2-T
corona.  The purpose is to provide a physical interpretation for
any trends in the spectral distribution with type of star or overall
X-ray activity level and also to estimate how any spectral changes
might influence the conversion from X-ray count rates into fluxes
(see Section~\ref{xrayactivity}).

Figure~\ref{hardnessfig} shows plots of the hardness ratios in the pn
detector versus colour and also versus count rate.
The plots for the MOS detectors are similar, but noisier, and are
omitted for brevity.  These
hardness ratios were simulated using a 2-T model, with fixed temperatures of
$T_{1}=0.6$\,keV, $T_{2}=1.5$\,keV, a fixed metallicity of $Z=0.3$ and
$N_{\rm H}=3.0\times10^{20}$\,cm$^{-2}$. This simple, though not
unique, model is justified by the spectral fitting results in
section~\ref{spectra}.  The emission measure ratio of the two
components is altered to generate a given hardness ratio. The required
emission measure ratio (hot/cold) is indicated against the y-axes of
Figure~\ref{hardnessfig}.

Recent work on high quality X-ray spectra of nearby stars with varying
activity levels has identified a trend of increasing emission
measure weighted mean coronal temperature with X-ray activity
(e.g. Telleschi et al. 2005). The same trend, albeit with poorer {\it ROSAT}
spectra, has also been identified among stars of the Pleiades cluster
(Gagn\'e, Caillault \& Stauffer 1995a). There is evidence for this in
Fig.~\ref{hardnessfig} in terms of increasing
hardness ratio and therefore hotter coronae as the pn count rate (and
hence X-ray luminosity) rises. A fitted relationship of the form
HR(pn)\,$=0.23(\pm0.04)\times \log_{10}({\rm pn rate}) - 0.07(\pm
0.06)$ (shown as a dashed line) appears a reasonable description.
However, there is also evidence for an
intrinsic scatter in this relationship, especially at pn count rates of
0.005--0.01\,s$^{-1}$. The reduced chi-squared is 2.53, 
indicating that further parameters may be
important. The plots of hardness ratio versus colour reveal why this
may be so. There are two groups of stars with the highest hardness
ratios: at $0.7<V-I<1$ and at $2.5<V-I<3.0$, corresponding to G and M
stars respectively. M stars are smaller than G stars and as we will see
in the next section, tend to have lower peak X-ray
luminosities. Nevertheless, many M stars appear to have hot coronae and
thus it seems that the variable that truly predicts average coronal
temperature may be an activity indicator that is normalised by stellar
surface area, such as the surface X-ray flux or the ratio of X-ray to
bolometric luminosity.

Three other features of the plots are worthy of comment. First, there
seem to be a lack of K stars ($1<V-I<2$) with large hardness ratios. In
fact, given the small number of such objects in the sample, that
conclusion is not statistically sound. The same could be said of the
apparent decrease in hardness ratios as we move from G to F stars
$0.4<B-V<0.6$ (recalling that NGC~2547 has $E(B-V)=0.06$) or
$0.5<V-I<0.7$. However, this trend would match observations in the
field and other clusters that F-stars have significantly cooler coronae
than G-M stars (e.g. Gagn\'e et al. 1995a; Panzera et al. 1999).  Third,
the hardness ratio versus $B-V$ plot adds a further 9 data points for
hotter stars. The distribution of hardness ratios for these stars is
not distinguishable from the rest of the sample. This is important
because these 9 A and B stars have no subphotospheric convection zones
and no massive winds (bar perhaps the hottest object in the sample)
that may be capable of generating X-ray activity. It is widely
hypothesised that the X-ray emission from these stars comes from
lower-mass companions. That the (crude) spectral properties of these
X-ray sources are similar to the bulk of the sample provides support
for this idea.

\section{X-ray activity levels in NGC 2547}
\label{xrayactivity}

\begin{figure}
\includegraphics[width=84mm]{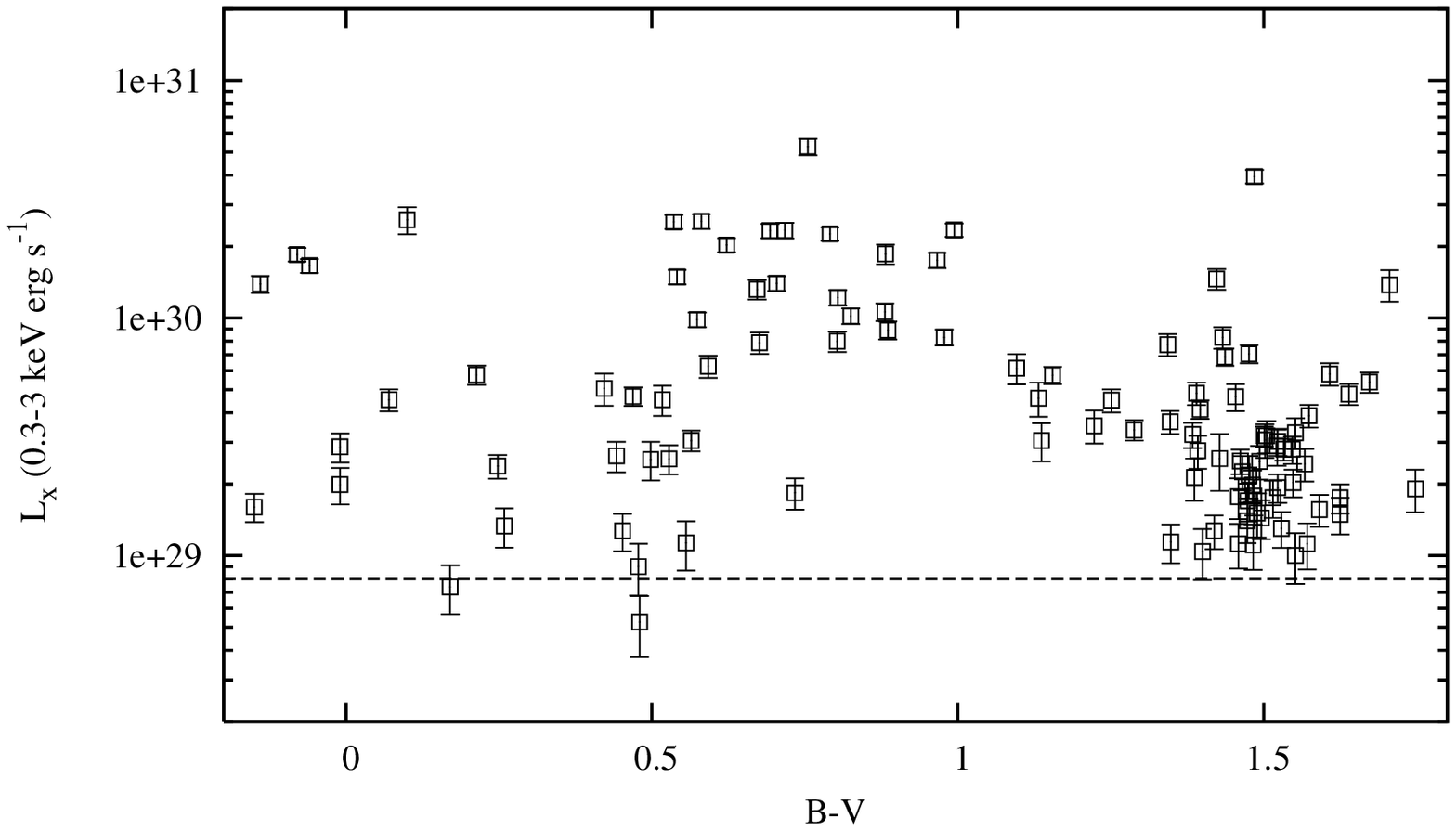}
\includegraphics[width=84mm]{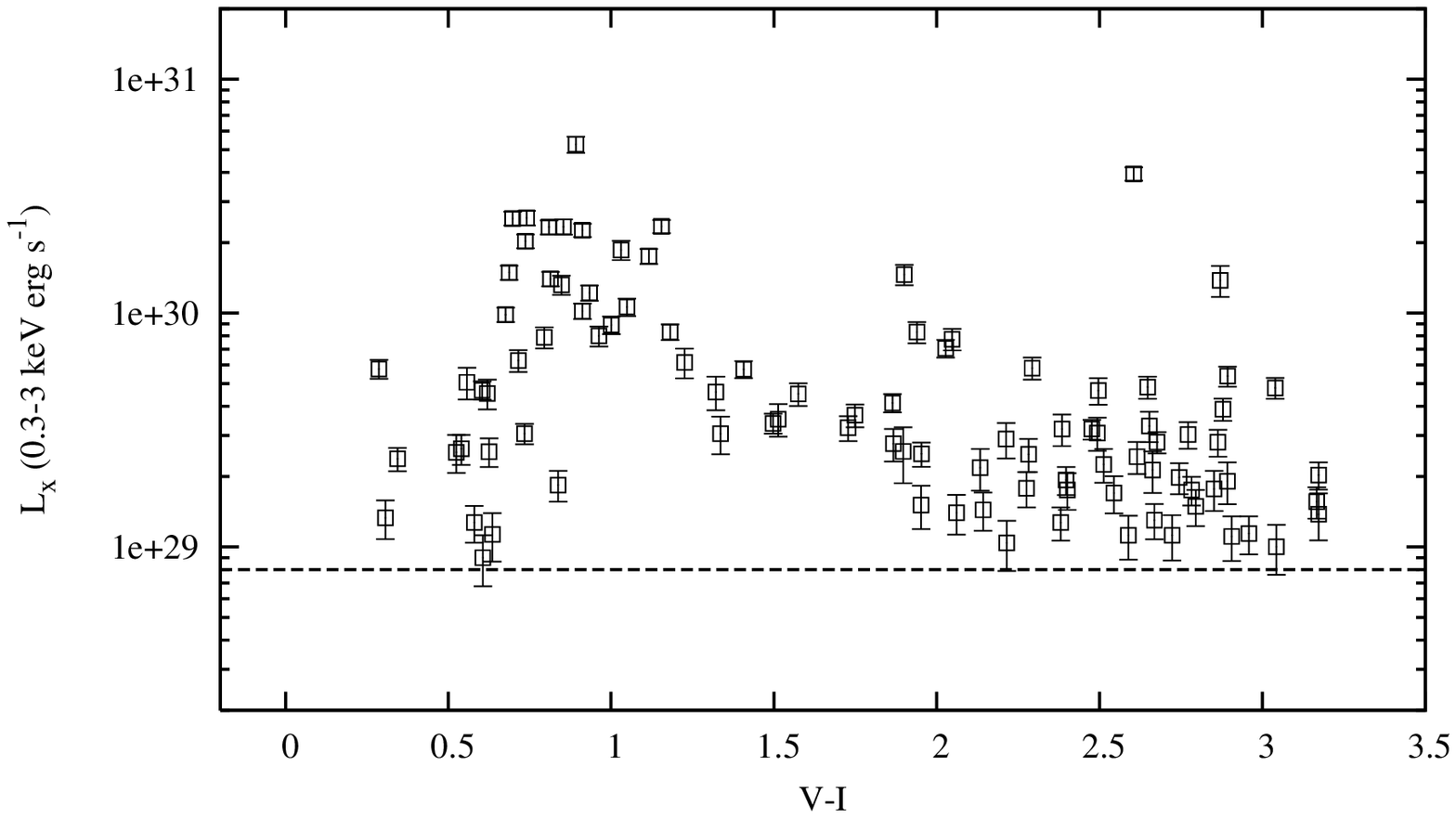}

\caption{X-ray luminosity (0.3--3.0\,keV, assuming a distance
  of 417\,pc) as a function of $B-V$ and $V-I$ for NGC 2547 candidate
  members. The dashed lines
  indicate the approximate sensitivity limit of the X-ray observations at the
  centre of the X-ray field.
}
\label{lxfig}
\end{figure}

\begin{figure}
\includegraphics[width=84mm]{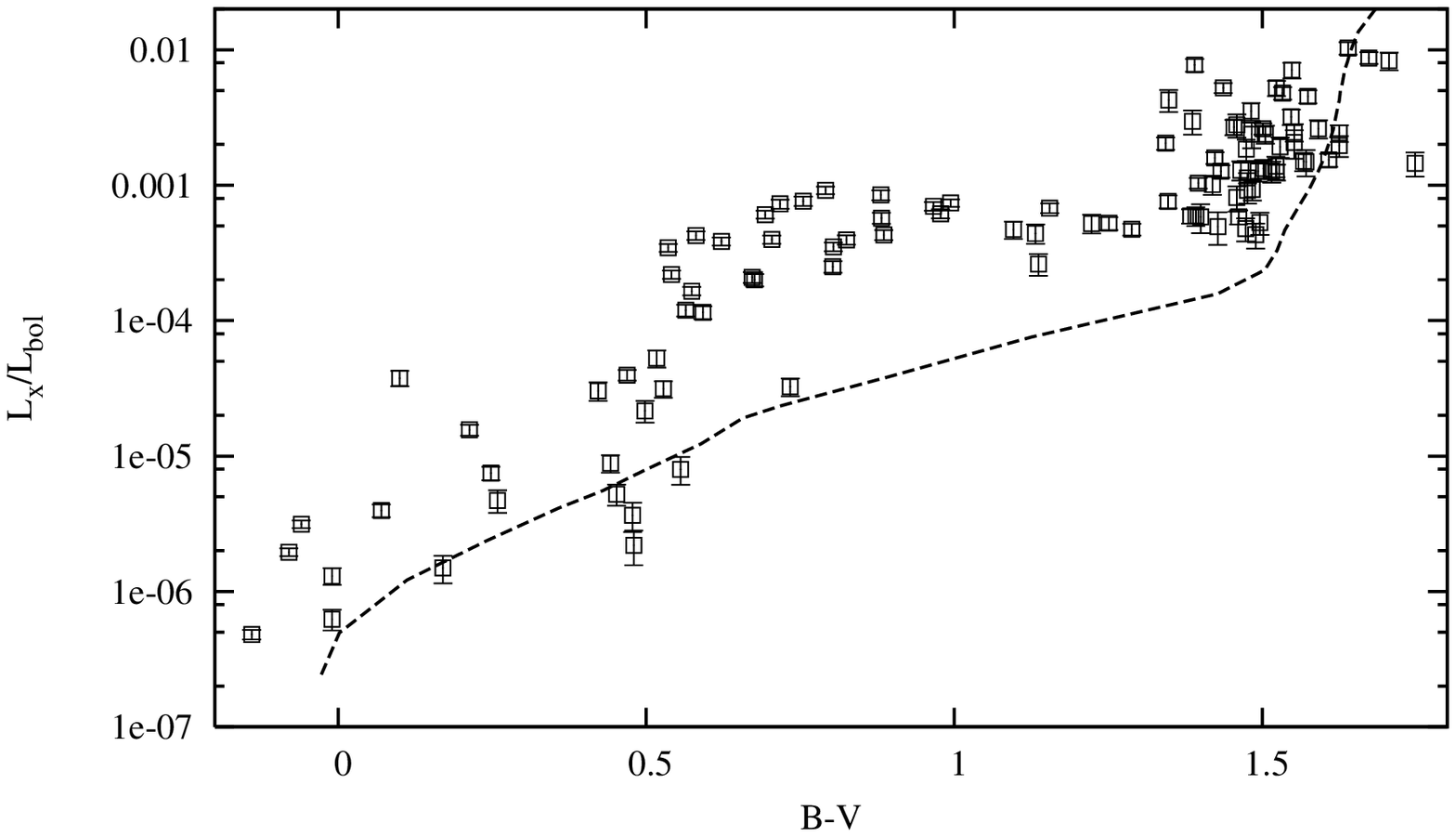}
\includegraphics[width=84mm]{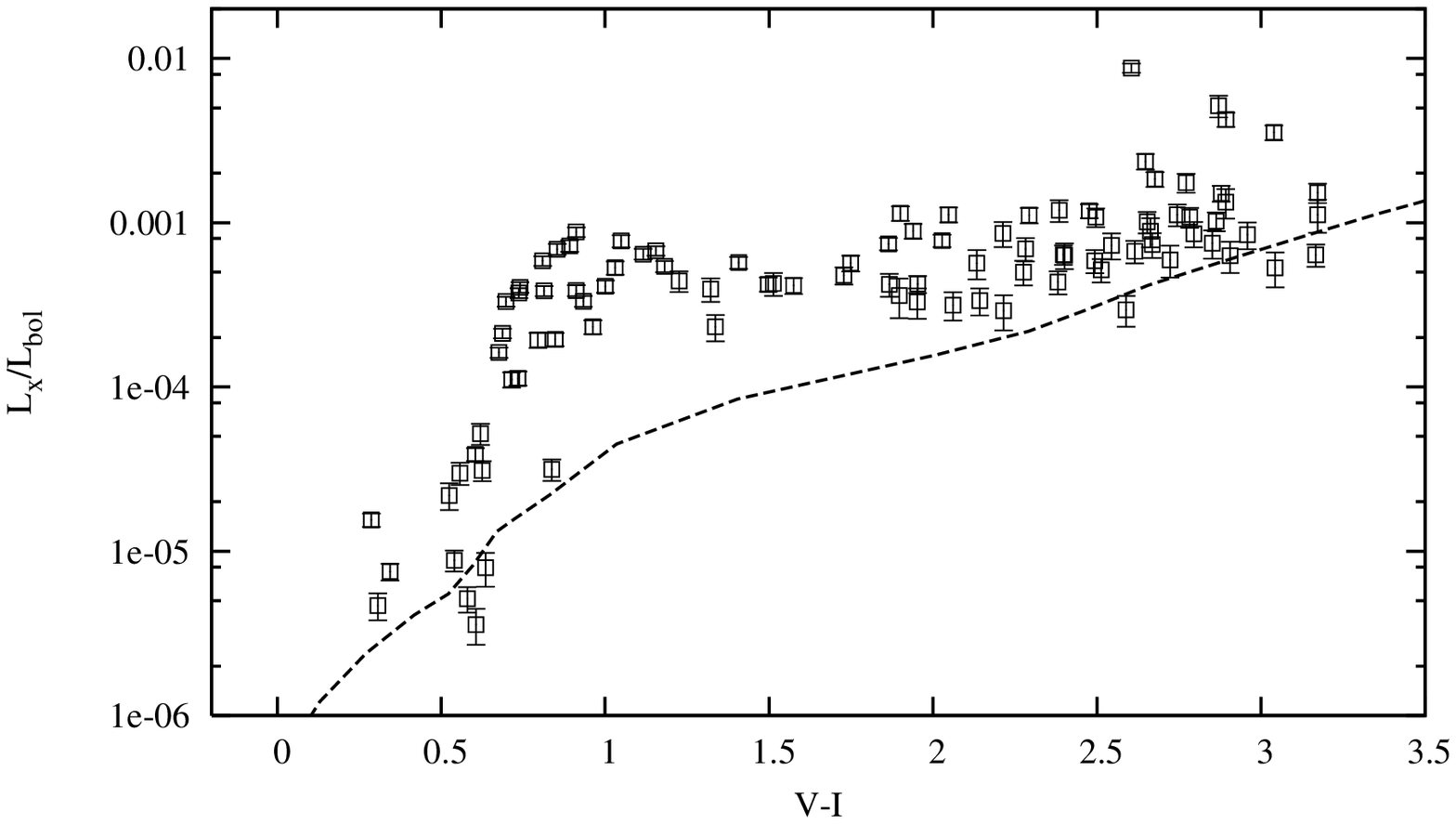}
\caption{The ratio of X-ray to bolometric luminosities for stars in NGC
  2547. The dashed lines representing the approximate sensitivity limit of the
  X-ray observations for a star on the cluster isochrone at the centre of the
  X-ray field.}
\label{lxlbolfig}
\end{figure}

\subsection{X-ray luminosity}
\label{xraylum}

To look at the overall level of activity in the cluster as a function
of colour/spectral type/$T_{\rm eff}$ requires a means of calculating X-ray
luminosity and the ratio of X-ray to bolometric luminosity.
For simplicity, and also because the available spectral information is
too sparse to do otherwise for most sources, a uniform
conversion factor between X-ray count-rate and unabsorbed X-ray flux in
the 0.3--3.0\,keV range was assumed for each of the three cameras. To do this,
the mean hardness ratio in each of the cameras was compared with
the two-component spectral model predictions discussed in
section~\ref{hardness}.  From this it seems appropriate to adopt an
average X-ray spectrum consisting of a 2-T model
with an emission measure ratio of hot/cool$=0.7$ and 
$Z=0.3$. This yields conversion factors (from counts in the
0.3--3.0\,keV range to a flux in the 0.3--3.0\,keV range) of
$1.86\times10^{-12}$, $6.72\times10^{-12}$ and
$6.59\times10^{-12}$\,erg\,cm$^{-2}$ per count for the pn, MOS1 and
MOS2 detectors respectively\footnote{To compare with luminosities given
in other flux bands in the literature, our $L_{\rm x}$ values should be
multiplied by 1.21 to obtain $L_{\rm x}$ in the 0.1--2.4\,keV range and
by 0.91 to obtain $L_{\rm x}$ in the 0.5--8.0\,keV range}.

There is little uncertainty injected by assuming a uniform conversion
factor regardless of hardness ratio. The variation in the conversion
factor is less than $\pm 6$ per cent for the most extreme hardness
ratios found in our data - with harder coronae leading to larger
conversion factors. We have also tested variations in the assumed metal
abundances between $Z=0.1$ and $Z=1.0$. Changing the metal abundance
while keeping other parameters fixed leads to conversion factor
variations of $\pm 5$ per cent, with more metal-poor coronae yielding
larger conversion factors.  Altering the absorbing column from our
assumed value of $3\times10^{20}$\,cm$^{-2}$ to $10^{20}$\,cm$^{-2}$ or
$10^{21}$\,cm$^{-2}$ leads to conversion factors that are only 6 per
cent smaller or 25 per cent larger respectively.

Figure~\ref{lxfig} shows the X-ray luminosity (0.3--3.0\,keV) of cluster
members versus $B-V$ and $V-I$. A weighted mean $L_{\rm x}$ was used where
measurements from more than one EPIC detector were
available. A distance of 417\,pc was assumed. 
Uncertainties in $L_{\rm x}$ have been estimated using the
count rate errors, but we added a further 10 per cent
systematic error in quadrature for each EPIC count rate to cover
uncertainties in the detector PSF modelling in the SAS {\sc
  edetect\_chain} task (e.g. Saxton 2003).  
Figure~\ref{lxlbolfig} shows similar plots for
the distance-independent ratio of X-ray to bolometric luminosity,
$L_{\rm x}/L_{\rm bol}$ as a function of $B-V$ and $V-I$. The
bolometric corrections are those used to produce the isochrones in
Figs.~\ref{vbvcmd} and~\ref{vvicmd} and are fully described in Naylor
et al. (2002). The bolometric correction--colour relationship becomes
very uncertain for $B-V>1.4$ and the $V-I$ plot is to be preferred in
those cases. $L_{\rm x}$ and $L_{\rm x}/L_{\rm bol}$ values are listed
in Table~\ref{counterparts}.

The dashed lines in Figs.~\ref{lxfig} and~\ref{lxlbolfig} represent an
approximate sensitivity limit for the EPIC observations, based on an
object observed by all 3 detectors at the centre of the field. This
limit is $L_{\rm x}\simeq 8.0\times10^{28}$\,erg\,s$^{-1}$,
corresponding to about 13 detected source counts in the MOS detectors
and about 23 source counts in the pn detector.  The detection threshold
rises to $\geq 1.5\times 10^{29}$\,erg\,s$^{-1}$ for sources more than
10 arcminutes from the centre of the X-ray image.

Figures~\ref{lxfig} and~\ref{lxlbolfig} reveal a pattern of X-ray
activity that has become familiar in older open clusters (e.g. the
Pleiades -- Stauffer et al. 1994; NGC~2516 -- Jeffries et al. 1997), but is
particularly well delineated here. X-ray emission is seen at all
spectral types. Early type stars have a wide spread in activity levels;
there is a trend of increasing $L_{\rm x}$ and $L_{\rm x}/L_{\rm bol}$
moving from F through to G stars.  The X-ray activity of G and K stars
has less than a factor of five spread and appears saturated at $L_{\rm
x}/L_{\rm bol}\leq 10^{-3}$. All the detected M stars ($V-I >2.2$)
have activity levels that are as high (or even higher), but it is not
immediately apparent whether a population of lower activity objects
might be present, as the least active detected objects lie at the X-ray
detection sensitivity threshold.

\subsection{Completeness and contamination}
\label{complete}

Whether the X-ray selected sample of NGC~2547 members is complete or
whether it is contaminated by X-ray active field stars, as a function
of spectral type, has been discussed at length by Jeffries \& Tolley
(1998) and Jeffries et
al. (2000). For completeness we briefly
re-visit those arguments which are based on the appearance of
Figs.~\ref{vbvcmd}, \ref{vvicmd} and~\ref{lxlbolfig}.

The first point to make is that Figs.~\ref{vbvcmd} and~\ref{vvicmd}
indicate that the vast majority of photometric cluster {\em candidates}
within the {\it  XMM-Newton} field of view are X-ray detected. The
exceptions are a handful of early-type objects, a clump of objects at
$B-V\simeq 1$, $V-I\simeq 1.2$ and an increasing number of undetected
objects at $V-I>2.8$. The latter group might simply be explained in
terms of a gradual decrease in the X-ray luminosity of cluster members
with decreasing $T_{\rm eff}$, which intercepts the detection sensitivity limit
(for the centre of the field) at around $V-I\simeq 3.2$. Hence the
average X-ray properties of the M-type NGC~2547 members will be biased upwards
by the non-detection of less active cluster members with $V-I>2.8$.

The X-ray quiet objects at intermediate colours are almost certainly
background giants which thrust a ``finger'' of contamination through
the NGC~2547 colour-magnitude loci at these colours. 
This interpretation is supported by
Fig.~\ref{lxfig} which shows that the X-ray census in NGC~2547 is very
likely to be complete for G and K stars ($0.6<B-V<1.3$, $0.7<V-I<2.2$),
because the sensitivity threshold is some way below the least active
detected cluster members at these colours. Of course there {\it could}
be a small tail of G/K stars with very low activity, but there would
only be scope for such a population at the colours occupied by the
giant contaminants. In particular, {\it all} photometric candidates
with $1.4<V-I<2.5$ were detected by {\it XMM-Newton}.

Contamination of our X-ray selected sample is likely to be very
light. Not only would the contaminants have to be at a small range of
distances consonant with appearing in colour-magnitude diagrams at the
same position as NGC~2547 members, but they would need to possess
comparable levels of X-ray activity. F-K type stars have X-ray activity
levels that are a sharply declining function of age (see
section~\ref{evolution}), so few field stars would jointly meet these
criteria. Jeffries et al. (2000) found that 23/24 photometrically
selected, X-ray active NGC~2547 candidates were probable members on the
basis of their radial velocities. Higher levels of contamination are
possible among M-stars as these have more slowly decaying X-ray
activity and a higher spatial density in the field.  However, Jeffries
\& Oliveira (2005) have shown that $\simeq 90$ per cent of
candidate cluster M-dwarfs selected by photometry alone turn out to be
members when investigated spectroscopically.  As the X-ray properties
of any small number of contaminants are likely to be similar to the
cluster members, they should not unduly bias the results and
conclusions.

\subsection{Coronal temperature variation with activity}
\label{tactivity}

\begin{figure}
\includegraphics[width=84mm]{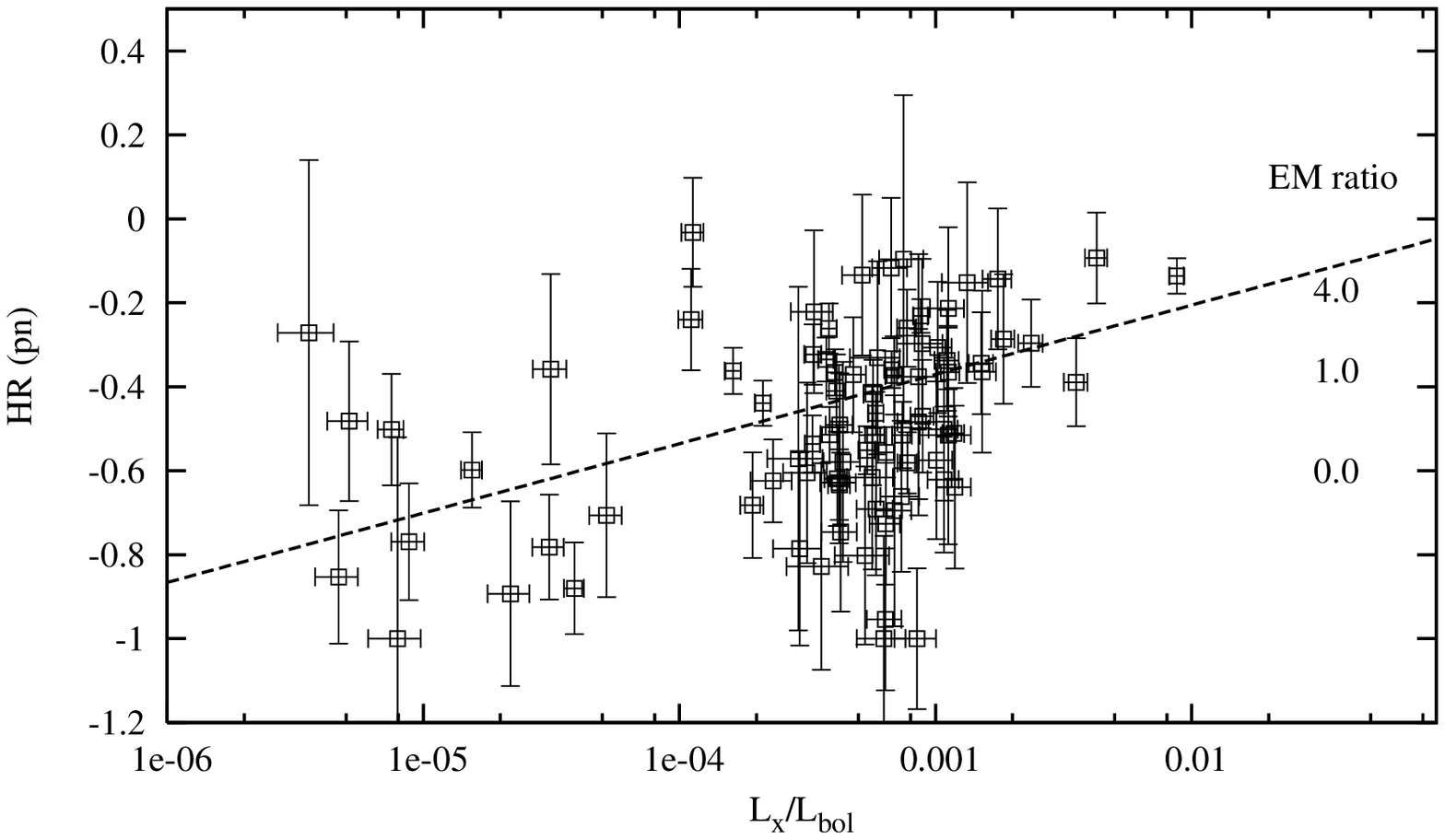}
\includegraphics[width=84mm]{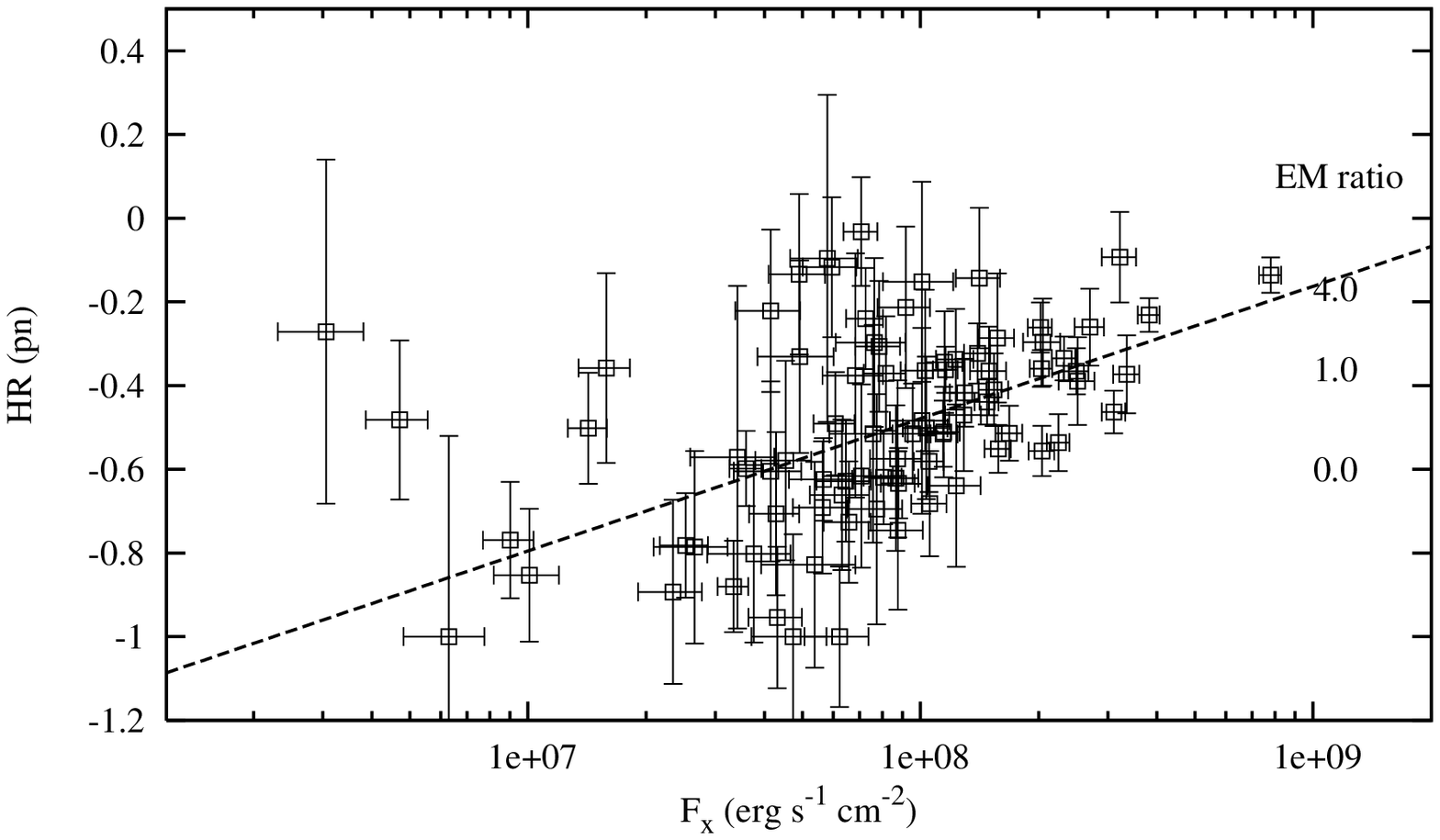}
\caption{Hardness ratio, measured by the pn detector, as a function of
  X-ray activity. The dashed lines are minimised chi-squared fits -- see text.
}
\label{hrlxlbol}
\end{figure}

The top panel of Fig.~\ref{hardnessfig} demonstrates that more luminous
X-ray coronae are hotter on average. However, the scatter in this plot
hints at a dependence on another parameter. Figure~\ref{hrlxlbol} shows
how the pn hardness ratio varies with the activity indicators $L_{\rm
x}/L_{\rm bol}$ and the surface X-ray flux, $F_{\rm x}$ (radii as a
function of colour were obtained from the evolutionary models used to
obtain the isochrone in Fig.~\ref{vvicmd}). Only those stars
with a $V-I$ measurement were used, to avoid any scatter introduced by the
inclusion of early-type stars in which the dominant optical source is
probably not the source of the X-rays.

Curves of the form HR(pn)\,$= 0.17(\pm 0.03)\times
\log_{10} (L_{\rm x}/L_{\rm bol}) + 0.13(\pm 0.10)$ and
HR(pn)\,$=0.32(\pm 0.04)\times \log_{10} F_{\rm x} - 3.00(\pm 0.33)$
were fitted,
which do indicate significant correlations between hardness ratio and
X-ray activity. According to the 2-T modelling of the
hardness ratio (see section~\ref{hardness}) this corresponds to an
emission measure ratio of hot to cool plasma which changes from about
zero at $L_{\rm x}/L_{\rm bol}\simeq 10^{-5}$, $F_{\rm x}\simeq
3\times10^{7}$~erg~s$^{-1}$~cm$^{-2}$ to unity at $L_{\rm x}/L_{\rm
bol}\simeq 10^{-3}$, $F_{\rm x}\simeq
3\times10^{8}$~erg~s$^{-1}$~cm$^{-2}$.  However, the reduced
chi-squared of these fits are 2.47 and 1.95 respectively (with 82
degrees of freedom). Both the significance of the correlations and the
scatter are almost the same as found for the hardness ratio versus pn
count-rate relationship in section~\ref{hardness}. Thus $L_{\rm x}$, $L_{\rm
x}/L_{\rm bol}$ or $F_{\rm x}$ could be used to predict how hot a coronae
will be, but are not deterministic in the sense that an rms hardness
ratio scatter of about 0.15 exists at a given $L_{\rm x}$, $L_{\rm
x}/L_{\rm bol}$ or $F_{\rm x}$.
This cannot be explained by the measurement
errors and corresponds roughly to a variation of a factor 2 in the
emission measure ratio of the hot to cool plasma in the 2-T
model.

\subsection{The rotation-activity connection}
\label{rotation}

\begin{figure}
\includegraphics[width=84mm]{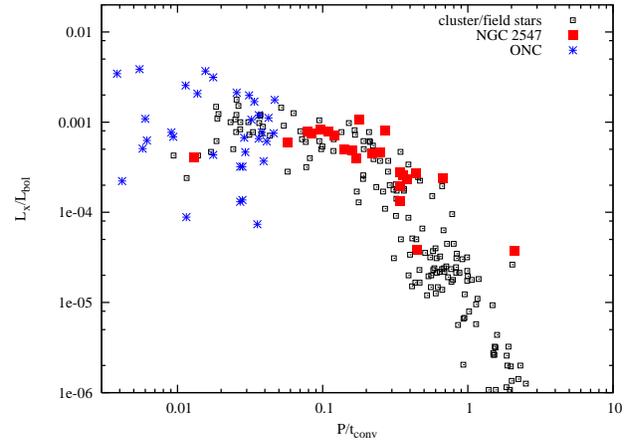}
\caption{X-ray activity (in the range 0.1-2.4\,keV -- see text) as a
function of Rossby number for NGC\,2547 (filled squares), a range of
solar-type stars from the field, Pleiades and other young clusters
(from Pizzolato et al. 2003) and PMS stars with $0.5<M<1.2\,M_{\odot}$ in
the Orion Nebula cluster (from Getman et al. 2005).
}
\label{rossbyplot}
\end{figure}

Jeffries et al. (2000) published projected equatorial velocities ($v
\sin i$) for 23 probable members of NGC~2547 which have $0.62< V-I <
1.33$ and hence spectral types from late F to early K. They found both
fast ($v \sin i > 50$\,km\,s$^{-1}$) and slow ($v \sin i <
10$\,km\,s$^{-1}$) rotators. Evidence for a rotation-activity
relationship was found among the cooler stars of this sample ($V-I >
0.78$) -- all such stars with $v\sin i >15$\,km\,s$^{-1}$ had saturated
levels of X-ray activity. The evidence was less clear among hotter
stars, probably because the rapidly changing convective zone depth as a
function of $T_{\rm eff}$ caused significant scatter in the dynamo
efficiencies at a given rotation rate.

This relationship was re-examined using the {\it XMM-Newton} data, but
adopting a more physical approach that incorporates both rotation
{\em and} the properties of the subphotospheric convection zone. An
approximate Rossby number, the ratio of rotation period to convective
turnover time at the base of the convection zone, was calculated for
each of the 23 NGC 2547 members in the Jeffries et al. (2000) sample.
The relationship between convective turnover time and $B-V$ from Noyes
et al. (1984) was used. This is more appropriate for main sequence
stars, but the NGC 2547 stars considered here are almost at the ZAMS so
this should not lead to serious errors (see Gilliland 1986 and
section~\ref{evolution}). The period was estimated from
the $v\sin i$ and a radius obtained from the D'Antona \& Mazzitelli
(1997) isochrone shown in Fig.~\ref{vbvcmd}. We divided each $v \sin
i$ by $\pi/4$ to correct for an average projection effect.

Figure~\ref{rossbyplot} shows X-ray activity, expressed as $L_{\rm
x}/L_{\rm bol}$ versus the Rossby number.  Similar data are plotted for
a large number of G/K field and cluster stars. These are from the
compilation of Pizzolato et al. (2003), who also used the Noyes et
al. (1984) estimation of convective turnover time, and for a group of
solar-type (0.5--1.2\,$M_{\odot}$) ONC stars with rotation periods from
Getman et al. (2005), where a mean convective turnover time of
250\,years has been assumed (see Preibisch et al. 2005). This latter
assumption may introduce at most a factor of two {\it horizontal}
scatter in the plot.  To provide an accurate comparison, the NGC~2547
X-ray fluxes have been increased by a factor of 1.21 to match the
0.1-2.4\,keV range of the Pizzolato et al. fluxes. The published
0.5--8.0\,keV ONC X-ray fluxes were increased by a factor of 1.38 (see
Preibisch \& Feigelson 2005).

The NGC~2547 objects fit the pattern defined by other young clusters
and field stars perfectly, with a scatter of about a factor of two
about a mean relationship. $L_{\rm x}/L_{\rm bol}$ increases from
$\simeq 5\times 10^{-5}$ in the slow rotators or those stars with thinner
convection zones and shorter convective turnover times, up to a peak
level (in the 0.1--2.4\,keV band) of $L_{\rm x}/L_{\rm
bol}\simeq 10^{-3}$.  This contrasts with the level of $10^{-3.3}$
found by Jeffries \& Tolley (1998), a discrepancy which 
is investigated and explained in section~\ref{comphri}.
There is one object in NGC~2547 which has a $v\sin i$ of at least
160\,km\,s$^{-1}$, a very small Rossby number, and which seems to have a
lower ``super-saturated'' level of X-ray activity. Similar objects have
been found in the IC~2391 and Alpha Per clusters (e.g. Prosser et
al. 1996; Randich 1998).  In contrast, PMS stars in the ONC
 all have Rossby numbers that would put them in the saturated or
super-saturated regime, but show a much larger scatter ($\simeq 2$ orders of
magnitude) in X-ray activity at a given Rossby number. Preibisch et
al. (2005) claim that much of this scatter is due to actively accreting
objects, of which there are none in the NGC 2547 sample (Jeffries et
al. 2000).

\section{X-ray variability}

\subsection{Short-term variability}
\label{flare}

\begin{figure*}
    \centering
    \begin{minipage}[t]{0.45\textwidth}
    \includegraphics[width=71mm]{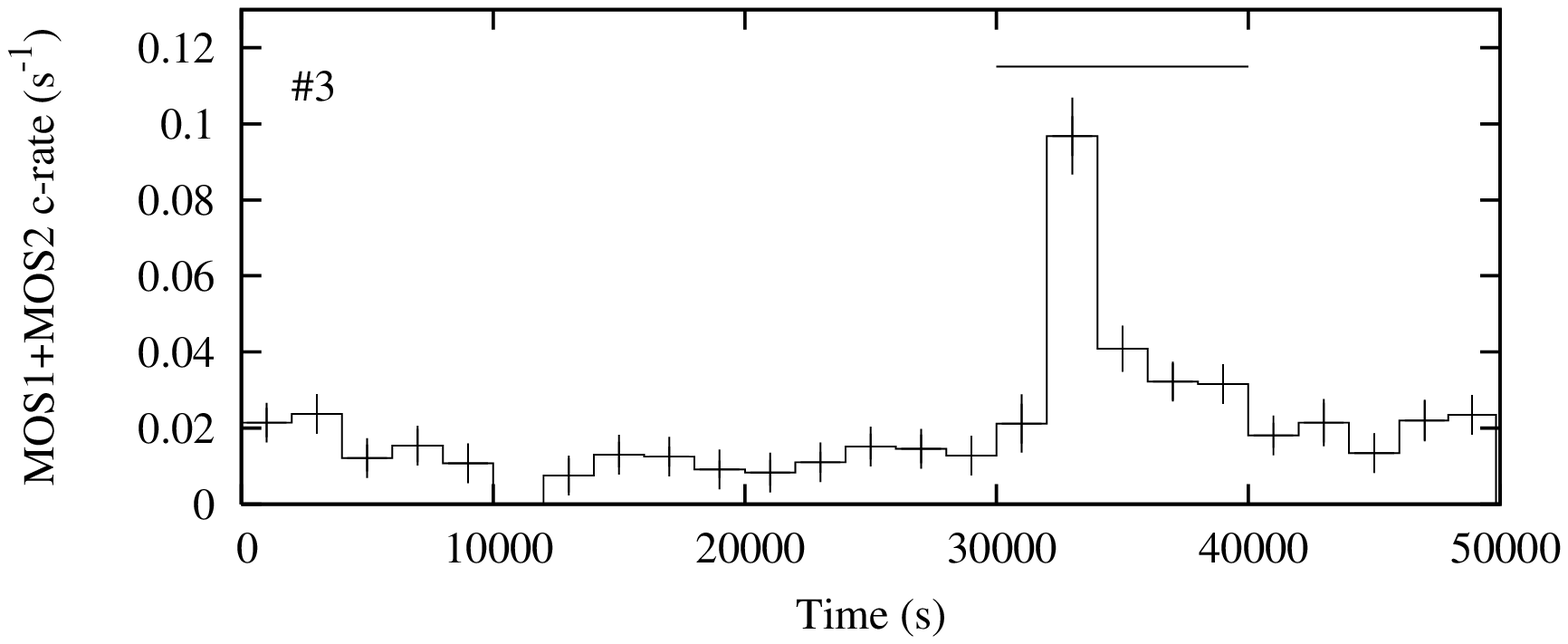}
    \includegraphics[width=71mm]{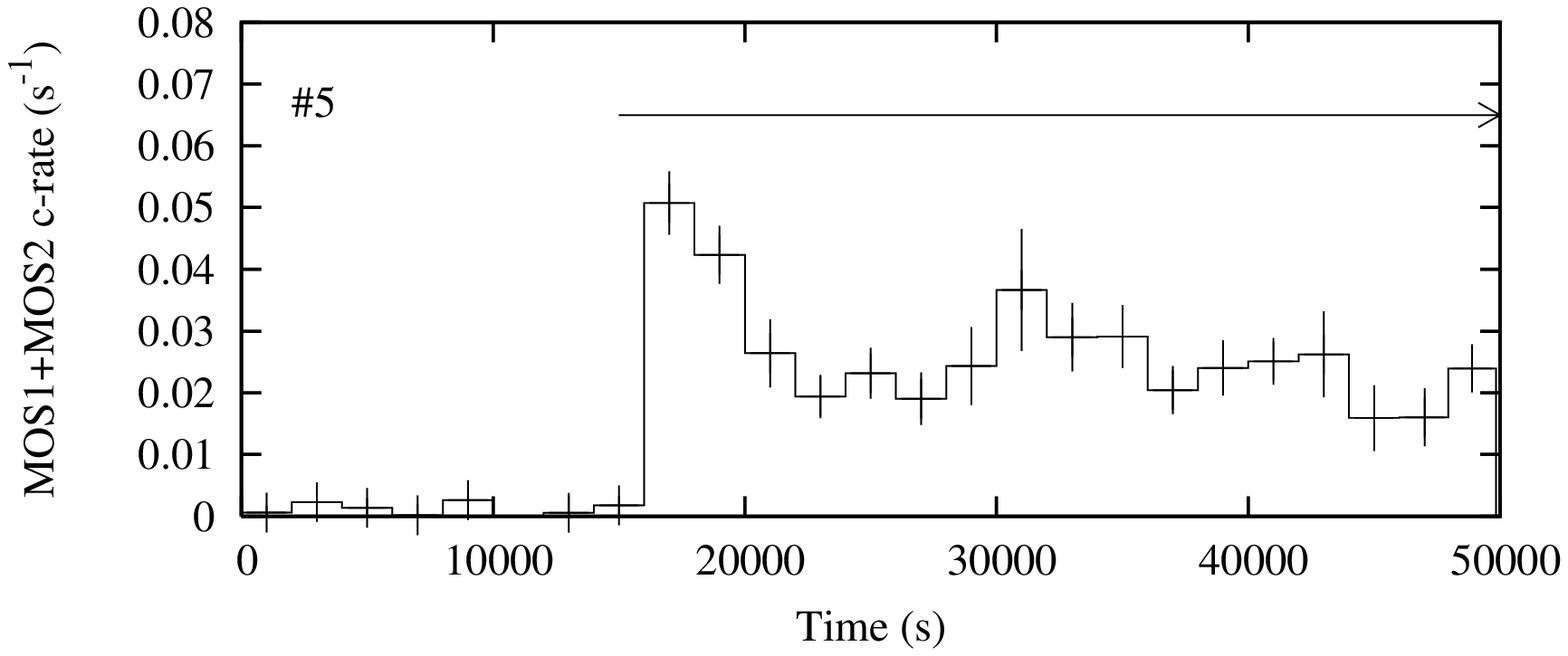}
    \includegraphics[width=71mm]{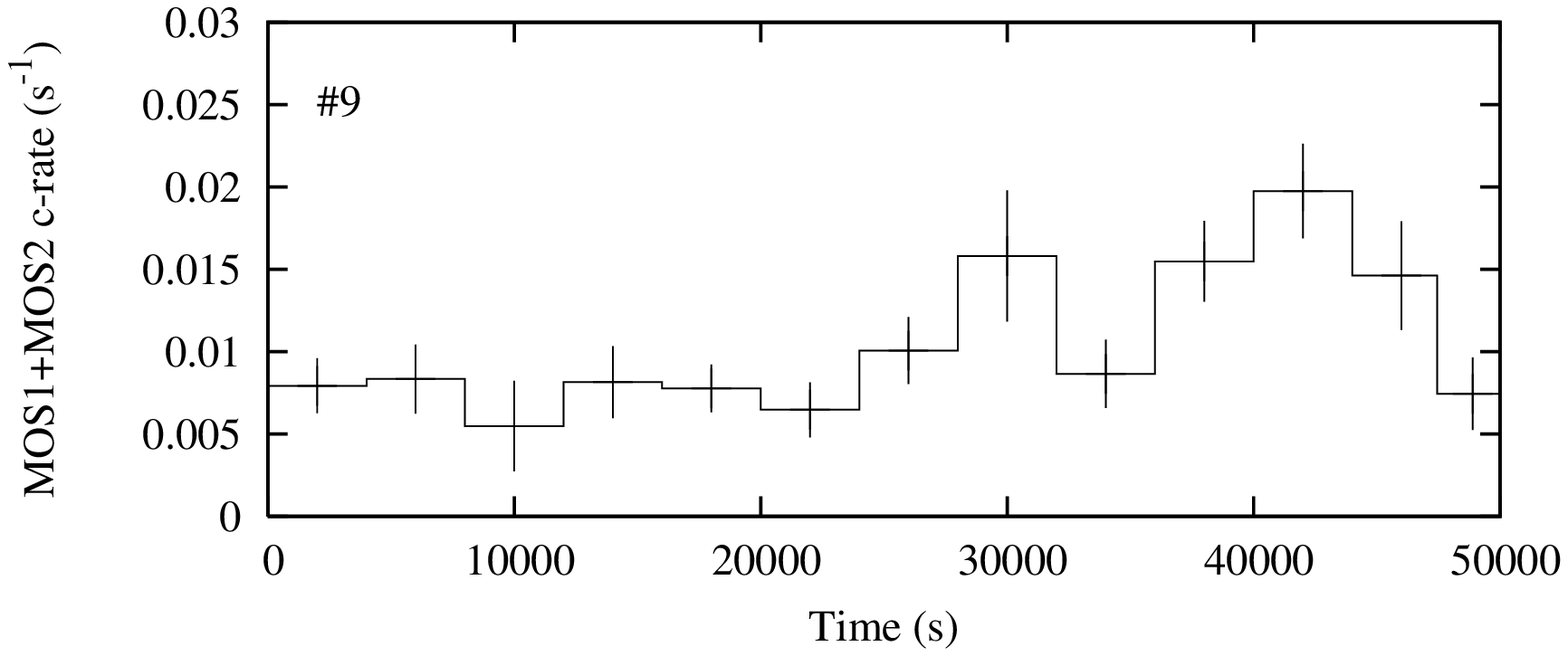}
    \includegraphics[width=71mm]{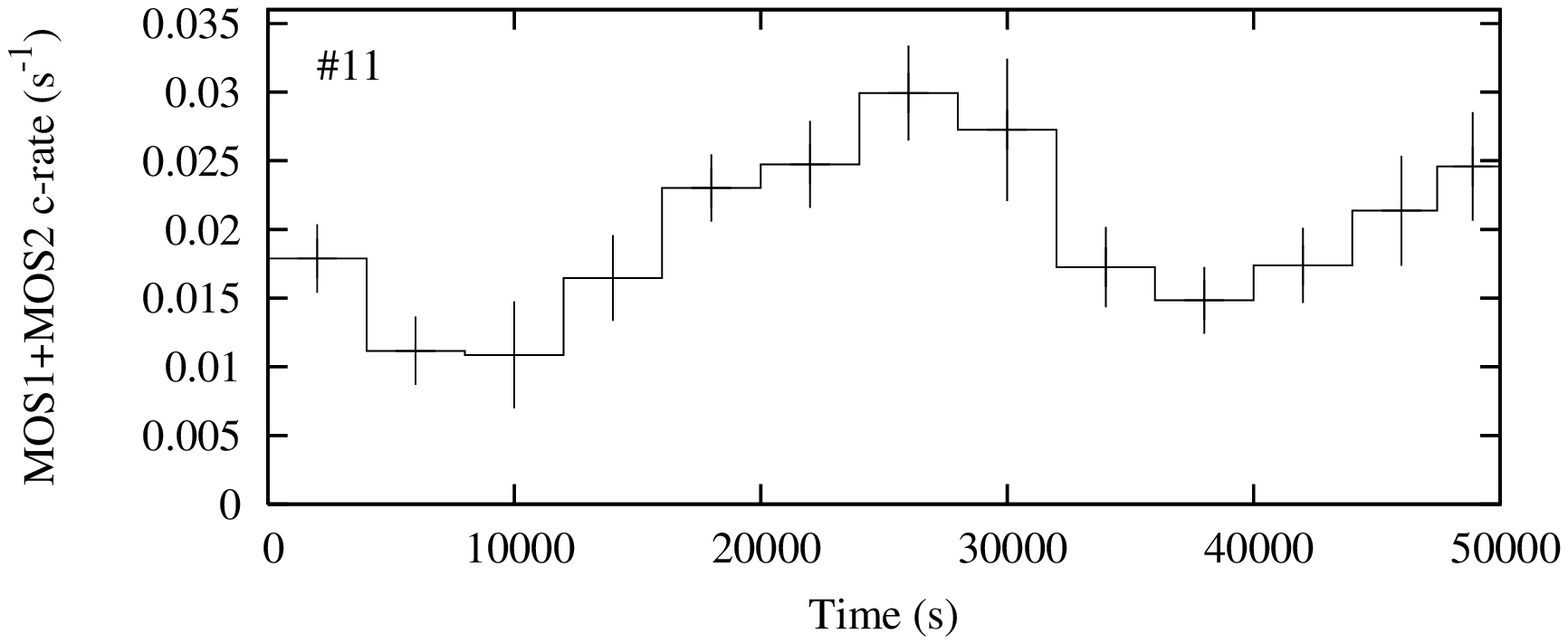}
    \includegraphics[width=71mm]{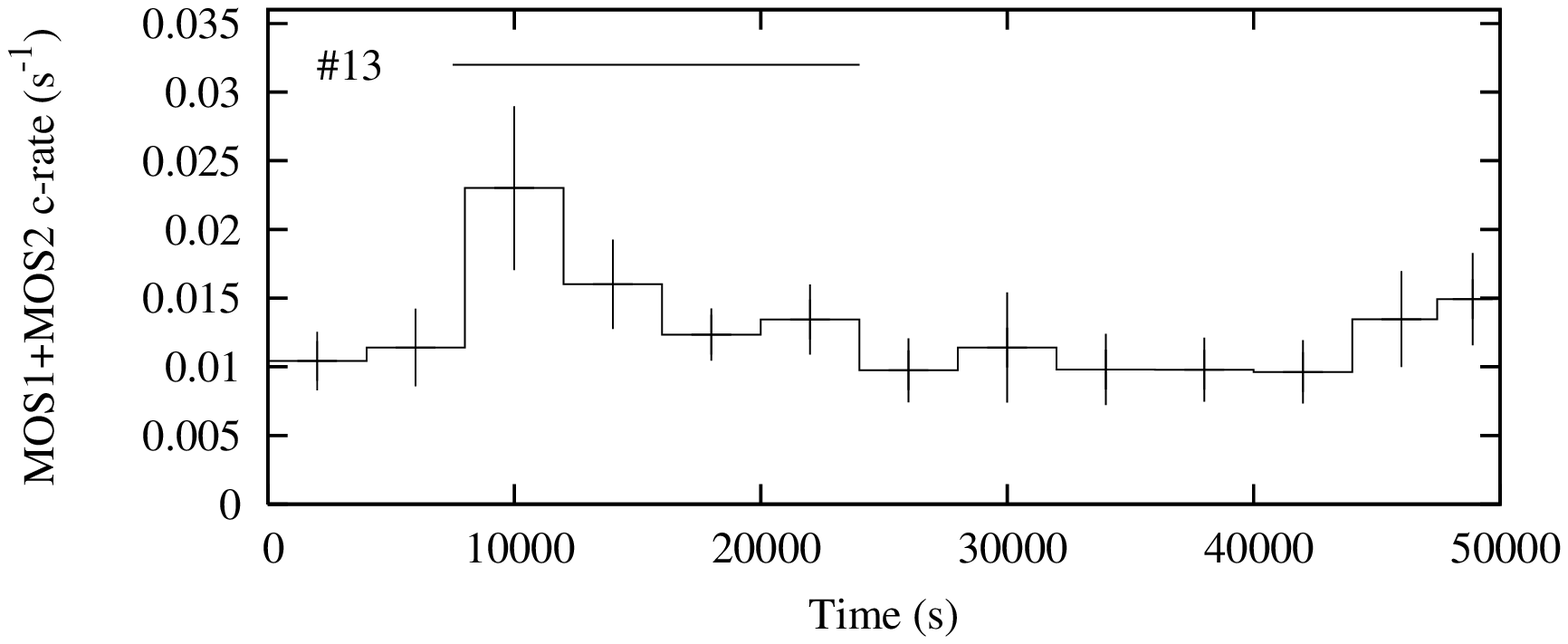}
    \includegraphics[width=71mm]{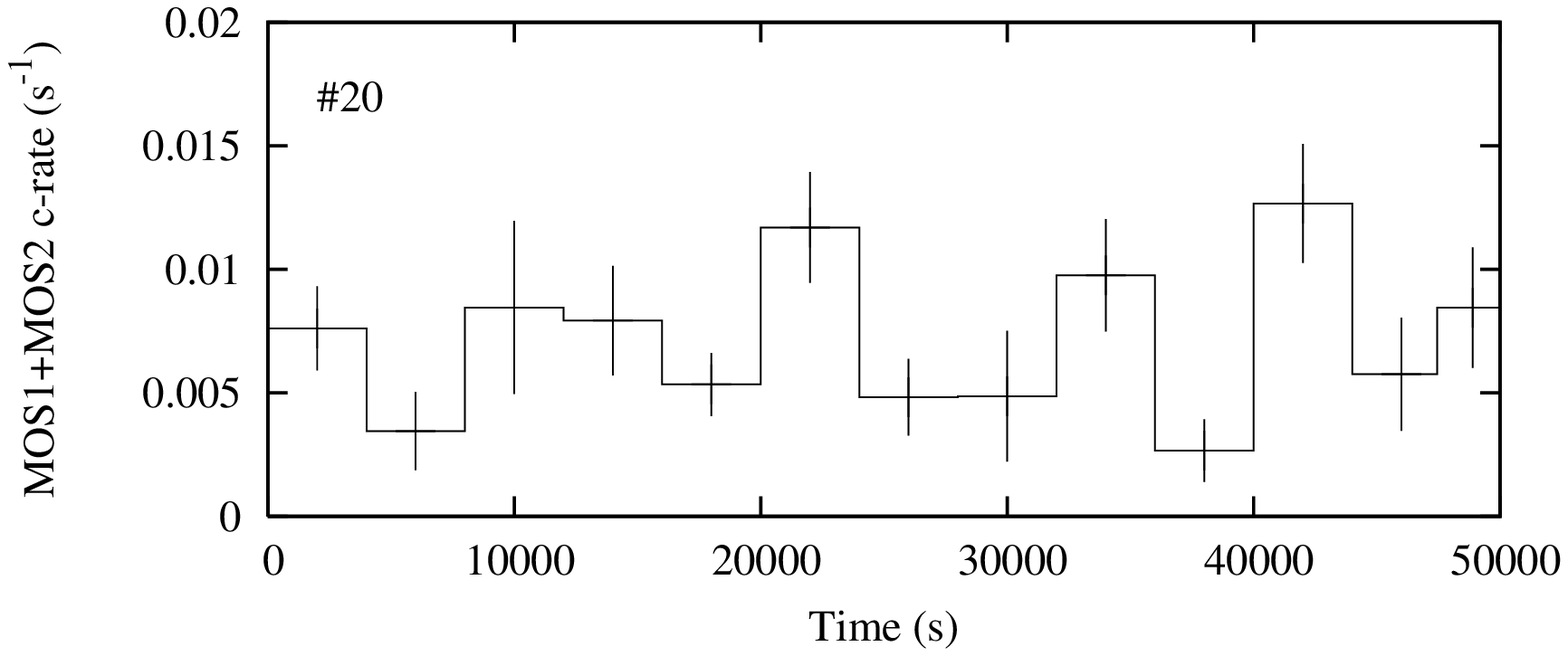}
    \includegraphics[width=71mm]{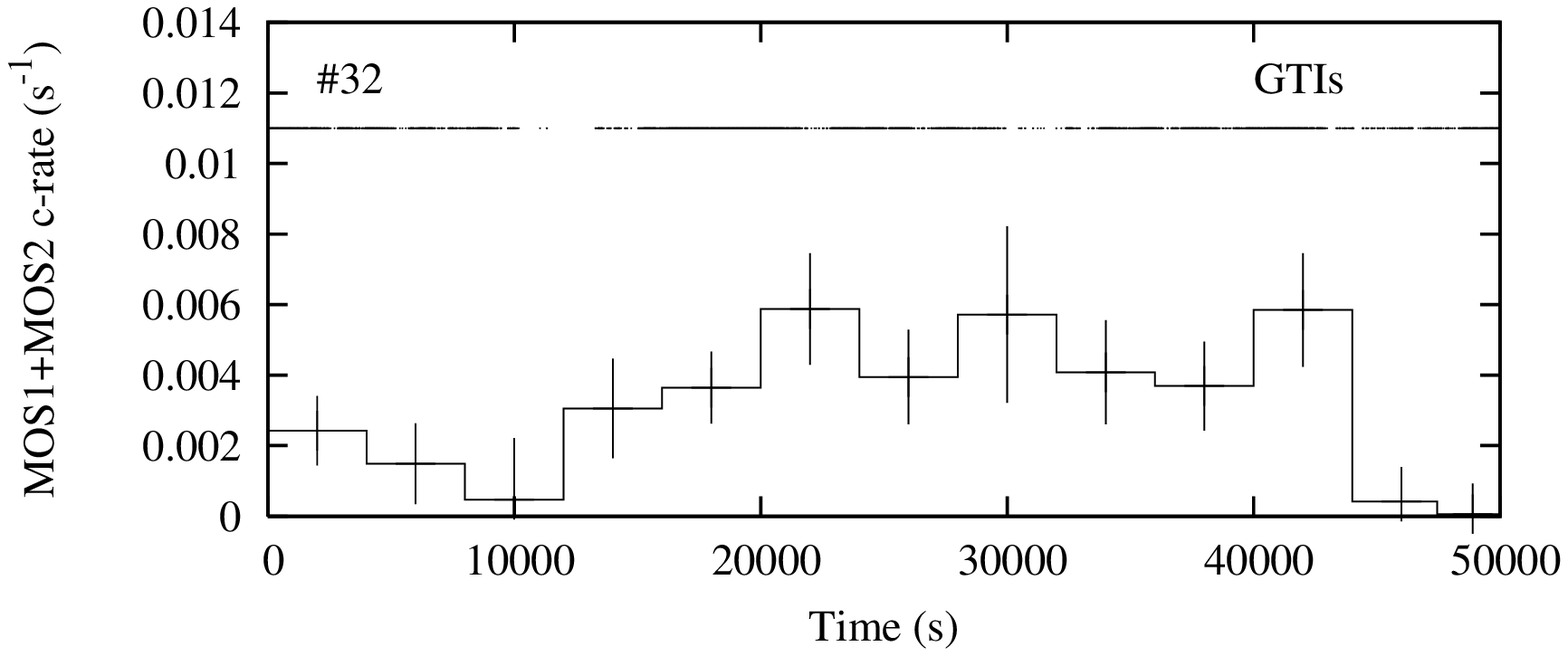}
    \end{minipage}
    \begin{minipage}[t]{0.45\textwidth}
    \includegraphics[width=71mm]{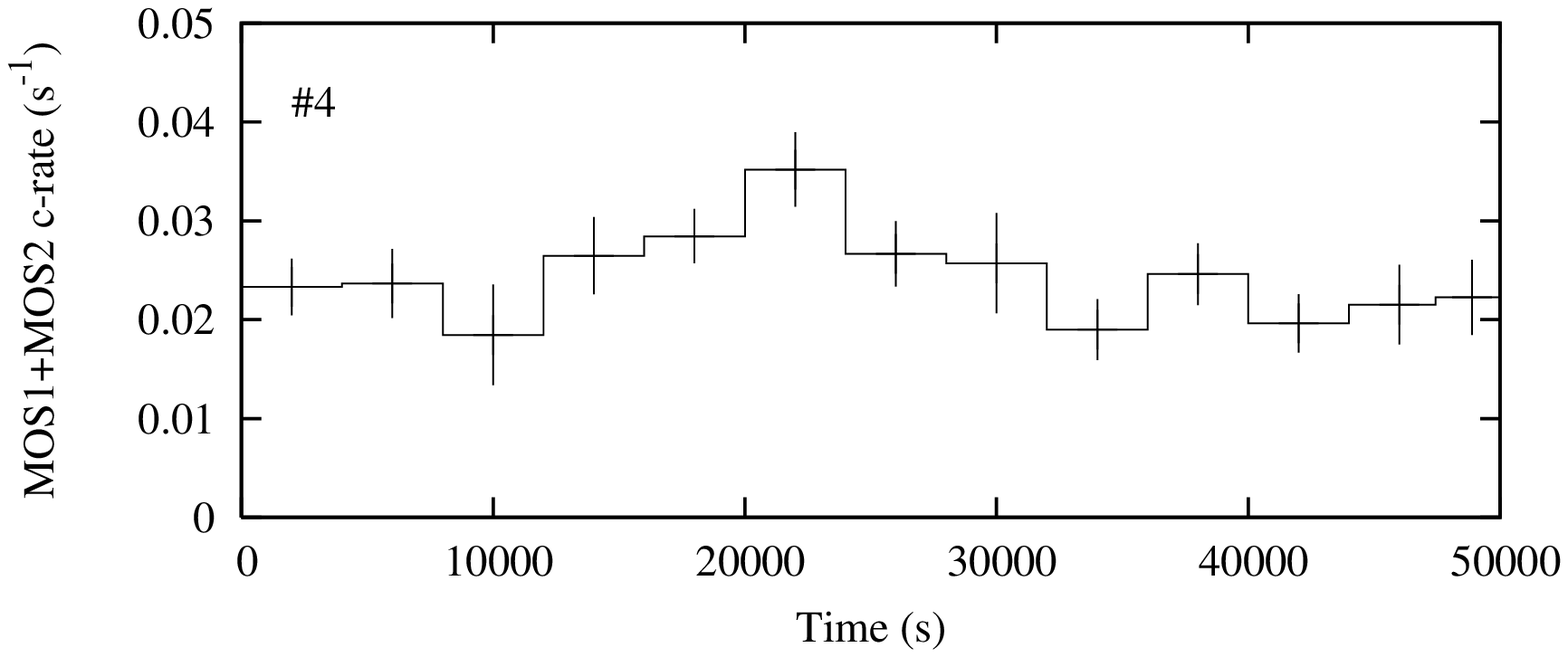}
    \includegraphics[width=71mm]{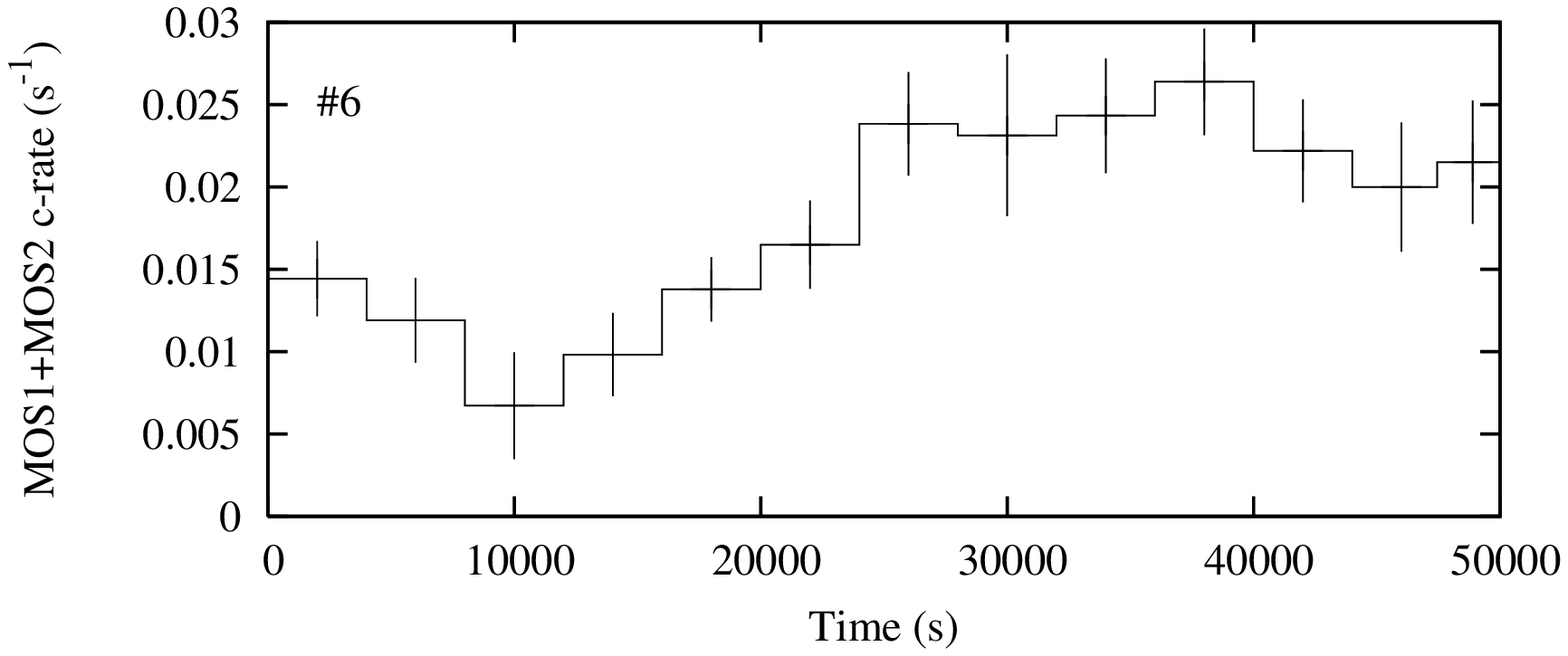}
    \includegraphics[width=71mm]{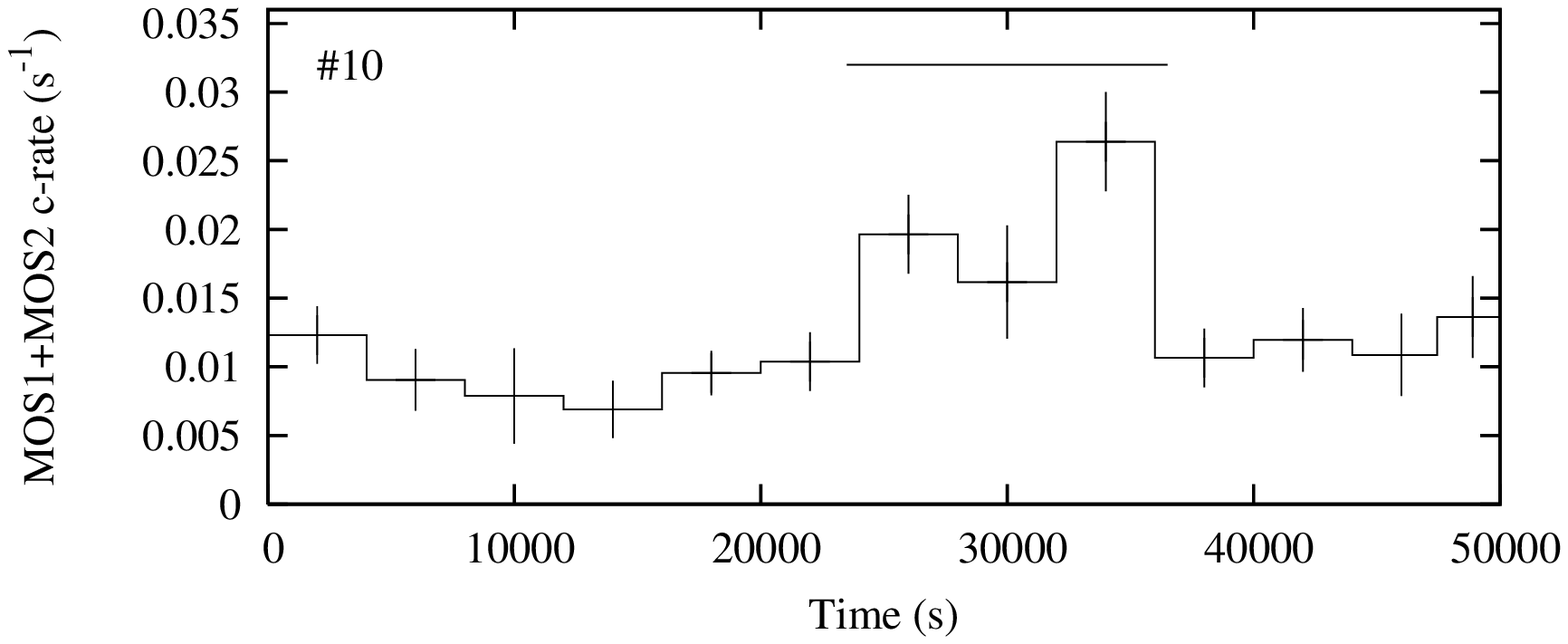}
    \includegraphics[width=71mm]{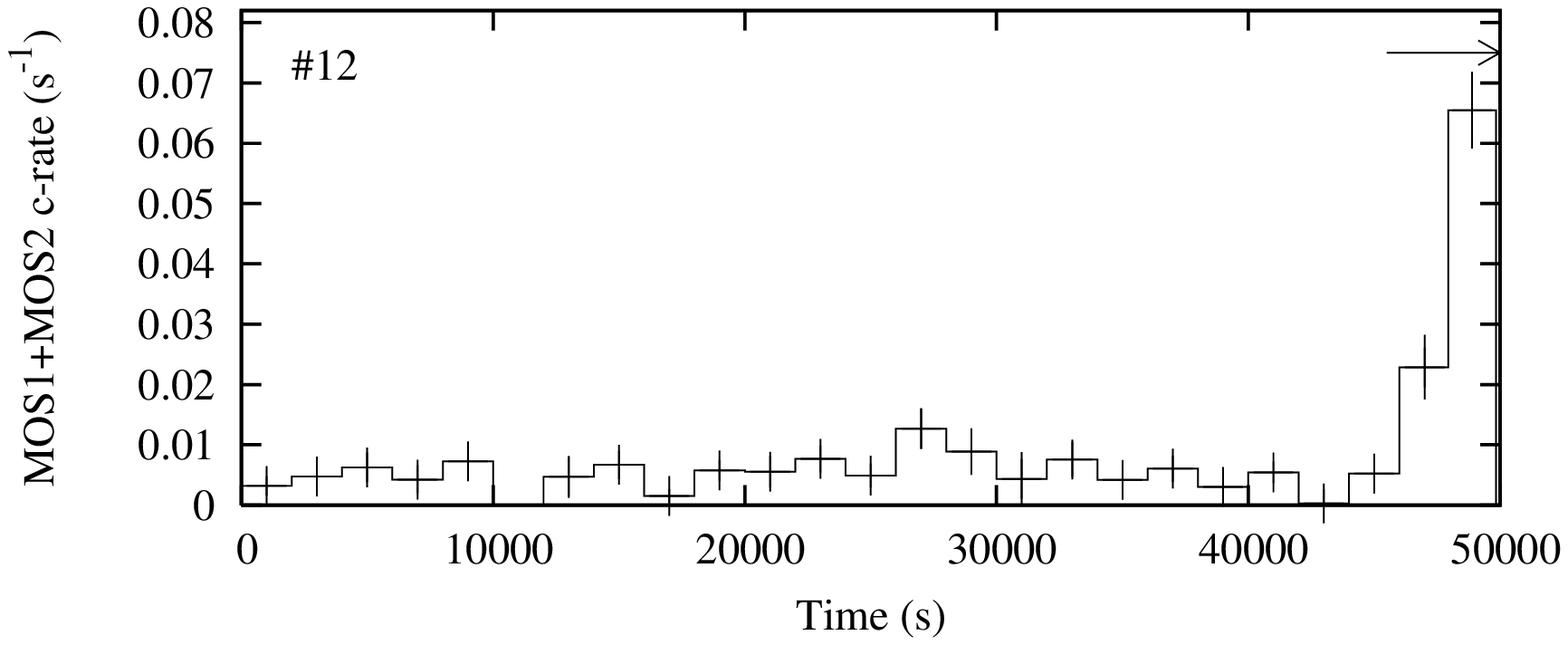}
    \includegraphics[width=71mm]{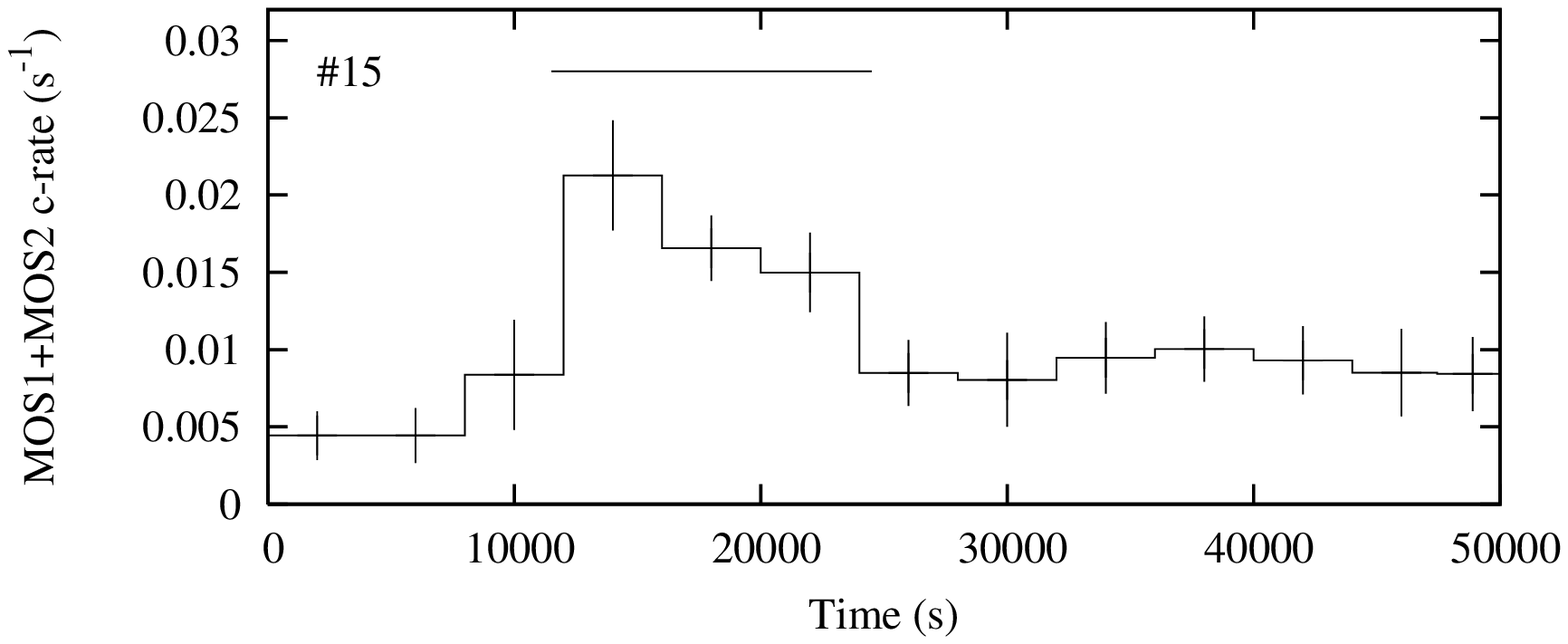}
    \includegraphics[width=71mm]{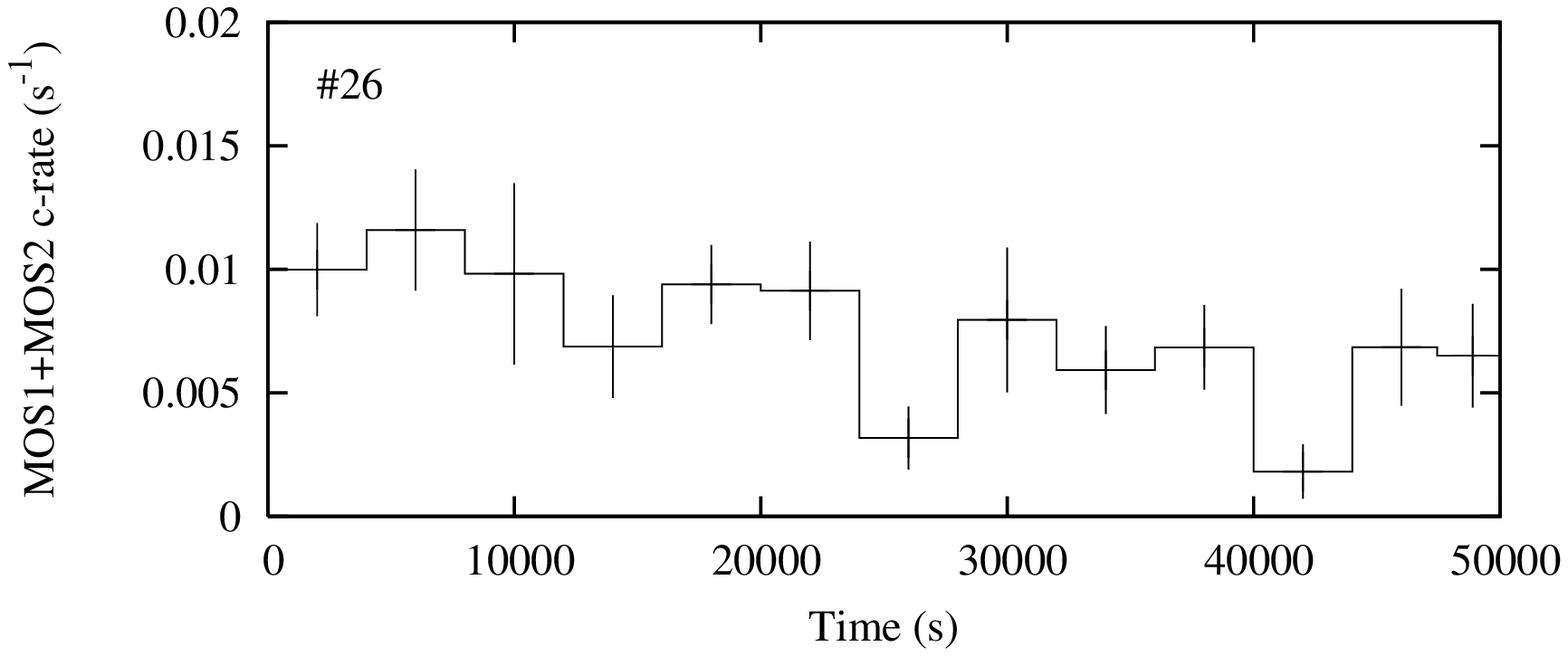}
    \includegraphics[width=71mm]{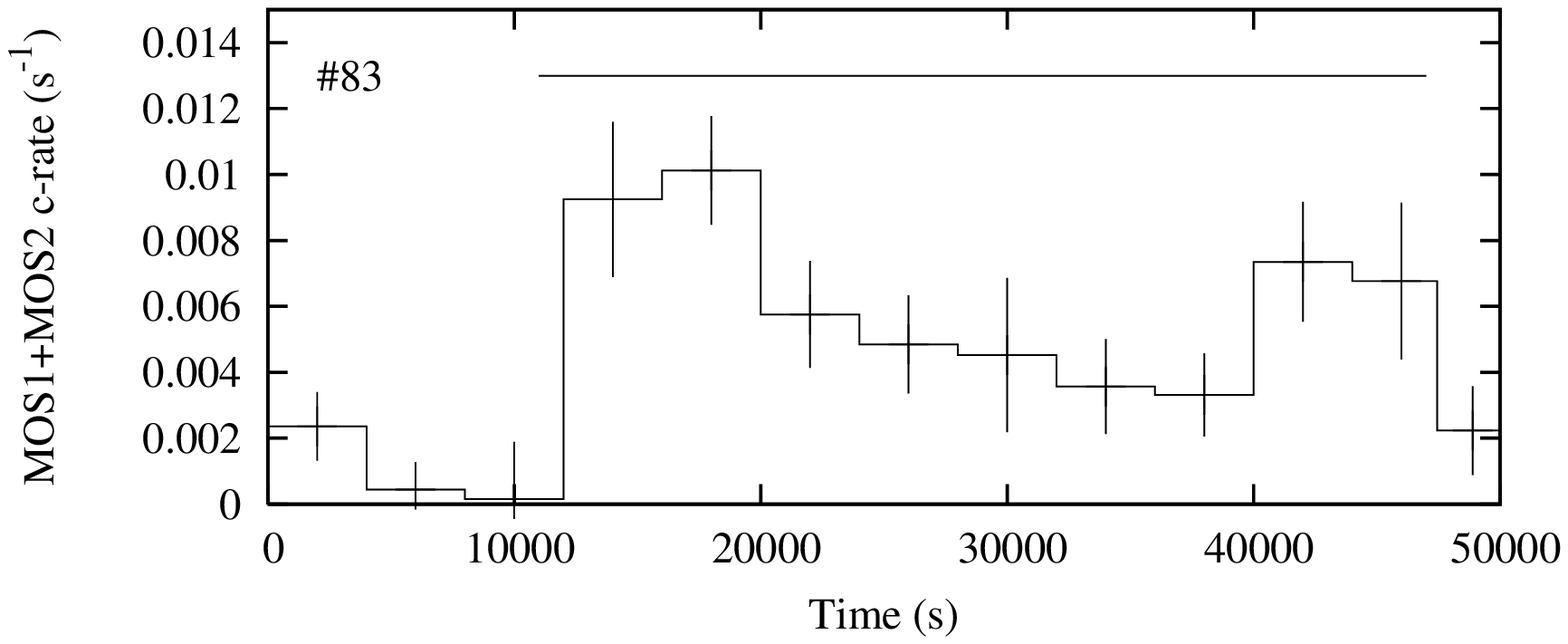}
    \end{minipage}
    \caption{X-ray light curves (0.3--3.0\,keV) for selected members of NGC~2547 that
      exhibit short-term variability (plus the ultra-fast rotator star 4 which
      appears not to show variability). Possible flaring periods are
      indicated with a horizontal line. All the curves begin at MJD
      2452367.4786, feature data from the MOS1 plus MOS2 detectors
      taken during good time intervals and are background
      subtracted. The reader should note that the light curves are
      based on data with frequent small coverage gaps. The data points
      in the plots represent the best estimate of the count rates over
      either 2\,ks or 4\,ks bins. The ``good'' time intervals are
      indicated by a broken line in the plot for star 32. There is only
      one extended time period of a few ks that has no good data.
}
\label{curveplot}
\end{figure*}

\begin{table}
\caption{Estimated flare parameters.}
\begin{tabular}{cccccc}
\hline
ID    & Quiescent $L_{\rm x}$ & Peak $L_{\rm x}$ & Duration & Energy \\
      &  \multicolumn{2}{c}{(erg s$^{-1}$)}      & (ks)     & (ergs) \\       
\hline
 3    &$1.8\times10^{30}$ &$8.6\times10^{30}$   & 8        &$3\times10^{34}$\\
 5    &$1.9\times10^{29}$ &$1.1\times10^{31}$   & \llap{$>$}36 &\llap{$>$}$2\times10^{35}$\\
10    &$2.0\times10^{30}$ &$4.8\times10^{30}$   & 12       &$2\times10^{34}$\\
12    &$9.6\times10^{29}$ &$8.5\times10^{30}$   & \llap{$>$}4  &\llap{$>$}$2\times10^{34}$\\
13    &$1.3\times10^{30}$ &$2.7\times10^{30}$   & 16       &$8\times10^{33}$\\
15    &$8.9\times10^{29}$ &$1.6\times10^{30}$   & 12       &$1\times10^{34}$\\
83    &$2.6\times10^{29}$ &$1.3\times10^{30}$   & 36       &$2\times10^{34}$\\
\hline
\end{tabular}
\label{flaretable}
\end{table}

Short term ($< 1$ day) X-ray variability in young low-mass stars is a
well known phenomenon and can be caused by changes in coronal
structures, rotational modulation, eclipses or flaring of a magnetic
origin (see G\"udel 2004 for a review).

For some of the X-ray bright NGC~2547 members, there are sufficient
detected counts to construct light curves. For the fainter sources we
can only hope to detect longer term variations in the X-ray activity
(see section~\ref{comphri}). Background subtracted light curves were
constructed for the 10 brightest NGC~2547 sources using the sum of
events in the MOS1 and MOS2 detectors. Only data from good (low
background) time
intervals (as discussed in section 2.1) were included, resulting in
numerous, but small, gaps in the coverage. We did not use the pn data
because of the much larger gaps in the coverage.  
The extraction and background regions were similar
to those used to obtain spectra. For the purposes of comparison with
published information on younger and older clusters (see
section~\ref{tevol}), light curves were extracted in a similar way for
all solar-type stars with $0.8\leq M <1.2\,M_{\odot}$, corresponding to
$1.34\geq V-I > 0.67$ according to the isochrone used in
Fig.~\ref{vvicmd}.

Initially the background subtracted light curves were put into 4\,ks
bins, a few of the brighter targets had sufficient counts to reduce
this to 2\,ks. The periods of high background were sufficiently
dispersed in the MOS data that there was adequate information to obtain
reasonably well sampled light curves that cover $\simeq 50$\,ks,
beginning at MJD 2452367.4786. The count rates shown of course take
account of the gaps in the coverage. A simple chi-squared test revealed 12/30
stars showing significant ($>99$ per cent confidence)
variability. X-ray light curves for these stars are shown in
Fig.~\ref{curveplot}. As expected, the variable stars tend to be the
brighter among our sample, with the best statistics. We suspect that,
given sufficient signal-to-noise ratios, all of the NGC~2547 targets
would show variability at some level.

Some of the variability seen may well be due to flares. Several of the
light curves showed the characteristic rapid rise and longer decay that
is often seen in X-ray flares on the Sun and other stars. A couple of
other ``events'' were more questionable. We chose to be reasonably
relaxed in our definition of a flare and indicate in
Fig.~\ref{curveplot} those stars and time intervals we considered
to represent flaring behaviour.  To estimate some crude energetic
parameters for these flares, the same count-rate to
intrinsic flux conversion factor was assumed to be valid throughout the flare even though
the coronal plasma will likely be hotter than average during flare
events. It was found in section~\ref{xraylum} that the conversion factor
was quite insensitive to spectral parameters, but as the conversion
factor does increase with temperature, the flare energies and luminosities are
probably under-estimates. The correction between
the count-rates in the light curves, which were extracted from a
small radius, and the total count-rate, was provided by comparing the
average light curve count-rate with those in Table~1 which include a
PSF correction. A quiescent luminosity was defined using the average flux
outside of the flare and a peak flare luminosity was esimated. The latter 
was quite uncertain and probably under-estimated because of the low
time resolution. Finally, a total flare energy above that which the star
would have emitted in its ``quiescent'' state was calculated.

The ``flare'' parameters, for the energy range 0.3--3.0\,keV, are listed
in Table~\ref{flaretable}. Given these values we are confident that we
would have detected all flares on the solar type stars with integrated
energies in excess of $\simeq 10^{34}$~erg\,s$^{-1}$. The exception
could be for the case of very long duration ($\ga 30$\,ks) flares which
might not be distinguishable from rotational modulation (see
below). Short flares ($\leq 4$\,ks), as seen in some young stars (see
Stelzer et al. 2000) would not be properly resolved but would easily be
seen as 1-2 very high points if the flare energy exceeded
$10^{34}$~erg.

For stars 6, 9, 11, 26 and 32 the variability does not appear to be
flare-like, but rather could be due to the rotational modulation of
reasonably compact coronal structures. In particular, the light curve
of star 11 could be repeating with a period of 30\,ks. Unfortunately
there is no published information on this star to decide whether a
period of 0.35\,days is plausible, but stars that rotate this rapidly
exist in other young clusters. At this rotation rate, the star would be
have a Rossby number of 0.024 -- approaching the super-saturated 
part of Fig.~\ref{rossbyplot}. Star 11
actually has $L_{\rm x}/L_{\rm bol} \simeq 7\times10^{-4}$, which could
indicate the onset of super-saturation, but is not conclusive.
Stars 9 and 32 are at least
moderately fast rotators with $v \sin i$ of 27 and 12~km\,s$^{-1}$ and
hence have rotation periods that are shorter than 1.9 and 3.6~days
respectively.  Star 6 is a slow rotator with $v \sin i =
6$~km\,s$^{-1}$ and star 26 is a close spectroscopic binary with two
slowly rotating components (Jeffries et al. 2000). 
It is worth noting that the fastest known rotators in
the cluster (stars 4 and 21 in this paper, labelled RX~30a and RX~35 in
Jeffries et al. 2000), with $v \sin i$ values of 86 and 160~km\,s$^{-1}$
respectively, show no signs of variability at the $\ga 20$ per cent
level. The X-ray light curve for star 4 is shown in Fig.~\ref{curveplot}.

\begin{figure}
\includegraphics[width=84mm]{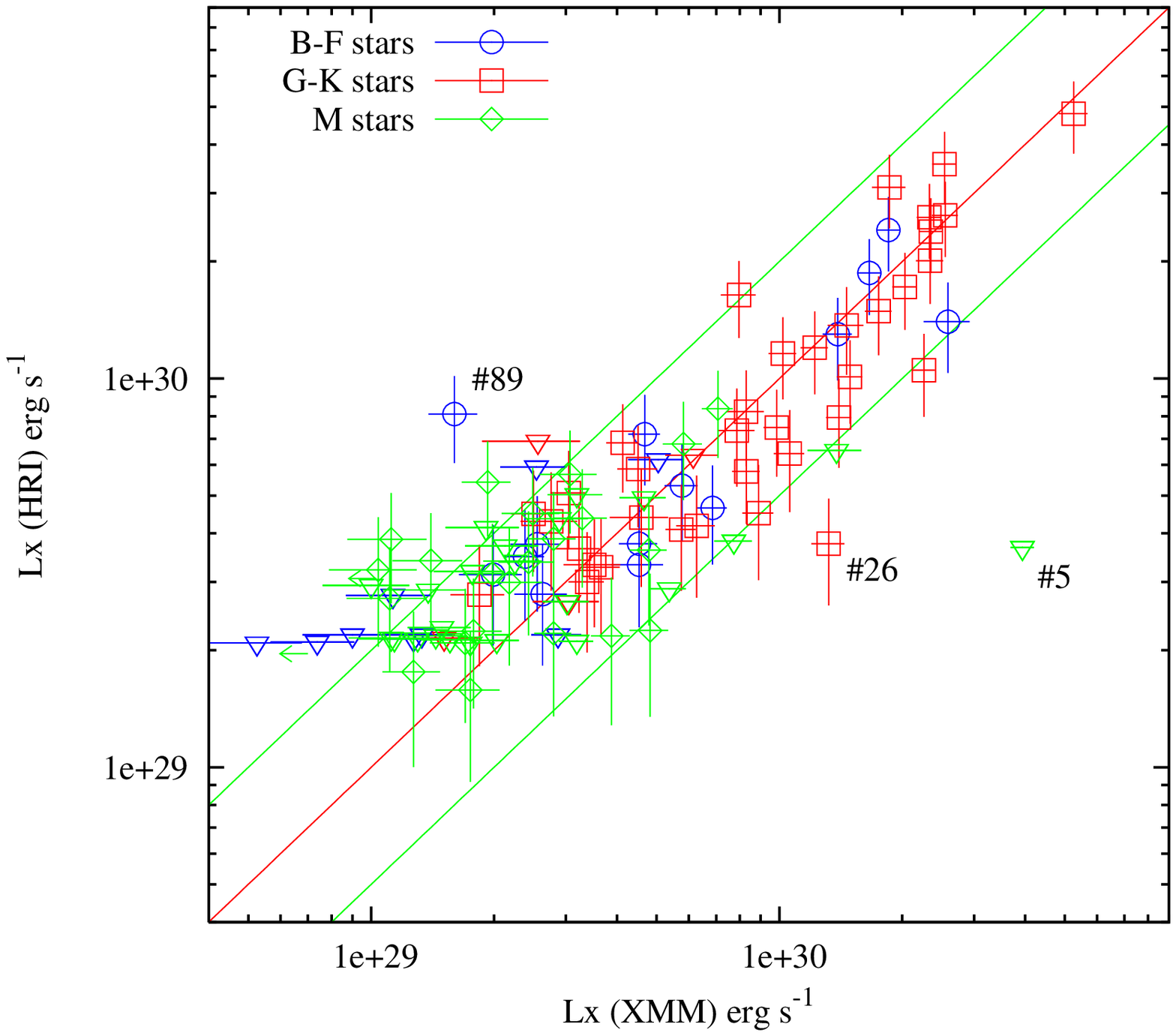}
\includegraphics[width=84mm]{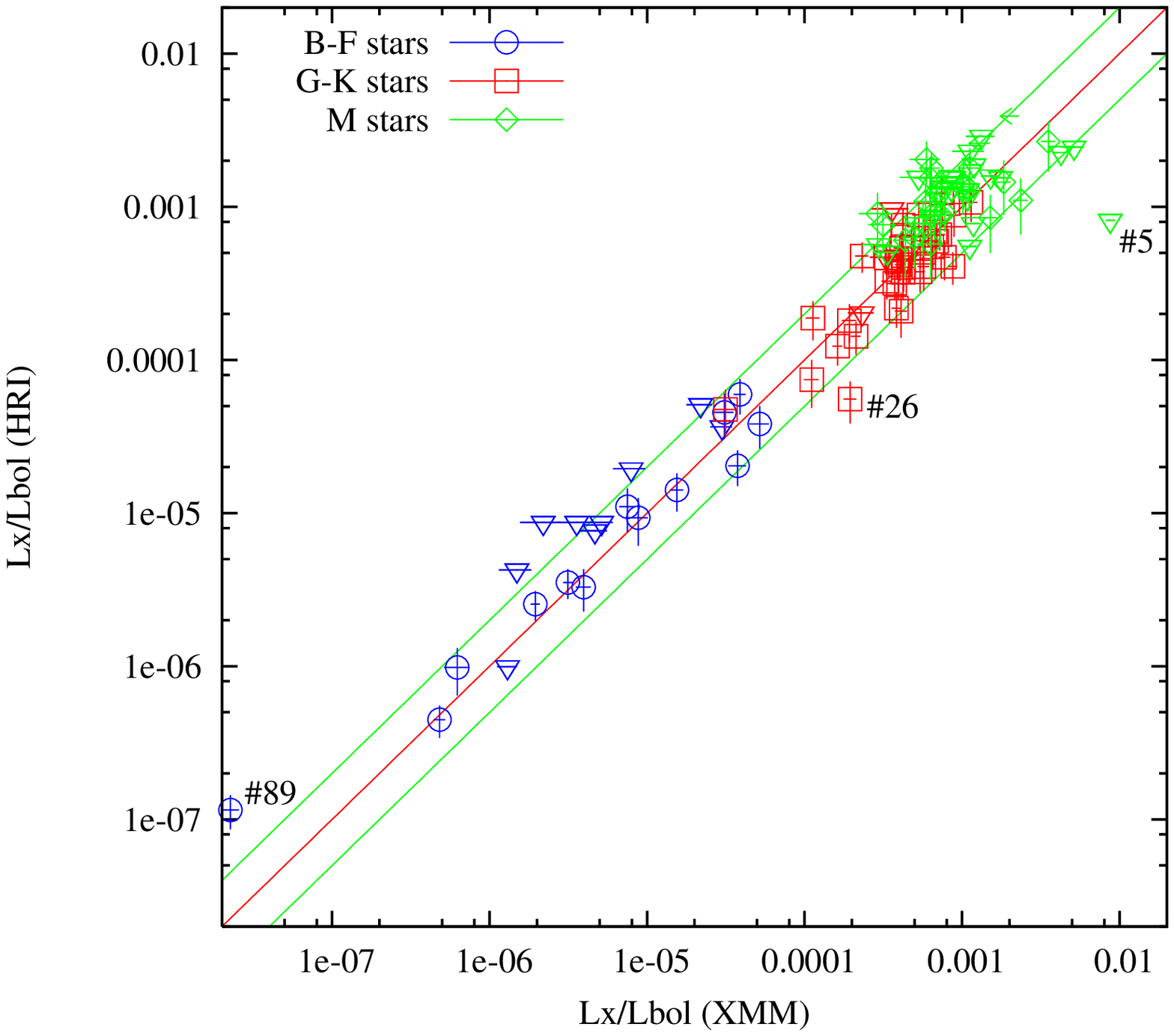}
\caption{A comparsion of X-ray activity (in the 0.3--3.0\,keV band)
  observed with {\it XMM-Newton} and the {\em ROSAT} HRI, separated by
  7 years. Different
  symbols (and colours in the electronic version) distinguish stars of
  F-type or earlier, G/K-type or M-type. Upper limits in the HRI are
  denoted by downward pointing triangles. Two upper limits in the {\it
  XMM-Newton} data are shown with leftward pointing arrows. The
  straight lines in the plots represent equality between the
  measurements and variations by factors of 0.5 and 2.0 respectively.
 Several significantly variable stars are identified and commented upon in the text.
}
\label{xmmhrifig}
\end{figure}

\subsection{Long term variability}
\label{comphri}

\begin{table*}
\caption{\ Correlations between {\em XMM-Newton} and {\it ROSAT} HRI
  detections (from Jeffries \& Tolley 1998) 
  of photometric cluster candidates. There are 108 entries
  in the Table, corresponding to Tables~\ref{xraymembers} and~\ref{counterparts}.
  The full table is only available electronically.
  The columns list (1) the running {\it XMM-Newton} source
  identification number (2) the RX identifier used by Jeffries \& Tolley, (3) the
  separation between the {\it XMM-Newton} and HRI positions, (4)--(5)
  the HRI countrate (modified from the Jeffries \& Tolley value as
  explained in section~\ref{comphri}) and its uncertainty,
  (6)--(8) three values of $L_{\rm x}/L_{\rm bol}$ (for the
  0.3--3\,keV energy range, using $B-V$, $V-I$ and $R-I$ indices) and
  (9) an estimate of $L_{\rm x}$ (0.3--3\,keV) using the HRI count rates
  in column (4). If there
  is no correlation with an HRI source (indicated by a zero in column
  (2)) then the last 4 columns are
  upper limits.}
\begin{flushleft}
\begin{tabular}{cccccccc}
\hline

No. & RX & Sep & HRI count rate (s$^{-1}$) &
\multicolumn{3}{c}{HRI $L_{\rm x}/L_{\rm bol}$} & HRI $L_{\rm x}$
(0.3--3\,keV) \\
    &    & arcsec &  & bc($B-V$) & bc($V-I$) & bc($R-I$) & erg\,s$^{-1}$ \\
(1) & (2) & (3) & (4) & (5) & (6) & (7) & (8) \\
\hline
3&  53& 2.5&1.28E-03$\pm$1.70E-04& 4.27E-04& 4.08E-04& 3.47E-04& 1.05E+30\\
4&  30& 2.7&4.21E-03$\pm$2.90E-04& 4.83E-04& 4.63E-04& 3.92E-04& 3.56E+30\\
\hline
\end{tabular}
\end{flushleft}
\label{hritable}
\end{table*}

\begin{table*}
\caption{HRI detected sources not found in the {\it XMM-Newton}
  data. The columns are as for Table~\ref{hritable} with the
  addition of the identification, and photometric properties, in the
  Naylor et al. (2002) catalogue.}
\begin{tabular}{ccccccccc}
\hline
RX & ID (N02) & $V$ & $V-I$ & HRI count rate (s$^{-1}$) & HRI $L_{\rm
    x}/L_{\rm bol}$ & HRI $L_{\rm x}$ (0.3--3\,keV) & XMM $L_{\rm
  x}/L_{\rm bol}$&
    XMM $L_{\rm x}$\\
\hline
20 & 14 2133 & 20.634 & 3.261 & 2.20E-04$\pm$8.00E-05 &3.91E-03&
    1.96E+29 & \llap{$<$}1.19E-03& \llap{$<$}5.98E+28 \\
23 & 18 2322 & 19.109 & 2.832 & 3.50E-04$\pm$1.00E-04 &2.60E-03&
    3.06E+29 & \llap{$<$}7.49E-04& \llap{$<$}8.82E+28 \\
\hline
\end{tabular}
\label{notinxmm}
\end{table*}

The previous observation of NGC~2547 with the {\it ROSAT} HRI (Jeffries
\& Tolley 1998), taken in December 1995, 
allows an investigation of variations in X-ray activity
on a $\simeq 7$ year timescale. The {\it XMM-Newton} list of cluster
members was cross-correlated against the HRI source positions allowing
a correlation radius of up to 16 arcseconds, reflecting the larger
positional uncertainties in the HRI data. There are 72 correlations
which are identified in Table~\ref{hritable} by their ``RX''
numbers from Jeffries \& Tolley. RX\,86 was correlated with, and 
lay mid-way between, {\it  XMM-Newton} sources 111 and 113. In the
absence of any additional information the X-ray flux in the HRI was
split equally between the two.

In Jeffries \& Tolley (1998) it was claimed that the most active stars
in NGC~2547 were under-active by nearly a factor of two (in terms of $L_{\rm
x}/L_{\rm bol}$) compared with other young clusters, especially among
the G and~K stars. This seems not to be the case in the {\it
XMM-Newton} observations (see sections~\ref{xraylum}
and~\ref{rotation}) and the discrepancy needs explaining. One
possibility that can now be ruled out is that the NGC~2547 stars have
peculiarly hot or cold coronae resulting in under-estimated fluxes from the HRI
count-rates.  Like {\em XMM-Newton}, the HRI count-rate to flux
conversion factor is quite insensitive to variations in temperature or column
density and the coronal parameters we have deduced here are close to
the 1\,keV coronal temperature assumed by Jeffries \& Tolley.  After
further careful investigation we have found two other effects that
resulted in under-estimated HRI fluxes.

First, the spectral response of the HRI was varying with time
throughout the {\it ROSAT} mission. Using an updated gain and response
matrix shift appropriate for the time of the NGC~2547 observation (see
David et al. 1999 for details) the count-rate to
flux conversion factor was re-calculated for (a) the 0.1--2.4\,keV range considered by
Jeffries \& Tolley (1998) and (b) the 0.3--3.0\,keV range for comparison with
the {\it XMM-Newton} data. In both cases we have assumed the mean
coronal model discussed in section~\ref{xraylum}. The case (a)
conversion factor is $4.12\times10^{-11}$~erg~cm$^{-2}$ per count,
which is 1.31 times larger than the conversion factor used by Jeffries
\& Tolley. The case (b) conversion factor which is used to estimate the $L_{\rm x}$ and
$L_{\rm x}/L_{\rm bol}$ (using the same distance and bolometric corrections
as in Table~\ref{counterparts}) values in Table~\ref{hritable} was
$3.40\times10^{-11}$~erg~cm$^{-2}$ per count.

Second, modelling of the HRI PSF was fairly crude at the time of the
analysis performed by Jeffries \& Tolley (1998). Their point source
parameterisation assumed a Gaussian PSF which got broader with off-axis
distance. This is a considerable simplification compared with the more
complex PSF modelling discussed by Campana et al. (1999) in the context
of constructing the Brera Multiscale Wavelet (BMW) {\it ROSAT} HRI
source catalogue (Panzera et al. 2003). A significant extended ``halo''
to the PSF means that the count-rates provided by Jeffries \& Tolley
(1998) may have been significantly underestimated. To test this 
the Jeffries \& Tolley catalogue was correlated against the BMW catalogue,
finding 78 matches. The Jeffries \& Tolley count-rates
were systematically lower by a factor of $1.152(\pm 0.042) +
0.012(\pm0.004)\,\theta$, where $\theta$ is the off-axis ange in
arcminutes. The scatter around this relationship was 0.15 (rms).  We
have chosen to correct the Jeffries \& Tolley count-rates by this
factor rather than use the BMW count-rates. The rationale for this is
that the BMW catalogue does not include 24 of the 102 sources found by
Jeffries \& Tolley, 15 of which have {\it XMM-Newton} detections of
cluster counterparts and 4 of the remaining 9 are also closely
correlated with photometric cluster candidates. It seems that Jeffries
\& Tolley were better at finding X-ray sources even if their
count-rates were systematically too low.

For the remaining 36 cluster candidates detected by {\it XMM-Newton}
which have no HRI counterparts an estimated upper limit
to the flux observed by the HRI was made by looking at the minimum detected HRI
count-rates (after the correction described above) as a function of
off-axis angle. Where no RX number is given in Table~\ref{hritable}
this indicates that the HRI count-rate given is an upper limit.

Two significant HRI sources (RX\,20 and RX\,23) were found which
correlate with photometric cluster candidates and lie within the EPIC
field of view, but which were not detected by {\it XMM-Newton}. Both of
these X-ray sources are closely correlated with M-dwarfs from the
membership lists of Naylor et al. (2002) and 2-sigma upper limits for
the count-rates and X-ray fluxes were calculated for these. The details
are given in Table~\ref{notinxmm}.

A comparison of the X-ray activity as judged by {\it XMM-Newton} and
the {\it ROSAT} HRI (as listed in Table~\ref{hritable}) 
in the form of both $L_{\rm x}$ and $L_{\rm
x}/L_{\rm bol}$ is shown in Fig.~\ref{xmmhrifig}. Different symbols
(and colours in the electronic version of the paper) are used to
represent stars of approximately F-type or earlier ($V-I<0.67$),
spectral types G or K ($0.67<V-I<2.00$) and M-type stars
($V-I>2.00$). Note that the additional error in the corrected HRI
count-rates, associated with uncertainties in the HRI PSF has been
added in quadrature and incorporated into these diagrams.  $L_{\rm
x}/L_{\rm bol}$ was calculated using the $V-I$ colour if available, or
using $B-V$ otherwise.

Figure~\ref{xmmhrifig} shows that there
is excellent agreement between the intrinsic X-ray fluxes of stars
observed with both {\it XMM-Newton} and the {\em ROSAT} HRI. The
majority of stars have varied by less than a factor of two between the
observations. There appears to be some evidence that stars with lower
X-ray luminosities were brighter at the time of the HRI
observations. Of course this is counterbalanced to some extent by the
upper limits on HRI fluxes in the same region of the diagram and by
the larger error bars for these sources. Another point to consider is
that for X-ray sources with low signal-to-noise, there is an inevitable
upward bias if (as is the case here and in Jeffries \& Tolley 1998) 
the position of the X-ray source is a free parameter in
the count-rate determination algorithm. This is because the (weak)
source tends, on average, to be located at the position of a positive
noise peak. This upward bias will predominantly affect the weaker sources in the less 
sensitive HRI observations. On that basis we don't believe there is strong
evidence that lower luminosity sources are more variable, or for any
systematic discrepancy between the HRI and EPIC fluxes as a function of
X-ray activity or spectral type.

Various ways of parameterising long-term variability can be found in
the literature. To make comparisons we have estimated the fraction of
stars that have varied by more than a factor of two and the mean (and
median) absolute deviation from equal luminosity. 
These statistics  were calculated for two limited samples: (A) $L_{\rm
x}>3\times10^{29}$~erg\,s$^{-1}$; (B) the G/K star sample as defined
above. These subsets were chosen to minimise the number of upper limits and to
avoid the weak X-ray source bias discussed above.  For samples A and B
there are 8/60 and 5/40 objects that varied by a factor of two or
more, treating the upper limits as detections. The fractions are more
likely to be 9/60 and 4/40 given a more thoughtful consideration of
where these upper limits lie. The high $L_{\rm x}$ M-stars could be more
variable than average; 5/15 of the M-dwarfs in sample A are
variables. However, there is an obvious selection effect favouring a
high fraction in this very incomplete sample. The mean (and median) absolute
deviations from equal luminosity of samples A and B are almost
identical at 0.098(0.095)\,dex and 0.107(0.094)\,dex respectively (i.e. a
factor of $\simeq 1.25$). The upper limits are treated as detections in this
estimate, but their inclusion does not affect the result significantly.

Assuming that the error bars in Fig.~\ref{xmmhrifig} represent a normal
distribution, we have made simulations under the additional assumption
that the two measured luminosities {\it are} equal. We find that we
would expect mean absolute deviations of 0.087 and 0.088 for samples A
and B respectively in any case. This strongly suggests that the
majority of sources have not varied at all and that the slightly larger
observed mean absolute deviations are attributable to a handful of
strongly varying objects, which are discussed below.

There are only three clear examples where
we are confident that variations of more than a factor of 2 have
occurred (i.e. with deviations
well in excess of the estimated errors). These are sources
5, 26 and 89, which are labelled in Fig~\ref{xmmhrifig}.

Source 5 is an M-dwarf that has clearly undergone an intense flare
during the {\it XMM-Newton} observation (see section~\ref{flare} and
Fig.~\ref{curveplot}). The ``quiescent'' $L_{\rm x}$ estimated in
Table~\ref{flaretable} is consistent with the upper limit derived from
the HRI observation.

Source 26 is a moderately active G-type star, identified as a short
period spectroscopic binary by Jeffries et al. (2000). This was
identified as variable in the EPIC data, but the light curve only shows
a gradual decrease in the X-ray flux during the observation.
It is however feasible that we have seen the decay phase of a long duration
($>50$\,ks) flare that is responsible for increasing the average
$L_{\rm x}$ as viewed by {\it XMM-Newton} by a factor of 3 compared
with {\it ROSAT}.

Source 89 corresponds to the optically brightest star in the cluster,
HD~68478 with a B3\,{\sc iv} spectral type. It is just possible that
this star is massive enough to generate X-rays in a stellar wind with
$L_{\rm x}/L_{\rm bol} \simeq 10^{-8}$--$10^{-7}$ (e.g. Cassinelli et
al. 1994). However, the strong variability of almost a factor of 4
between the EPIC and HRI observations is not expected from early-type
X-ray sources (Bergh\"ofer et al. 1997), so we hypothesise that it is
more likely that the X-ray emission arises from an as-yet-undiscovered 
late-type companion that flared during the HRI observation.

\section{Coronal activity at 30\,Myr}

The main focus of this paper is to gather information on the X-ray
coronae of PMS stars at $\simeq 30$\,Myr and put them in context with
what we know about X-ray activity in younger star forming regions, like
the ONC and older, well-studied clusters such as the Pleiades (age
$\simeq 120$\,Myr) and Hyades (age $\simeq 600$\,Myr). We can then ask
whether the X-ray activity of such stars conforms to our expectations
based on the age-rotation-activity paradigm (ARAP). Or, are other
factors besides the changing stellar structure and decreasing rotation
rate important, such as the apparent suppression of X-ray activity by
accretion in the ONC?

Thanks to the relative insensitivity of estimated coronal fluxes to
assumptions about the temperature structure of the coronae, we can
split our discussion into a consideration of the evolution of the
overall coronal energy losses followed by the evolution of coronal
temperatures and variability.

\subsection{The Evolution of X-ray activity}
\label{evolution}

\begin{figure*}
    \centering
    \begin{minipage}[t]{0.45\textwidth}
    \includegraphics[width=71mm]{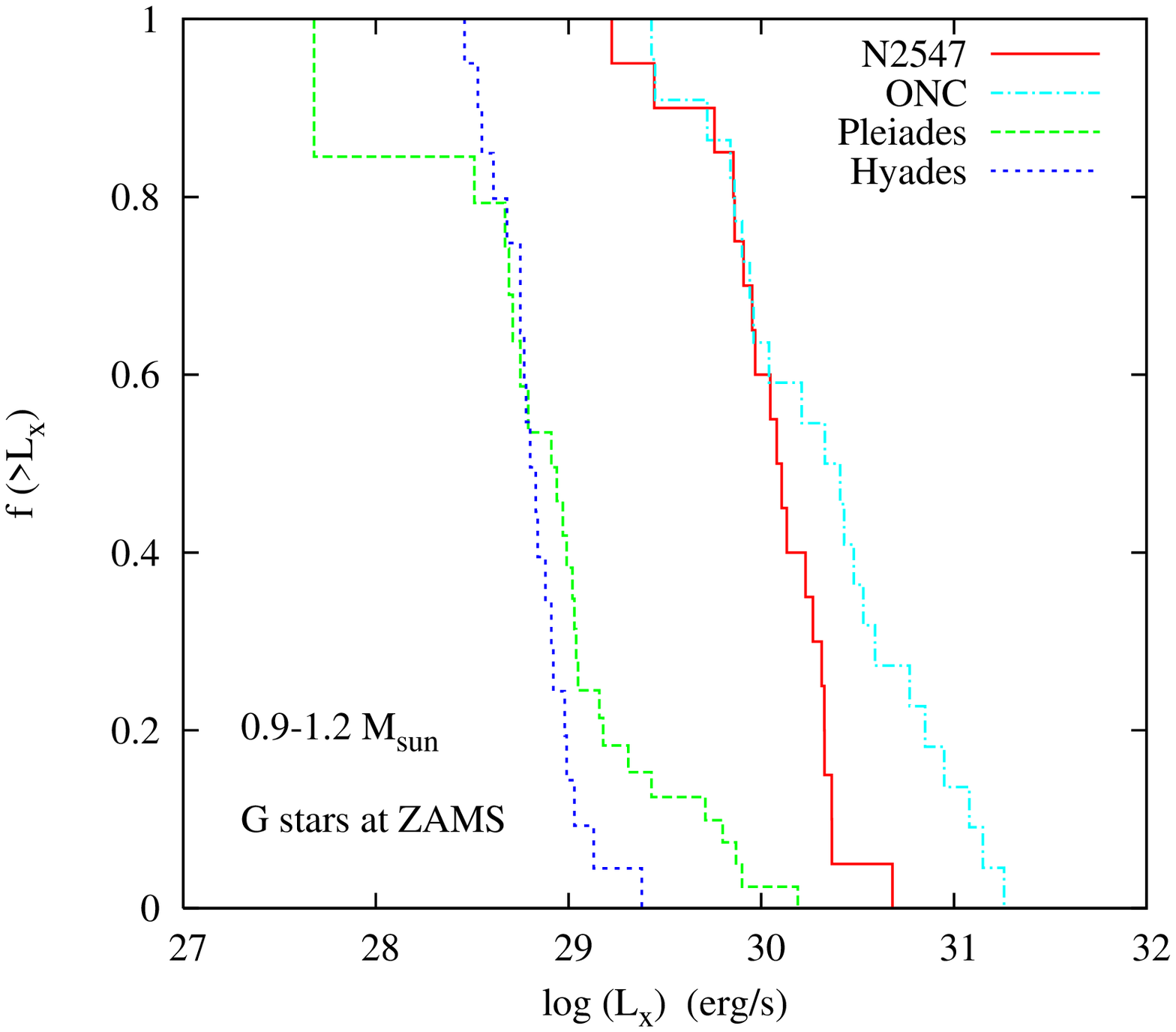}
    \includegraphics[width=71mm]{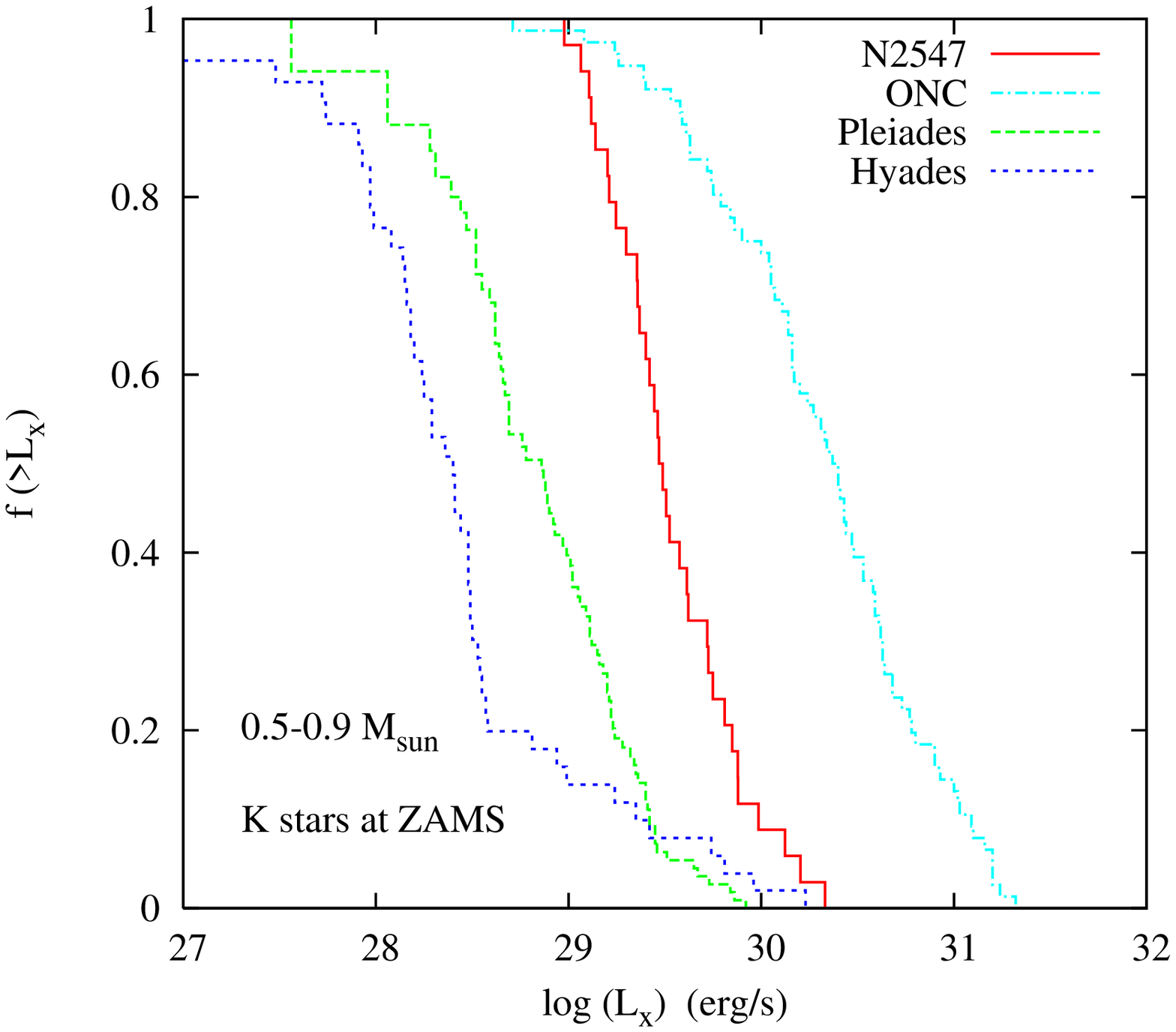}
    \includegraphics[width=71mm]{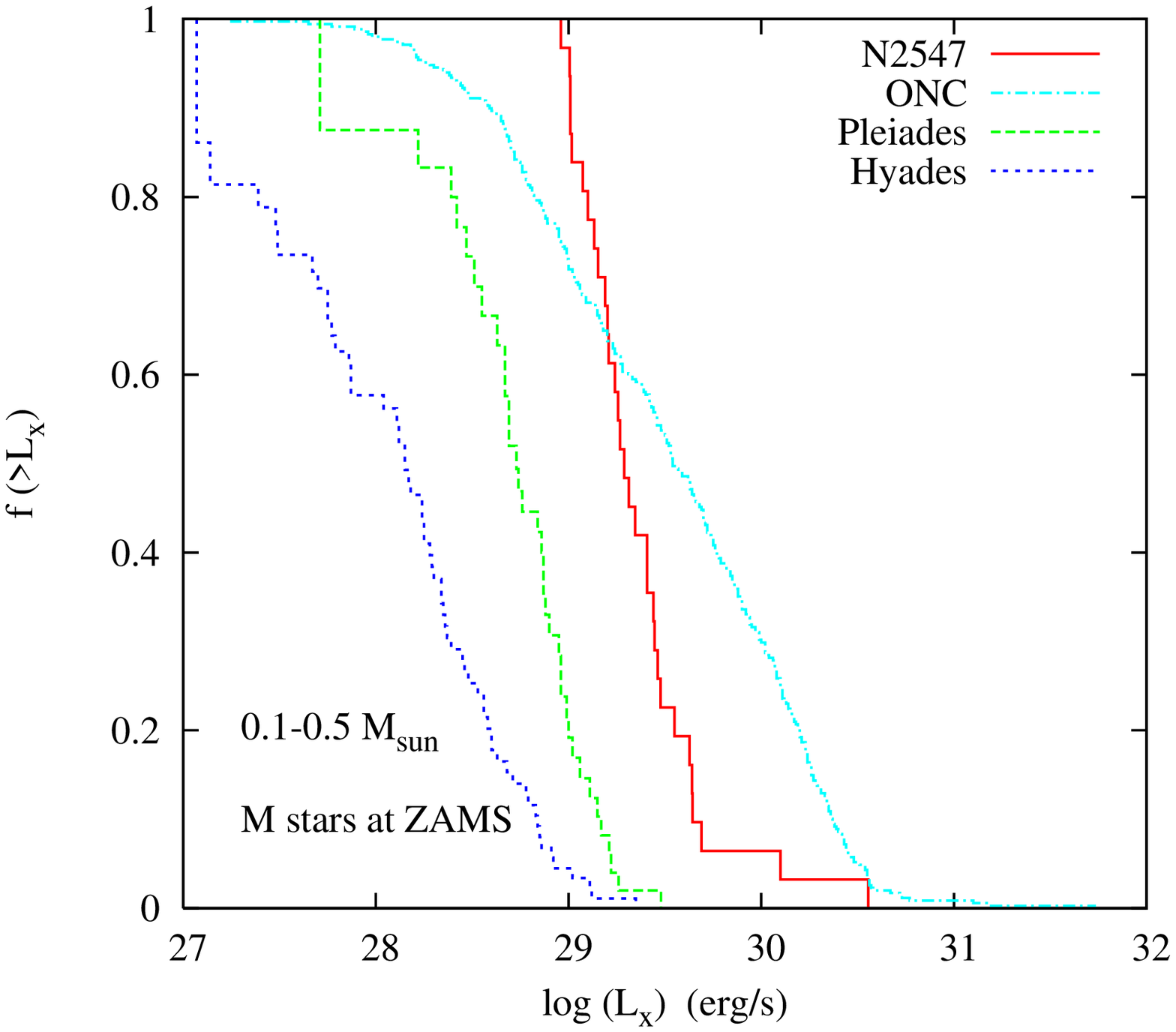}
    \end{minipage}
    \begin{minipage}[t]{0.45\textwidth}
    \includegraphics[width=71mm]{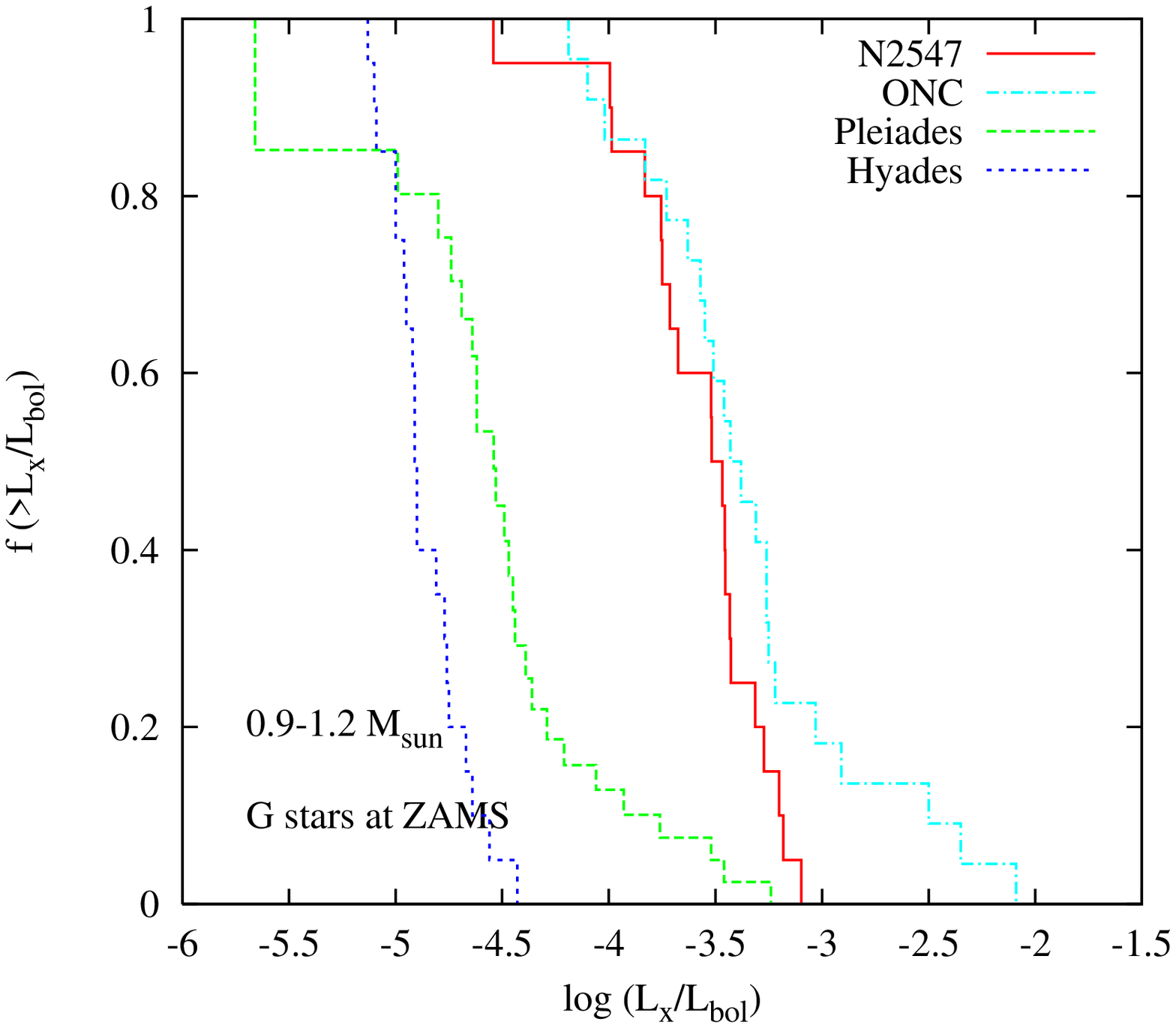}
    \includegraphics[width=71mm]{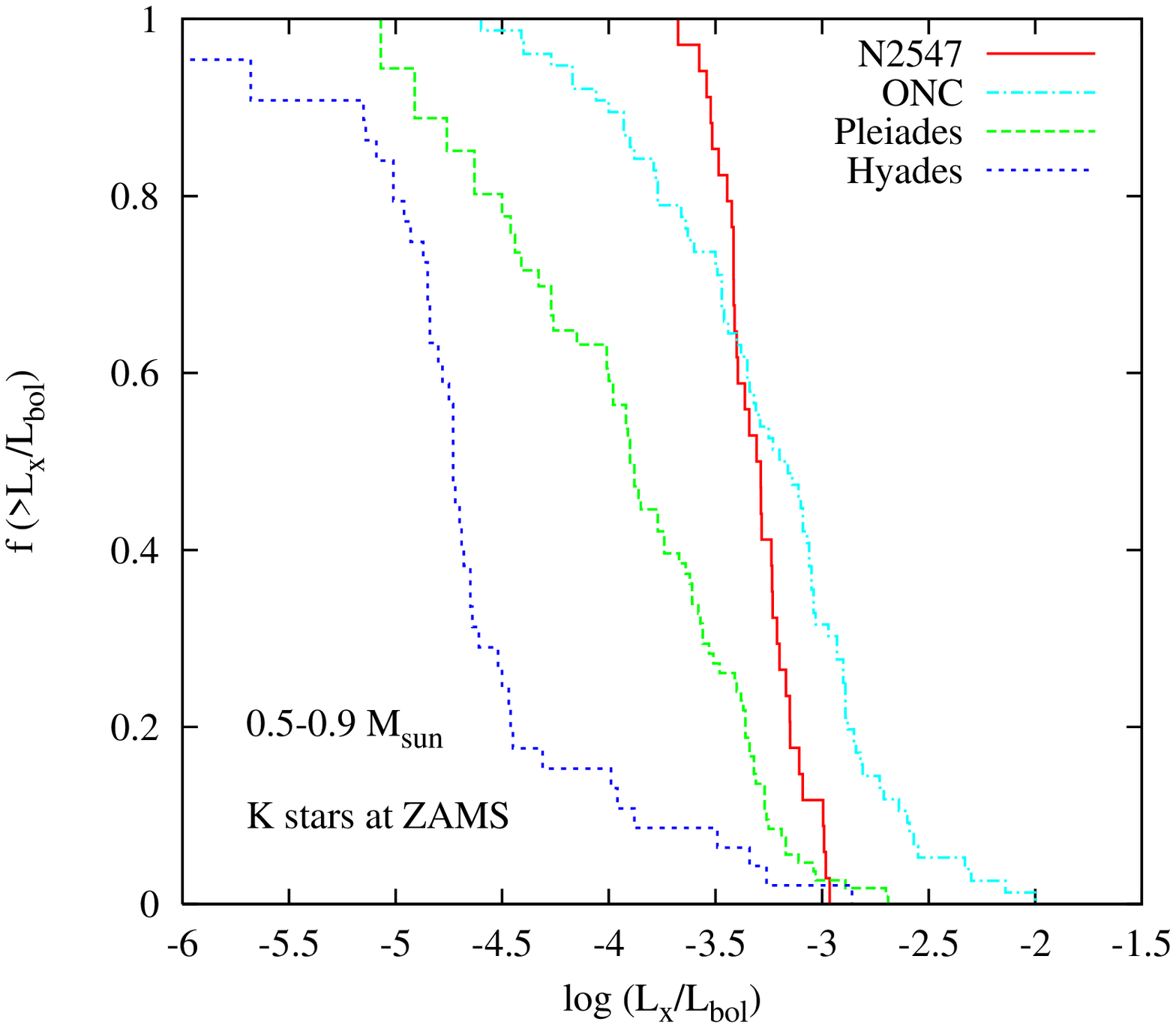}
    \includegraphics[width=71mm]{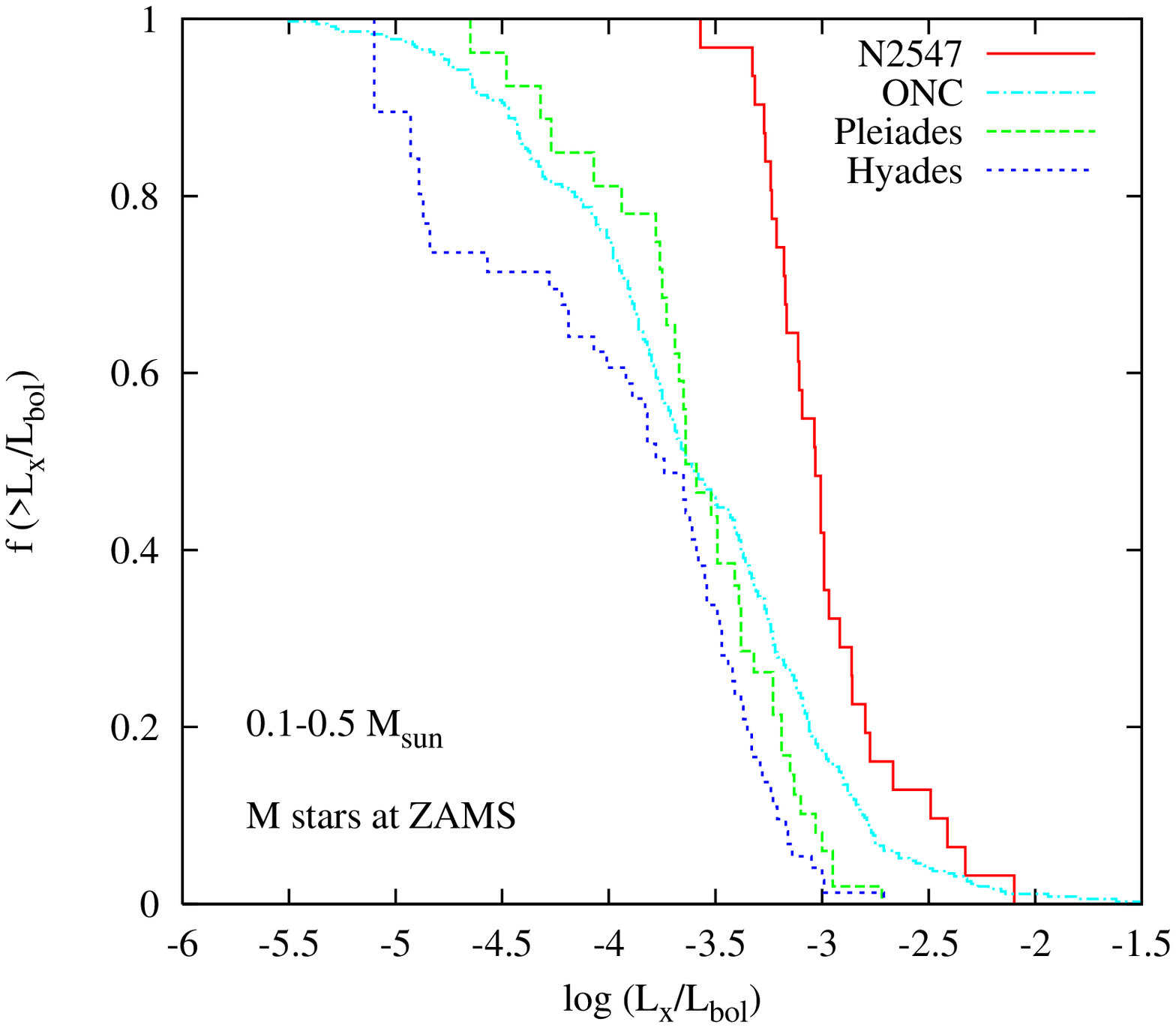}
    \end{minipage}
    \caption{Evolution of X-ray luminosity and ratio of X-ray to
      bolometric luminosity for stars in three mass ranges from the
      very early PMS in the Orion nebula cluster through NGC\,2547 to
      ZAMS stars in the Pleiades and Hyades clusters. X-ray
      luminosities are presented in the 0.5--8.0\,keV band. Data for
      other clusters comes from Stelzer \& Neuh\"auser (2001) and
      Preibisch \& Feigelson (2005). Note that for the lowest mass
      range in NGC~2547, the X-ray data are incomplete and therefore
      the X-ray luminosity functions are biased upwards (see text).}
\label{xlfplot}
\end{figure*}

Figure~\ref{xlfplot} shows the cumulative X-ray luminosity functions
(XLFs) for NGC~2547 and the equivalent functions for $L_{\rm x}/L_{\rm
bol}$.  Comparison plots are shown for the ONC, Pleiades and
Hyades. These latter samples were obtained from the works of Stelzer \&
Neuh\"{a}user (2001) and Preibisch \& Feigelson (2005). To provide a
fair comparison, all the X-ray fluxes have been adjusted to correspond
to the 0.5--8.0\,keV band considered by Preibisch \& Feigelson
(2005). This has been achieved by subtracting 0.14\,dex from the
0.1--2.4\,keV Pleiades and Hyades fluxes (see Preibisch \& Feigelson)
and by subtracting 0.04\,dex from the NGC\,2547 0.3--3.0\,keV fluxes
(see section~\ref{xraylum}).  

The Pleiades and Hyades XLFs account for X-ray flux upper limits in
non-detected cluster members using ``survival analysis'' techniques. No
account of upper limits is taken for the ONC and NGC~2547 data. In the
ONC essentially all cluster members were detected at X-ray
wavelengths (see Getman et al. 2005). 
This is also true for NGC~2547 members in certain mass or
colour ranges, but not in others (see section~\ref{complete}). It is
not easy to take account of upper limits to the X-ray fluxes of
undetected NGC~2547 members because, outside of photometric candidacy,
there is no list of confirmed cluster members. Hence the inclusion of
X-ray upper limits for photometric candidates which turn out to be
non-members could bias the results in a way that is very difficult to
assess. Instead we deal with the XLF of detected photometric
candidates, accepting that where the X-ray census is incomplete then
the XLF will be overestimated.

X-ray emission is known to be mass-dependent, or at least to depend on
the structural properties of a star -- which are both mass and age
dependent. Fig.~\ref{xlfplot} provides XLFs for three different mass
subsets. These are not the same as spectral type subsets because the
ONC stars are sufficently young that most stars are on their Hayashi
tracks and significantly cooler than they will appear when they reach
the ZAMS. The mass estimates for the ONC sample are discussed by Getman
et al. (2005).  For NGC~2547 we use a relationship between $V-I$ and
mass from the isochrone adopted in Fig.~\ref{vvicmd} to to make similar
mass-range selections; that $0.9\leq M < 1.2\,M_{\odot}$
corresponds to $1.035\geq V-I > 0.670$; that $0.5\leq M <
0.9\,M_{\odot}$ corresponds to $2.47\geq V-I > 1.035$; and that
$0.1\leq M < 0.5\,M_{\odot}$ corresponds to $3.94 \geq V-I > 2.47$. In
section~\ref{complete} arguments were presented to suggest that the
former two samples are likely to be complete for NGC~2547 and suffer
little contamination. The lower-mass sample is likely to be very
incomplete for NGC~2547. There are no X-ray detections with $V-I>3.2$
and a significant number of X-ray undetected candidates with $2.8< V-I
< 3.2$.  The simple G, K, M  spectral type divisions for the XLFs of
the Pleiades and Hyades used by Stelzer \& Neuh\"{a}user are assumed to
correspond approximately with the three mass ranges defined for
NGC~2547.

\subsubsection{G stars at the ZAMS}

The XLFs of stars that are or will become G-stars at the ZAMS show
large changes with time. These changes could be caused by decreases in
the convective turnover time ($\tau_{\rm conv}$) 
and decreases in the rotation period, but
may also be connected with the disappearance of circumstellar material.
For a $1\,M_{\odot}$ star, Gilliland (1986) shows that $\tau_{\rm
conv}$ decreases by a factor of 10 between 1 and 30\,Myr, but hardly
changes thereafter. The same evolutionary models also show the total
moment of inertia of the star decreasing by a factor of 10 between 1
and 30\,Myr and staying constant thereafter. Without angular
momentum loss (AML) the rotation periods would decrease with the 
moment of inertia and little change in the Rossby number would be
expected. The other major structural change is of course the
development of a radiative core which takes place after a few Myr.
Looking at the upper panels of Fig.~\ref{xlfplot} we see that
whilst the upper levels of X-ray luminosity and $L_{\rm x}/L_{\rm bol}$
are higher in the ONC than NGC~2547, the median and minimum levels are
very close. This is an important result -- indicating that the
high level of X-ray irradiation seen in the vicinity of the young ONC
stars is maintained for at least $\simeq 30$\,Myr prior to a quite
significant decrease by the age of the Pleiades.

In terms of the ARAP we might conclude that there was very little AML
between 1 and 30\,Myr so that stars in NGC~2547 and the ONC had similar
Rossby numbers. However, we know that this is not true as rotational
velocities have been measured for reasonable samples of solar-type
stars in both clusters. The Rossby numbers in the ONC are significantly
smaller (see Fig.~\ref{rossbyplot}). It is still an unsolved puzzle as
to how this angular momentum is lost but probably involves interactions
between circumstellar material and the coronal magnetic fields as well
as losses through a magnetised stellar wind.  X-ray activity levels are
maintained at high levels in NGC~2547 because many (but not all) stars
have not yet slowed sufficiently to place them on the declining portion
of the relationship between activity and Rossby number in
Fig.~\ref{rossbyplot}. On the other hand, the {\em range} of activity
levels in the ONC is too large to be explained by the ARAP. This is
discussed at length by Stassun et al. (2004) and Preibisch et al. (2005)
who conclude that stars without active accretion have
saturated or super-saturated X-ray activity that is equivalent to
fast-rotating ZAMS stars, but that active accretors often have
significantly suppressed X-ray activity. Active accretion is not found
in the NGC~2547 stars and the spread in X-ray activity is much smaller
and quite consistent with the spread in rotation rates and a small
amount (less than a factor of two -- see section~\ref{comphri}) of
variability.

Between NGC~2547 and the Pleiades there is an order of magnitude
decline in the median X-ray activity, a similar and perhaps even larger
decline in the minimum activity levels, but comparatively little
decline in the peak levels of X-ray activity. In terms of the ARAP,
this can be understood if the fastest rotating G-stars in the Pleiades
have yet to spin down sufficiently that their X-ray activity falls
below the saturation level. Queloz et al. (1998) find that about 14 per cent of
0.9--1.1\,$M_{\odot}$ Pleiads have equatorial velocities exceeding
$20$\,km\,s$^{-1}$ and hence Rossby numbers $\leq 0.2$ which lead to
saturated X-ray activity. Conversely, to explain their much lower X-ray
activity, the slowest rotators in the Pleiades must rotate at rates
that are factors of $\simeq 3$ lower than the slowest rotators in
NGC~2547. This tallies with the available observations. From the
projected equatorial velocity measurements in Jeffries et al. (2000) we
see that only 4 out of 20 G-type NGC~2547 members have $v \sin i
<10$\,km\,s$^{-1}$, whereas that fraction is about 50 per cent in the
Pleiades, with the slowest 10 per cent having rotational velocities
similar to those in the even older Hyades (Queloz et al. 1998).  Hence
the solar-type stars undergo a period of rapid spindown during the
first 100\,Myr on the ZAMS, followed by a much longer ``plateau''
phase. This behaviour can plausibly be explained if the convective
envelope and radiative core are rotationally coupled on a timescale
somewhere between the age of the Pleiades and Hyades (Sills,
Pinsonneault \& Terndrup 2000).

\subsubsection{K stars at the ZAMS}

The situation is somewhat different in the lower mass stars. Here, the
structural evolution means that whilst $\tau_{\rm conv}$ 
decreases by a factor of 2--6 for 0.5--0.9\,$M_{\odot}$ stars between 1
and 30\,Myr, there is a corresponding decrease in the moment of inertia
by factors of 6--8 for the whole star. In the absence of AML we
might expect the Rossby number to decrease by factors of 1.3--3. Beyond 30\,Myr
there is little evolution of either $\tau_{\rm conv}$ or
moment of inertia, so the Rossby number would remain nearly constant in
the absence of AML (see Gilliland 1986). The K-stars in the NGC~2547
sample will have developed a radiative core during the last $\simeq 20$\,Myr.

The most notable property of the NGC~2547 XLF for K-stars is the very
narrow spread in $L_{\rm x}$ and $L_{\rm x}/L_{\rm bol}$. In
terms of the ARAP this is not explained by a narrow spread in
rotational rates, but instead by the majority of objects rotating
fast enough to lie on the saturated, or even super-saturated portions
of the relationship between X-ray activity and Rossby number. It is
therefore startling to see that a significant fraction ($\simeq 30$ per
cent) of ONC stars in this mass range have lower $L_{\rm x}/L_{\rm
bol}$ values than even the least active NGC~2547 members, although they
do still have higher values of $L_{\rm x}$. There is no explanation of
this in terms of the ARAP unless the K-stars of NGC~2547 have an
angular momentum distribution that is skewed to higher values compared
with the ONC. But even if that were true, there are clear examples of
ONC objects with very small Rossby numbers that have $L_{\rm x}/L_{\rm
bol}\simeq 10^{-4}$ or even lower. Again, it is accreting objects in the ONC
that tend to have these lower levels of $L_{\rm x}/L_{\rm bol}$. If
anything, the bolometric luminosities derived by Getman et al. (2005)
for these stars may be underestimates, making the result more significant
(see discussion in Hillenbrand 1997).

Between NGC~2547 and the Pleiades there is an order of magnitude
decrease in the minimum and median levels of X-ray activity, but
comparable levels of maximum X-ray activity. Again, this can be interpreted
in terms of the rotation distributions in the two clusters. A
fraction ($\simeq 20$ per cent) of K-type Pleiades have equatorial
velocities exceeding 15\,km\,s$^{-1}$ that would result in
saturated levels of coronal emission. Almost half have spun down to
less than 7.5\,km\,s$^{-1}$ (Queloz et al. 1998).  Unfortunately
only three $v \sin i$ measurements exist for NGC~2547 stars in the same
mass range (13.4, 9.3 and 19.2\,km\,s$^{-1}$), so we cannot test this
hypothesis in detail. On the basis of the NGC~2547 XLF we predict
that NGC~2547 members with masses of 0.5--0.9\,$M_{\odot}$ all have
Rossby numbers less than 0.3 and equatorial velocities in excess of
10\,km\,s$^{-1}$.

\subsubsection{M stars at the ZAMS}

The lowest mass stars we have considered have a much slower
evolutionary timescale on the PMS. Hence changes in 
$\tau_{\rm conv}$ and moment of inertia are expected both before and
after the age of NGC~2547. In the absence of AML there should be a
decrease in the Rossby number by factors of a few between 1 and
30\,Myr, but only small changes after that (Gilliland 1986).
A notable difference between this subsample and the higher mass
stars considered previously is that many of the NGC~2547 stars may
still be fully convective. In the D'Antona \& Mazzitelli (1997) models,
this transition takes place at 0.4\,$M_{\odot}$ at
30\,Myr, corresponding to $V-I=2.58$. There is no great change apparent
in Figs~\ref{lxfig} and~\ref{lxlbolfig} at this colour, except perhaps
for the development of a small tail of high activity objects.

Unlike the higher mass subsamples, the X-ray census of M stars in
NGC~2547 is definitely not complete and so the XLFs will be biased
upwards. There are no detections of any NGC~2547 stars with
$M<0.2\,M_{\odot}$, although confirmed cluster members (see Jeffries \&
Oliveira 2005) are certainly within the {\it XMM-Newton} field of view.
All we can say is that the majority of
NGC~2547 M stars are expected to exhibit saturated or even super-saturated levels of
X-ray emission and this is consistent with the observed XLFs. The peak
levels of X-ray luminosity and activity are close to those in the ONC
and several times higher than in the Pleiades and Hyades. It is
hard to be sure that this is a significant difference though, because
approaching the flux sensitivity limit of the survey one is bound to
upwardly bias the fluxes of the detected objects simply through
fluctuations in the background and the intrinsic X-ray variability of
the stars. It would be fascinating to do a deeper survey in order to
establish whether NGC~2547 (like the ONC 0.1--1.2\,$M_{\odot}$ stars)
exhibits a significant fraction of objects with $L_{\rm x}/L_{\rm
bol}>10^{-2.5}$ and objects with $L_{\rm x}/L_{\rm bol}<10^{-4}$. The
confirmed existence of either of these two populations might lend
support to the suggestion that young, fully convective stars might not
conform to the ARAP possibly through the operation of a dynamo that is
distributed throughout the convective interior rather than at the
interface between the radiative core and convective envelope (Feigelson
et al. 2003).

\subsection{The evolution of coronal temperatures}
\label{specevol}

\begin{figure}
\includegraphics[width=84mm]{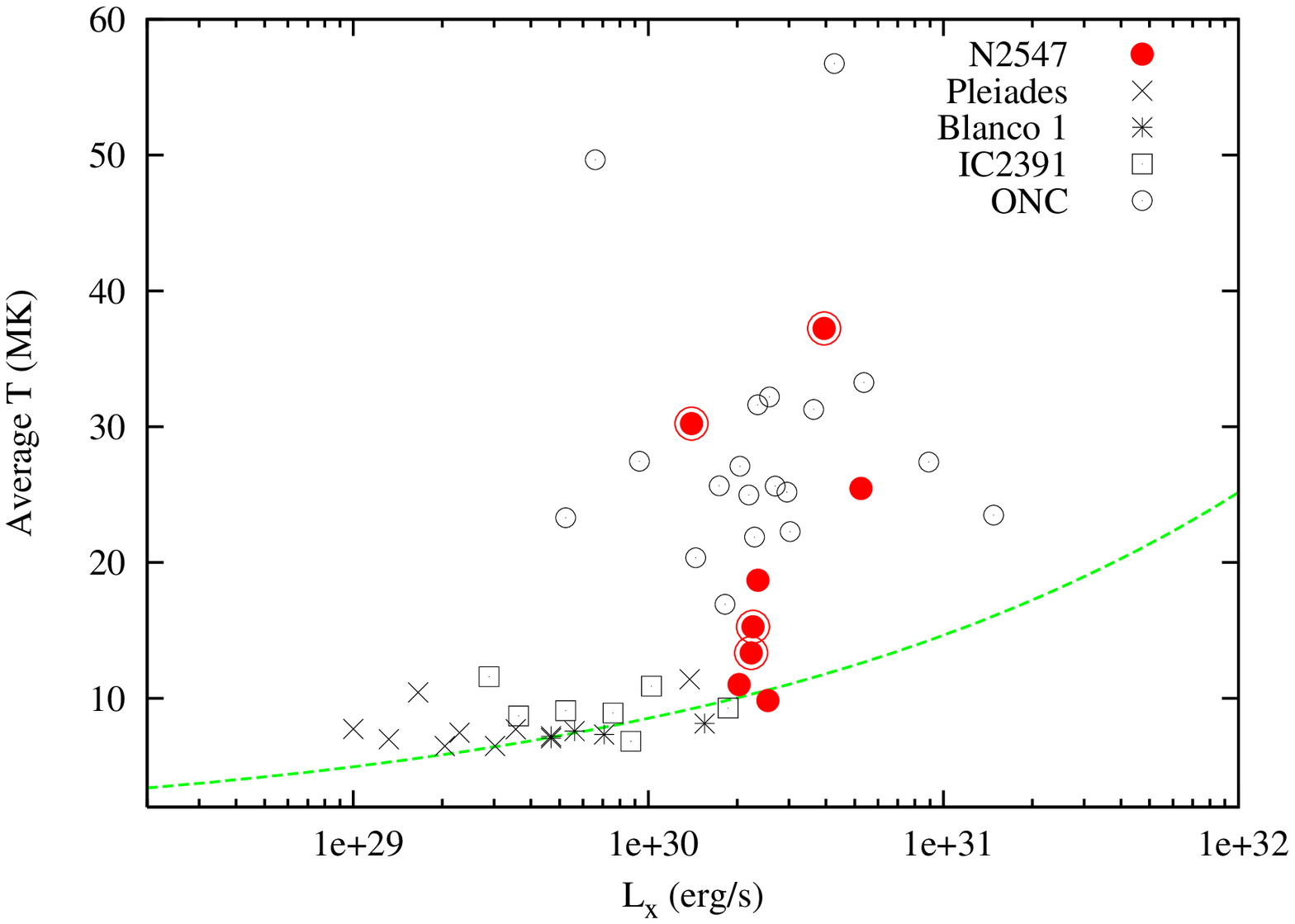}
\includegraphics[width=84mm]{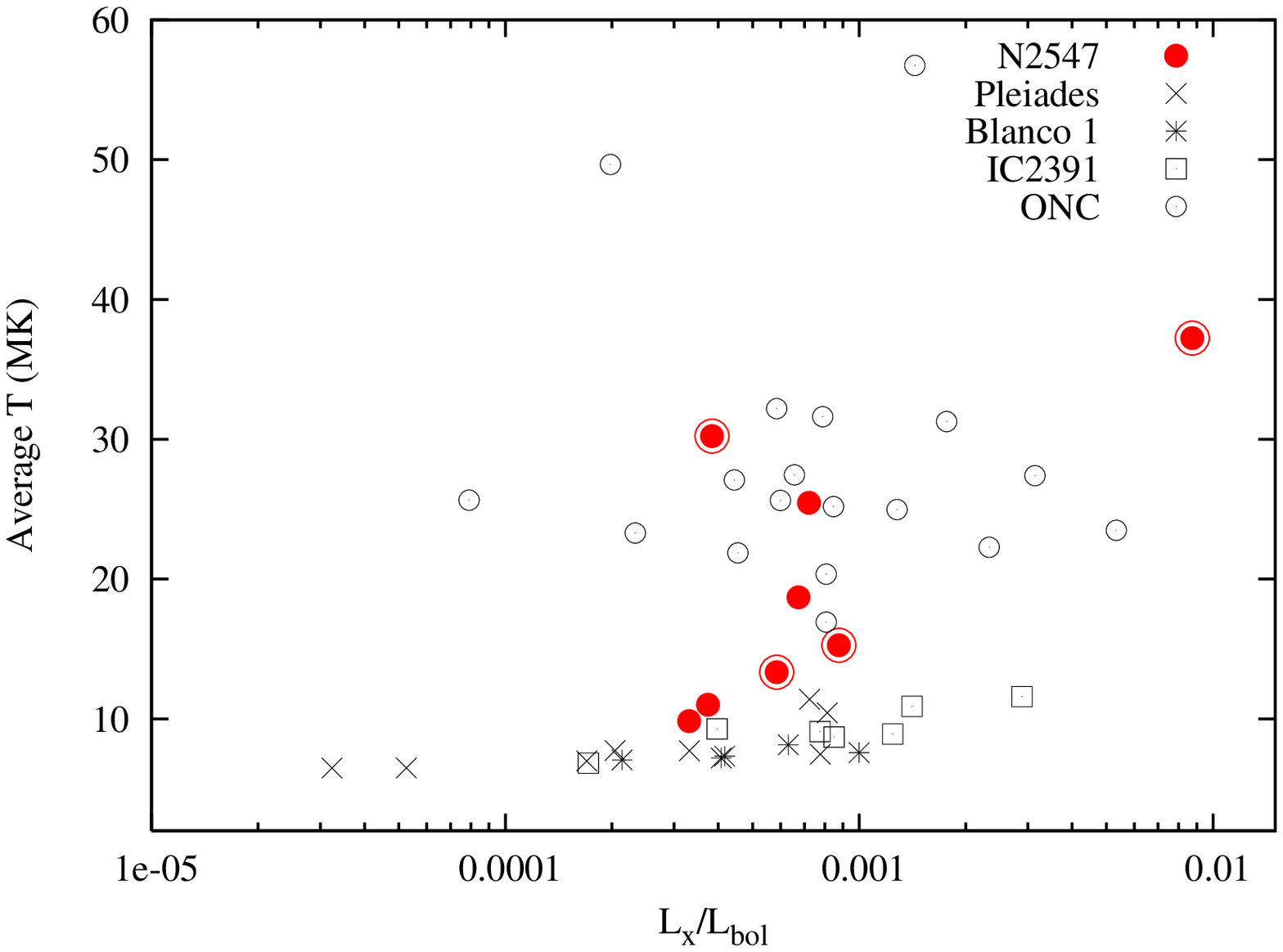}
\caption{Emission measure weighted mean coronal temperature as function
  of $L_{\rm x}$ and $L_{\rm x}/L_{\rm bol}$ for low-mass stars in
  NGC~2547 and several other young clusters (see text). The dashed line
  is a relationship derived for field G-stars by Telleschi et
  al. (2005) that has been extrapolated beyond
  $2\times10^{30}$~erg\,s$^{-1}$. Objects in NGC~2547 which are
  ``contaminated'' with flaring emission are encircled.
}
\label{lxmeant}
\end{figure}

In section~\ref{tactivity} we showed that average coronal temperatures,
indicated by crude hardness ratios, increased with $L_{\rm x}$, $L_{\rm
x}/L_{\rm bol}$ and $F_{\rm x}$, although there was considerable
scatter in these relationships. As these activity
indicators decline with time (also with considerable scatter) an
interesting question is whether coronal temperatures, and by
implication coronal structures and heating mechanisms, change solely
with coronal activity, or whether there is some other time-dependent
variable involved. It has been known for some time that this is the
case for older stars, with average coronal temperatures decreasing
among solar type stars between ages of $\simeq 70$\,Myr and 9\,Gyr. This
evolution appears to be largely governed by the gradual disappearance
of the hotter ($\ga 10$\,MK) coronal plasma (e.g. G\"udel, Guinan \&
Skinner 1997; Telleschi et al. 2005). Telleschi et al. (2005) propose a
relationship of the form $L_{\rm x}\propto T^{4.26}$ between coronal
luminosity and mean coronal temperature,


This dependence of coronal temperature on activity and age apparently
continues to younger ages. Whilst the most active field stars studied
by G\"udel et al. (1997) have a coronal temperature of about 10\,MK,
detailed X-ray spectroscopy of T-Tauri stars (both accreting and
non-accreting) finds that their coronae are dominated by a very hot
component ($>20$\,MK) that is seen only in the largest flares in the
solar corona (Skinner \& Walter 1998; Grosso et al. 2004). What is not
clear is whether these hotter coronae are merely an extension of the
trend defined by older stars, which is presumably driven by changes in
rotation rate, or whether the significant increase in coronal
temperatures is attributable to differences in the stellar structure or
the presence of discs in PMS stars.

Figure~\ref{lxmeant} illustrates where the stars of NGC~2547, at an age
of $\simeq 30$\,Myr, fit into this progression. The emission measure
weighted mean coronal temperature is shown as a function of $L_{\rm x}$
and $L_{\rm x}/L_{\rm bol}$ for the 8 stars in Table~\ref{tab:fits2}
with $B-V\geq 0.536$ and which are unambiguously low-mass stars. The
dashed line indicates the relationship derived for 0.07--9\,Gyr G-type
stars in the field, that have $2\times10^{28}\leq L{\rm x}\leq
2\times10^{30}$~erg\,s$^{-1}$ (Telleschi et al. 2005). Comparable data
were gathered from the literature for low-mass stars in IC~2391 (age
50\,Myr, Marino et al. 2005), Blanco~1 (age $\simeq 100$\,Myr,
Pillitteri et al. 2004) and the Pleiades (age 120\,Myr, Briggs \& Pye
2003). These comparison samples were chosen on the basis that the
spectra were also obtained with the {\it XMM-Newton} EPIC instrument
and modelled with two thermal components with abundance as a free
parameter. Finally, data were added for a sample young PMS stars in the
ONC (Getman et al. 2005). These spectra were obtained with the {\it
Chandra} ACIS instrument, but modelled in a similar manner to the
NGC~2547 stars.  Objects were selected that had $0.3<M<1.2\,M_{\odot}$,
that had X-ray sepectra which were modelled with two thermal components
and for which no problems were reported in the spectral fits (see
Getman et al. for details).

It appears that the PMS stars of the ONC have coronae that are {\it
much} hotter than a simple extension of the trend defined by older
stars with lower $L_{\rm x}$. This becomes even more apparent when
considering mean coronal temperature versus $L_{\rm x}/L_{\rm bol}$,
where the higher coronal temperatures of the ONC appear to be a
consequence of youth rather than a higher overall level of magnetic
activity.  The ONC stars are photospherically
cooler than the young cluster samples
because of their younger evolutionary stage and therefore using X-ray
surface flux ($F_{\rm x} \propto T_{\rm eff}^{4}\,L_{\rm x}/L_{\rm
bol}$) as an activity indicator would yield an even worse correlation.

The stars of NGC~2547 form an intermediate population. They lie
significantly above the mean relationship defined by the field G-stars.
Of course there is a bias towards the most active stars in NGC~2547,
and the objects with the highest mean temperatures (stars 5 and 12)
were seen to flare during the observation, but this is also the case in
the comparison stars in the other clusters. When compared at a given
activity level (defined by $L_{\rm x}/L_{\rm bol}$) the NGC~2547 stars
lie between the older clusters and the ONC. We emphasize that the
Blanco~1, Pleiades and IC~2391 samples were analysed with the same
instrument and in an entirely consistent manner with our approach.  We
also checked whether our assumed hydrogen column density of
$3\times10^{20}$~cm$^{-2}$ could influence this result. Doubling the
column density (see section 3.1) reduces the mean coronal temperature
in our NGC~2547 sample by only $\simeq 10$ per cent -- insufficient to
account for the observed differences.

``Activity'', either judged by $L_{\rm x}$, $L_{\rm
x}/L_{\rm bol}$ or $F_{\rm x}$, is not capable of predicting what the
average coronal temperature will be, so it is difficult to attribute a
causal effect to these quantities.  Instead there must be an additional
age-related factor which should be taken into account. It is unlikely
that this factor has anything to do with active accretion or the
presence of circumstellar material. We can confidently rule these out
in the case of NGC~2547 and of 19 objects included in the ONC sample only
4 are classical T-Tauri stars and about half show evidence of a near-IR
excess, but these do not appear to have systematically higher
temperatures.  This leaves changes in the interior structure of the PMS
stars, the absolute rotation rate or perhaps the surface gravity as the
most likely causative effects of the higher coronal temperatures.

One possibility is that the higher coronal temperatures in the younger
stars are a manifestation of a more turbulent dynamo distributed
throughout the entirely convective interior of the young ONC stars
(e.g. see the discussion in Feigelson et al. 2003 and references
therein). This could lead to differing coronal structures and higher
temperatures, perhaps as a result of a less ordered magnetic field and
more frequent flaring interactions on a variety of scales.
Intermediate coronal temperatures in the older NGC~2547 stars may be
due to the
beginning of strong magnetic braking, differential rotation between the
developing radiative core and envelope and hence the presence of a
tachocline region, and initiation of an $\alpha\, \Omega$ dynamo
(e.g. Parker 1993). The influence of a distributed convective dynamo
that depends on a deep convection zone would diminish. 

There are (at least) two problems with these ideas: First, there is no
clear evidence that large flares are more common in ONC stars than the
most active stars in NGC~2547 or indeed older ZAMS cluster (see section
6.3.1). Second, if being young and fully convective is the recipe for
hotter coronal temperatures then $>20$\,MK coronae should be present in
all stars with $M<0.4\,M_{\odot}$ ($V-I>2.65$) in NGC~2547. Although we
have no detailed spectra with which to test this, the hardness ratios
are reasonably indicative.  Simulating a 2\,keV thermal plasma with a
similar column density and abundance to the other NGC~2547 stars yields
a pn hardness ratio of 0.00. There {\em is} a hint that coronae get
hotter for $V-I>2.5$, but hardness ratios are comfortably below zero
for stars of any spectral type (see Fig.~\ref{hardnessfig}). 
Of course this analysis neglects any cool coronal component and it is
true that there are a few fully convective stars with hardness
ratios as large as the hottest stars with detailed spectral fits.
More detailed spectral modelling of better
data is required.


An alternative explanation for hotter coronae in PMS stars arises from
simple scaling laws (see also Jordan \& Montesinos 1991).  Suppose that
the coronal plasma and magnetic field approach pressure equilibrium,
such that $p \propto B^{2}$, where $p$ is the coronal pressure and $B$
the coronal magnetic field. We could further assume that X-ray emission
is observed from magnetic loops that extend to some fraction of a
pressure scale height, so that $L \propto T g^{-1}$, where $L$ is the
loop semi-length and $g$ the surface gravity. The final ingredient is the
scaling law between loop length, pressure and temperature derived for
hydrostatic loops by Rosner, Tucker \& Vaiana (1978), who show that $T
\propto p^{1/3}\, L^{1/3}$. Combining these we find
\be
T \propto B\, g^{-1/2}\, .
\label{tr0}
\ee
For large (unsaturated) Rossby numbers, mean field dynamo theory
suggests that $B$ should increase with decreasing Rossby number (Durney
\& Robinson 1982), but for small (saturated) Rossby numbers it is quite
possible that feedback effects limit any further growth in $B$ or
magnetic activity. Hence at large Rossby numbers and for stars that
have already reached the main sequence with similar gravities, coronal
temperature will chiefly depend on Rossby number and hence rotation (or
$L{\rm x}/L_{\rm bol}$ or $L_{\rm x}$ for a given spectral type) -- as
observed in older G dwarfs (Telleschi et al. 2005). For small Rossby
numbers and saturated coronae the dependence should be dominated by
gravity variations.  The models of D'Antona \& Mazzitelli (1997)
suggest that the surface gravity of a 0.8\,$M_{\odot}$ star changes
from $\log g = 3.74$ to $\log g = 4.48$ (in cgs units) between 1 and
30\,Myr, with only a small further decrease thereafter. Accordingly,
equation~\ref{tr0} predicts that stars in the saturated regime will
exhibit a corresponding drop in coronal temperature of a factor 2.3
between 1 and 30\,Myr -- approximately what is seen. A probable
complicating factor is that the scale heights in the ONC stars will be
large enough for centrifugal forces to play a role. Indeed, Jardine \&
Unruh (1999) have shown that coronal loops extending beyond the
Keplerian co-rotation radius would be unstable and the simple pressure
scale height equals the co-rotation radius at $T\simeq 30$\,MK for a
star with $\log g = 3.5$ and rotation period of 5 days.

If coronal temperature just scales with gravity as described above then
this is readily testable by determining the relative coronal heights
ONC and ZAMS stars. The lower gravity ONC coronae should be
approximately 10 times more extended ($T$ is 2.5 times hotter and $g$
is 4 times smaller) than in active ZAMS stars\footnote{Note this is in
absolute units, not relative to the stellar radius.}.  In practice,
there is little strong evidence for this. Whilst the coronae of ZAMS
stars have usually been found to be consistent with an extent of a
pressure scale height (or less) from eclipse mapping or combinations of
emission measure and density estimates (see G\"udel 2004 for a review),
the same diagnostics are not yet available for PMS stars.  Perhaps the
only technique applied to both classes of objects is the modelling of
strong X-ray flares with loop models. Favata et al. (2005) have used
the analysis of strong flares to determine loop half lengths of
$10^{9}$--$10^{10}$\,m in several ONC stars.  Equivalent analyses of
strong flares on very active ZAMS stars do imply significantly smaller
loop lengths ($\simeq 3\times10^{8}$\,m -- G\"udel et al. 2001; Scelsi
et al. 2005) with an extent of about a pressure scale height. It must
be emphasised though that these approaches to interpreting spatially
unresolved flares are highly model dependent.

\subsection{The evolution of coronal variability}
\label{tevol}

\subsubsection{Flaring}

In principle we could attempt to compare the ``flare rate'' in NGC~2547
stars with objects of similar mass in older and younger clusters. In
practice this turns out to be difficult because there is little
uniformity in the literature about how a flare should be defined and
numerous selection effects concerning the detection of flaring events
that are difficult to account for. In particular, care must be taken in
comparing data with differing sensitivities (for instance from clusters
at different distances) or from instruments with different energy
band-passes.

In section~\ref{flare} and Table~\ref{flaretable} a number of possible
flaring events were identified. Although we have identified one major
flare on an M-star in NGC~2547, the majority of such sources are so
weak in our data that a proper assessment of the flaring frequency in
M-stars is impossible. The situation is better for solar-type stars,
where we would claim to have been able to detect all flares with
integrated energies exceeding $10^{34}$~erg, provided that
they had total durations $\la 30$\,ks. We have detected 4 such flares
and 2 others that fall just outside these thresholds from a sample of 28
solar-type stars with masses $0.8<M<1.2\,M_{\odot}$. Each source was
monitored for approximately 50\,ks, with no gaps large enough for us to
have missed any significant flaring. Hence the flare rate (for flares
with total energy $>10^{34}$~erg and duration $<30$\,ks) is
approximately 1 every $350^{+350}_{-120}$ \,ks.

Equivalent statistics are presented or can be calculated from a few
sources in the literature for older and younger stars.
Gagn\'{e} et al. (1995a) used {\em ROSAT} observations of the Pleiades
to find 12 flares with peak $L_{\rm x}$ of
$10^{30}$--$2.5\times10^{31}$~erg\,s$^{-1}$.  We estimate that they
observed 3 flares from G-stars with energies in excess of $10^{34}$~erg
(making a small approximate correction for the differences in
band-passes) from an effective monitoring time of 60\,ks on 33
G-stars. Hence the comparable flaring statistic to our NGC~2547
measurement is 1 flare every 660\,ks. Wolk et al. (2005) have presented
a detailed search for flares in ONC stars with $0.9<M<1.2\,M_{\odot}$
using the {\it Chandra} COUP dataset. They conclude that the equivalent
flare rate statistic is 1 flare every 650\,ks.  In terms of duration
and peak luminosity, the flares in NGC~2547 are quite comparable with
the large flares seen in the ONC and Pleiades. Larger flares
($>10^{35}$~erg) are seen occasionally in the ONC, but they are
reasonably rare and it is not necessarily significant that none were
observed in the solar-type stars of NGC~2547.

Hence there appears to be little or no change in the rate of occurrence
of large flares in solar-type stars during their first $\sim 100$\,Myr.
If anything the flare rate seems larger in NGC~2547 than the ONC, but
the reader should note that 2 of the 4 flares in NGC~2547, those in
stars 10 and 15, may not have been classified as flares in the ONC
study of Wolk et al. (2005), because their peak luminosities were less
than three times their quiescent luminosity. On the other hand, some of
the ONC flares were so long ($>30$\,ks) that such variability may not
have been classed as a flare in this paper. A longer observation of
NGC~2547 is required for a more reliable comparison.

A similar flare rate in NGC~2547, the ONC and the Pleiades seems at
odds with explanations of the enhanced coronal temperatures in very
young PMS stars which rely on enhanced flaring rates. On the other
hand, only the rate of occurrence of the very brightest flares have
been measured here. It is still possible that the flare energy spectrum
has a shallower slope in the ONC than NGC~2547 or the Pleiades,
resulting in coronal heating by more numerous smaller flares.

\subsubsection{Long term variations}

Comparisons of long term variability as a function of age needs to be
restricted to those studies which (i) define variabilty in terms of an
amplitude, rather than stating that some fraction of stars are
``variable'' which will just be an increasing function of the
sensitivity of the observations; (ii) consider upper limits where
necessary and do not restrict themselves to stars detected in all
observations and (iii) adequately account for systematic differences in
the estimated luminosities from different instruments.  Several papers
have been found that fulfil these criteria.

In section~\ref{comphri} it was established that only 10--15 per cent
of G/K stars or stars with $L_{\rm x}>3\times10^{29}$~erg\,s$^{-1}$ in
NGC~2547 show variations of a factor $\geq 2$ on timescales of 7 years.
Similarly, Simon \& Patten (1998) find only 2/28 members of IC~2391
(age $\simeq 50$\,Myr) have X-ray luminosities that vary by more than a
factor of two on a 2 year timescale.

In older clusters Gagn\'e et al. (1995a) and Micela et al. (1996) found
that $\simeq 25$ per cent of low-mass Pleiads (age $\simeq 120$\,Myr)
showed variations of a factor two or more on timescales of 1-2 years
and perhaps 40 per cent on timescales of 11 years.  Pillitteri et
al. (2005) found that at least 8/22 G- and K-type stars in Blanco~1
(age $\simeq 100$\,Myr) varied by just over a factor of two on a 6 year
timescale.  Marino et al. (2003) showed that older (but still
relatively X-ray active) solar-type (F7-K2) field stars were even more
variable than their counterparts in the Pleiades.  Micela \& Marino
(2003) have shown that the Sun would exhibit variations by factors of
$\simeq 10$ if observed in a similar manner over the course of its
magnetic activity cycle.  Favata et al. (2004) have presented the first
compelling evidence for solar-like X-ray variations in an old field G2
star, where $L_{\rm x}$ varied by a factor of 10 in 2.5 years.

Detailed studies of the long-term variability of 
younger stars are scarce. There are some
indications that X-ray variability may be much more common in young PMS
stars. Gagn\'e, Caillault \& Stauffer (1995b) claim that {\em at least}
25 per cent of X-ray luminous late-type ONC stars vary by more than a factor of two on
timescales of 1 year. Preibisch et al. (2005) quote a median absolute
deviation from equal luminosities of 0.31\,dex between their {\it
Chandra} observations of ONC stars and another long {\it Chandra}
dataset taken 3 years earlier. This is much higher than the equivalent
variability of 0.1\,dex reported here for NGC~2547, suggesting that
variations of a factor of 2 are common in very young PMS stars.

In summary the long term X-ray variability presents a confused
picture. Very young PMS stars may be very variable, but
clusters at the age of NGC~2547 show very little long term variability
on timescales of 7\,Myr and so certainly nothing like a solar magnetic
activity cycle with a period less than about 20 years.  Any variability
which exists may be entirely attributable to large flares which can
significantly alter the average X-ray luminosity seen in relatively
short ($<1$ day) observations.  The amount of long-term variability
then appears to increase for older clusters, but it is not clear
whether this marks the development of activity cycles or whether it is
due to the maintenance of a reasonably high rate of large flares
contrasting with rapidly decreasing ``quiescent'' X-ray luminosity with
age (see Fig.~\ref{xlfplot}).

\section{Summary}

With a reasonably long and sensitive {\it XMM-Newton} observation of
NGC~2547 we have been able to go some way towards characterising the
X-ray emission from analogues of the Sun (and at lower masses) at ages
of $\simeq 30$\,Myr. The main findings of this study can be summarised
as follows:
\begin{enumerate}
\item
Candidate cluster counterparts have been found for 108 X-ray sources.
X-ray emission is seen from stars of all spectral types. X-ray
luminosities peak among the G-stars at $10^{30.5}$~erg\,s$^{-1}$
declining to $\leq 10^{29.0}$~erg\,s$^{-1}$ among M stars with masses as low as
$0.2\,M_{\odot}$. The ratio of X-ray to bolometric luminosity increases
rapidly with decreasing $T_{\rm eff}$ among F-type stars, 
reaching a saturated peak of $L_{\rm
x}/L_{\rm bol} \simeq 10^{-3}$ for G stars and cooler.

\item
Optically thin thermal plasma models have been used to fit the X-ray
spectra of the most active stars in NGC~2547. Where there were
sufficient detected X-ray photons, a multi-temperature model fares
better than a single temperature model. Coronal metal abundances are clearly
sub-solar. There are reasonable correlations between increasing coronal
temperature, indicated by a hardness ratio, and X-ray
luminosity, surface flux or $L_{\rm x}/L_{\rm bol}$. However, none of
these relations is entirely deterministic, with a significant scatter
in hardness ratio at any activity level.

\item
The relationship between X-ray activity and Rossby number among the G-
and K-type stars of NGC~2547 follows that established in a number of
older clusters and field stars. The majority of the solar-type stars in
NGC~2547 have saturated or even super-saturated levels of coronal
activity that are commensurate with their rotation rates. Based on
their saturated activity levels, we predict
that none of the K stars in NGC~2547 have spun down below equatorial
velocities of 10\,km\,s$^{-1}$.

\item
The X-ray luminosity functions (XLFs) of NGC~2547 have been compared
with younger and older stars in several mass ranges. For
G- and K-type stars the
median activity levels in NGC~2547, indicated by $L_{\rm x}/L_{\rm bol}$, are very
similar to stars of equivalent mass in the Orion Nebula cluster (ONC),
but nearly an order of magnitude higher than in the Pleiades at $\simeq
120$\,Myr. The X-ray luminosities of G-type stars in NGC~2547 are only
marginally lower than their counterparts in the ONC. The spreads in the
XLFs of NGC~2547 are much smaller than in older clusters and the ONC.
These phenomena are readily interpreted in terms of a
magnetic dynamo and the 
age-rotation-activity paradigm, providing account is taken of the evolution
of convection zone parameters and angular momentum loss using Rossby numbers.
The reported low activity levels of some very young accreting PMS stars
in the ONC do not straightforwardly fit into this scenario.

\item
Coronal temperatures in the most active solar-type stars in NGC~2547
are intermediate between those of older clusters and the very young ONC
stars. We show that correlations proposed between coronal temperature
and X-ray activity for field stars do not predict the temperatures in
NGC~2547 or the much higher temperatures in the ONC.  Simple scaling
law arguments predict that coronal temperatures should increase as
surface gravity decreases for stars with saturated levels of magnetic
activity. These arguments require that the scale heights of coronae in
the ONC stars are an order of magnitude larger than in ZAMS
stars. Alternative explanations such as enhanced flaring or different
modes of dynamo action in younger, fully convective PMS stars are
possible, but there is no clear evidence for very hot coronae in the
fully convective stars of NGC~2547 and the flaring rate (at least for
strong flares) is comparable in the solar-type stars of the ONC,
NGC~2547 and the Pleiades (see below). However, the hottest coronal
temperatures in our sample do arise from clear flaring behaviour.

\item
A number of candidate flares were detected among the low-mass stars of
NGC~2547, with peak luminosities (0.3--3.0\,keV) exceeding
$10^{30}$~erg\,s$^{-1}$ and integrated flare energies of
$10^{34}$--$10^{35}$~erg. The occurrence rate for flares with energy
$>10^{34}$~erg is approximately 1 every $350^{+350}_{-120}$\,ks 
and comparable to flare frequencies in the ONC and Pleiades.

\item
By comparison with earlier {\it ROSAT} HRI observations we find that
only 10--15 per cent of G- and K-type stars, or stars of any spectral
types with $L_{\rm x}>3\times10^{29}$~erg\,s$^{-1}$, show variations of
a factor of two or more on timescales of 7 years. The level of long
term variability is incompatible with solar-like magnetic activity
cycles with periods of 20 years or less. This is comparable with
variability seen in similarly aged clusters (IC~2391) but less than
observed in older clusters and field stars {\it and} in the younger PMS
stars of the ONC.

\end{enumerate}

\section*{Acknowledgments}
This work is based on observations obtained by {\it XMM-Newton}, an ESA
science mission with instruments and contributions directly funded by
ESA member states and the USA (NASA).
We would like to thank Dr. T. Preibisch for supplying electronic tables
containing the X-ray luminosity functions of the Orion nebula cluster.
We thank an anonymous referee for helping us clarify the paper in a
number of places.

\nocite{gilliland86}
\nocite{queloz98}
\nocite{cutri03}
\nocite{jeffries04}
\nocite{naylor02}
\nocite{jeffries98n2547}
\nocite{jeffries00}
\nocite{dantona97}
\nocite{claria82}
\nocite{stauffer94}
\nocite{preibisch05a}
\nocite{preibisch05b}
\nocite{feigelson03}
\nocite{telleschi05}
\nocite{gudel04}
\nocite{jeffries99xrayrev}
\nocite{randichxrayrev00}
\nocite{flaccomio03a}
\nocite{flaccomio03c}
\nocite{stassun04}
\nocite{siess00}
\nocite{young04}
\nocite{turner01}
\nocite{baraffe02}
\nocite{kirsch05}
\nocite{jeffries05}
\nocite{mekal95}
\nocite{anders89}
\nocite{briggs03}
\nocite{gagne95}
\nocite{panzera99}
\nocite{saxton03}
\nocite{noyes84}
\nocite{pizzolato03}
\nocite{prosser96}
\nocite{randich98supersat}
\nocite{stelzer00}
\nocite{campana99}
\nocite{panzera03}
\nocite{cassinelli94}
\nocite{berghofer97}
\nocite{stelzer01}
\nocite{getman05}
\nocite{sills00}
\nocite{gudel97}
\nocite{skinner98}
\nocite{grosso04}
\nocite{marino05}
\nocite{pillitteri04}
\nocite{pillitteri05}
\nocite{parker93}
\nocite{rosner78}
\nocite{durney82}
\nocite{audard04}
\nocite{jardine99}
\nocite{favata05}
\nocite{gudel01}
\nocite{scelsi05}
\nocite{favata04}
\nocite{simon98}
\nocite{marino03a}
\nocite{micela03}
\nocite{micela96}
\nocite{jeffriesn251697}
\nocite{stern95}
\nocite{struder01}
\nocite{david99}
\nocite{gagne95b}
\nocite{wolk05}
\nocite{jordan91}
\nocite{hillenbrand97}

\bibliographystyle{mn2e}  
\bibliography{iau_journals,master}

\begin{thebibliography}{}

\bibitem[\protect\citeauthoryear{Anders \& Grevesse}{Anders \&
  Grevesse}{1989}]{anders89}
Anders E.,  Grevesse N.,  1989, Geochimica et Cosmochimica Acta, 53, 197

\bibitem[\protect\citeauthoryear{Audard, Telleschi, Gudel, Skinner, Pallavicini
  \& Mitra-Kraev}{Audard et~al.}{2004}]{audard04}
Audard M.,  Telleschi A.,  Gudel M.,  Skinner S.~L.,  Pallavicini R.,
  Mitra-Kraev U.,  2004, ApJ, 617, 531

\bibitem[\protect\citeauthoryear{Baraffe, Chabrier, Allard \&
  Hauschildt}{Baraffe et~al.}{2002}]{baraffe02}
Baraffe I.,  Chabrier G.,  Allard F.,    Hauschildt P.~H.,  2002, A\&A, 382,
  563

\bibitem[\protect\citeauthoryear{Berghofer, Schmitt, Danner \&
  Cassinelli}{Berghofer et~al.}{1997}]{berghofer97}
Berghofer T.~W.,  Schmitt J. H. M.~M.,  Danner R.,    Cassinelli J.~P.,  1997,
  A\&A, 322, 167

\bibitem[\protect\citeauthoryear{Briggs \& Pye}{Briggs \& Pye}{2003}]{briggs03}
Briggs K.,  Pye J.~P.,  2003, MNRAS, 345, 714

\bibitem[\protect\citeauthoryear{Campana, Lazzati, Panzera \&
  Tagliaferri}{Campana et~al.}{1999}]{campana99}
Campana S.,  Lazzati D.,  Panzera M.~R.,    Tagliaferri G.,  1999, A\&A, 524,
  423

\bibitem[\protect\citeauthoryear{Cassinelli, Cohen, Macfarlane, Sanders \&
  Welsh}{Cassinelli et~al.}{1994}]{cassinelli94}
Cassinelli J.~P.,  Cohen D.~H.,  Macfarlane J.~J.,  Sanders W.~T.,    Welsh
  B.~Y.,  1994, ApJ, 421, 705

\bibitem[\protect\citeauthoryear{Clari\'{a}}{Clari\'{a}}{1982}]{claria82}
Clari\'{a} J.~J.,  1982, A\&AS, 47, 323

\bibitem[\protect\citeauthoryear{{Cutri, R. M. et al.}}{{Cutri, R. M. et
  al.}}{2003}]{cutri03}
{Cutri, R. M. et al.} 2003, Technical report, Explanatory supplement to the
  2MASS All Sky data release.
http://www.ipac.caltech.edu/2mass/

\bibitem[\protect\citeauthoryear{D'Antona \& Mazzitelli}{D'Antona \&
  Mazzitelli}{1997}]{dantona97}
D'Antona F.,  Mazzitelli I.,  1997, Mem. Soc. Astr. It., 68, 807

\bibitem[\protect\citeauthoryear{David, Harnden, Kearns, Zombeck, Harries,
  Mossman, Prestwich, Primini, Silverman \& Snowden}{David
  et~al.}{1999}]{david99}
David L.~P.,  Harnden F. R.~J.,  Kearns K.~R.,  Zombeck M.~V.,  Harries D.~E.,
  Mossman A.~E.,  Prestwich A.,  Primini F.~A.,  Silverman J.~D.,    Snowden
  S.~L.,  1999, Technical report, The ROSAT High Resolution Imager Calibration
  Report.
U.S. ROSAT Science Data Center/SAO

\bibitem[\protect\citeauthoryear{Durney \& Robinson}{Durney \&
  Robinson}{1982}]{durney82}
Durney B.~R.,  Robinson R.~D.,  1982, ApJ, 253, 290

\bibitem[\protect\citeauthoryear{Favata, Flaccomio, Reale, Micela, Sciortino,
  Shang, Stassun \& Feigelson}{Favata et~al.}{2005}]{favata05}
Favata F.,  Flaccomio E.,  Reale F.,  Micela G.,  Sciortino S.,  Shang H.,
  Stassun K.~G.,    Feigelson E.~D.,  2005, ApJS, 160, 469

\bibitem[\protect\citeauthoryear{Favata, Micela, Baliunas, Schmitt, Gudel,
  Harnden, Sciortino \& Stern}{Favata et~al.}{2004}]{favata04}
Favata F.,  Micela G.,  Baliunas S.~L.,  Schmitt J. H. M.~M.,  Gudel M.,
  Harnden F. R.~J.,  Sciortino S.,    Stern R.~A.,  2004, A\&A, 418, L13

\bibitem[\protect\citeauthoryear{Feigelson, Gaffney, Garmire, Hillenbrand \&
  Townsley}{Feigelson et~al.}{2003}]{feigelson03}
Feigelson E.~D.,  Gaffney J.~A.,  Garmire G.,  Hillenbrand L.~A.,    Townsley
  L.,  2003, ApJ, 584, 911

\bibitem[\protect\citeauthoryear{Flaccomio, Damiani, Micela, Sciortino,
  Harnden, Murray \& Wolk}{Flaccomio et~al.}{2003}]{flaccomio03c}
Flaccomio E.,  Damiani F.,  Micela G.,  Sciortino S.,  Harnden F. R.~J.,
  Murray S.~S.,    Wolk S.~J.,  2003, ApJ, 582, 398

\bibitem[\protect\citeauthoryear{Flaccomio, Micela \& Sciortino}{Flaccomio
  et~al.}{2003}]{flaccomio03a}
Flaccomio E.,  Micela G.,    Sciortino S.,  2003, A\&A, 402, 477

\bibitem[\protect\citeauthoryear{Gagn\'{e}, Caillault \& Stauffer}{Gagn\'{e}
  et~al.}{1995}]{gagne95}
Gagn\'{e} M.,  Caillault J.~P.,    Stauffer J.~R.,  1995, ApJ, 450, 217

\bibitem[\protect\citeauthoryear{Gagn\'e, Caillault \& Stauffer}{Gagn\'e
  et~al.}{1995}]{gagne95b}
Gagn\'e M.,  Caillault J.~P.,    Stauffer J.~R.,  1995, ApJ, 445, 280

\bibitem[\protect\citeauthoryear{Getman, A, A, A, A, A, A, A, A, A, A, A, A, A,
  A, A \& A}{Getman et~al.}{2005}]{getman05}
Getman K.,  A B.,  A B.,  A B.,  A B.,  A B.,  A B.,  A B.,  A B.,  A B.,  A
  B.,  A B.,  A B.,  A B.,  A B.,  A B.,    A B.,  2005, ApJS, 160, 319

\bibitem[\protect\citeauthoryear{Gilliland}{Gilliland}{1986}]{gilliland86}
Gilliland R.~L.,  1986, ApJ, 300, 339

\bibitem[\protect\citeauthoryear{Grosso, Montmerle, Feigelson \& Forbes}{Grosso
  et~al.}{2004}]{grosso04}
Grosso N.,  Montmerle T.,  Feigelson E.~D.,    Forbes T.~G.,  2004, 419, p.~653

\bibitem[\protect\citeauthoryear{Gudel}{Gudel}{2004}]{gudel04}
Gudel M.,  2004, ARA\&A, 12, 71

\bibitem[\protect\citeauthoryear{Gudel, Audard, Briggs, Haberl, Magee, Maggio,
  Mewe, Pallavicini \& Pye}{Gudel et~al.}{2001}]{gudel01}
Gudel M.,  Audard M.,  Briggs K.,  Haberl F.,  Magee H.,  Maggio A.,  Mewe R.,
  Pallavicini R.,    Pye J.,  2001, A\&A, 365, L336

\bibitem[\protect\citeauthoryear{Gudel, Guinan \& Skinner}{Gudel
  et~al.}{1997}]{gudel97}
Gudel M.,  Guinan E.~F.,    Skinner S.~L.,  1997, ApJ, 483, 947

\bibitem[\protect\citeauthoryear{Hillenbrand}{Hillenbrand}{1997}]{hillenbrand9%
7}
Hillenbrand L.~A.,  1997, AJ, 113, 1733

\bibitem[\protect\citeauthoryear{Jardine \& Unruh}{Jardine \&
  Unruh}{1999}]{jardine99}
Jardine M.,  Unruh Y.~C.,  1999, A\&A, 346, 883

\bibitem[\protect\citeauthoryear{Jeffries}{Jeffries}{1999}]{jeffries99xrayrev}
Jeffries R.~D.,  1999, in Butler C.~J.,  Doyle J.~G.,  eds, Solar and stellar
  activity: similarities and differences: ASP Conference series Vol.158 ASP,
  San Francisco, p.~75

\bibitem[\protect\citeauthoryear{Jeffries, Naylor, Devey \& Totten}{Jeffries
  et~al.}{2004}]{jeffries04}
Jeffries R.~D.,  Naylor T.,  Devey C.~R.,    Totten E.~J.,  2004, MNRAS, 351,
  1401

\bibitem[\protect\citeauthoryear{Jeffries \& Oliveira}{Jeffries \&
  Oliveira}{2005}]{jeffries05}
Jeffries R.~D.,  Oliveira J.~M.,  2005, MNRAS, 358, 13

\bibitem[\protect\citeauthoryear{Jeffries, Thurston \& Pye}{Jeffries
  et~al.}{1997}]{jeffriesn251697}
Jeffries R.~D.,  Thurston M.~R.,    Pye J.~P.,  1997, MNRAS, 287, 350

\bibitem[\protect\citeauthoryear{Jeffries \& Tolley}{Jeffries \&
  Tolley}{1998}]{jeffries98n2547}
Jeffries R.~D.,  Tolley A.~J.,  1998, MNRAS, 300, 331

\bibitem[\protect\citeauthoryear{Jeffries, Totten \& James}{Jeffries
  et~al.}{2000}]{jeffries00}
Jeffries R.~D.,  Totten E.~J.,    James D.~J.,  2000, MNRAS, 316, 950

\bibitem[\protect\citeauthoryear{Jordan \& Montesinos}{Jordan \&
  Montesinos}{1991}]{jordan91}
Jordan C.,  Montesinos B.,  1991, MNRAS, 252, 21P

\bibitem[\protect\citeauthoryear{Kirsch}{Kirsch}{2005}]{kirsch05}
Kirsch M.,  2005, Technical Report XMM-SOC-CAL-TN-0018, EPIS status of
  calibration.
XMM-Newton SOC

\bibitem[\protect\citeauthoryear{Marino, Micela, Peres, Pillitteri \&
  Sciortino}{Marino et~al.}{2005}]{marino05}
Marino A.,  Micela G.,  Peres G.,  Pillitteri I.,    Sciortino S.,  2005, A\&A,
  430, 287

\bibitem[\protect\citeauthoryear{Marino, Micela, Peres \& Sciortino}{Marino
  et~al.}{2003}]{marino03a}
Marino A.,  Micela G.,  Peres G.,    Sciortino S.,  2003, A\&A, 406, 629

\bibitem[\protect\citeauthoryear{Mewe, Kaastra \& Leidahl}{Mewe
  et~al.}{1995}]{mekal95}
Mewe R.,  Kaastra J.~S.,    Leidahl D.~A.,  1995, Technical report, The HEASARC
  Journal, Legacy Volume 6.
Goddard Space Flight Center

\bibitem[\protect\citeauthoryear{Micela \& Marino}{Micela \&
  Marino}{2003}]{micela03}
Micela G.,  Marino A.,  2003, A\&A, 404, 637

\bibitem[\protect\citeauthoryear{Micela, Sciortino, Kashyap, Harnden \&
  Rosner}{Micela et~al.}{1996}]{micela96}
Micela G.,  Sciortino S.,  Kashyap V.,  Harnden F.~R.,    Rosner R.,  1996,
  ApJS, 102, 75

\bibitem[\protect\citeauthoryear{Naylor, Totten, Jeffries, Pozzo, Devey \&
  Thompson}{Naylor et~al.}{2002}]{naylor02}
Naylor T.,  Totten E.~J.,  Jeffries R.~D.,  Pozzo M.,  Devey C.~R.,    Thompson
  S.~A.,  2002, MNRAS, 335, 291

\bibitem[\protect\citeauthoryear{Noyes, Hartmann, Baliunas, Duncan \&
  Vaughan}{Noyes et~al.}{1984}]{noyes84}
Noyes R.~W.,  Hartmann L.,  Baliunas S.~L.,  Duncan D.~K.,    Vaughan A.~H.,
  1984, ApJ, 279, 763

\bibitem[\protect\citeauthoryear{Panzera, Campana, Covino, Lazzati, Mignani,
  Moretti \& Tagliaferri}{Panzera et~al.}{2003}]{panzera03}
Panzera M.~R.,  Campana S.,  Covino S.,  Lazzati D.,  Mignani R.~P.,  Moretti
  A.,    Tagliaferri G.,  2003, A\&A, 399, 351

\bibitem[\protect\citeauthoryear{Panzera, Tagliaferri, Pasinetti \&
  Antonello}{Panzera et~al.}{1999}]{panzera99}
Panzera M.~R.,  Tagliaferri G.,  Pasinetti L.,    Antonello E.,  1999, A\&A,
  348, 161

\bibitem[\protect\citeauthoryear{Parker}{Parker}{1993}]{parker93}
Parker E.,  1993, ApJ, 408, 707

\bibitem[\protect\citeauthoryear{Pillitteri, Micela, Reale \&
  Sciortino}{Pillitteri et~al.}{2005}]{pillitteri05}
Pillitteri I.,  Micela G.,  Reale F.,    Sciortino S.,  2005, A\&A, 430, 155

\bibitem[\protect\citeauthoryear{Pillitteri, Micela, Sciortino, Damiani \&
  Harnden}{Pillitteri et~al.}{2004}]{pillitteri04}
Pillitteri I.,  Micela G.,  Sciortino S.,  Damiani F.,    Harnden F. R.~J.,
  2004, A\&A, 421, 175

\bibitem[\protect\citeauthoryear{Pizzolato, Maggio, Micela, Sciortino \&
  Ventura}{Pizzolato et~al.}{2003}]{pizzolato03}
Pizzolato N.,  Maggio A.,  Micela G.,  Sciortino S.,    Ventura P.,  2003,
  A\&A, 397, 147

\bibitem[\protect\citeauthoryear{Preibisch \& Feigelson}{Preibisch \&
  Feigelson}{2005}]{preibisch05a}
Preibisch T.,  Feigelson E.~D.,  2005, ApJS, 160, 390

\bibitem[\protect\citeauthoryear{Preibisch, Kim, Favata, Feigelson, Flaccomio,
  Getman, Micela, Sciortino, Stassun, Stelzer \& Zinnecker}{Preibisch
  et~al.}{2005}]{preibisch05b}
Preibisch T.,  Kim Y.~C.,  Favata F.,  Feigelson E.~D.,  Flaccomio E.,  Getman
  K.,  Micela G.,  Sciortino S.,  Stassun K.,  Stelzer B.,    Zinnecker H.,
  2005, ApJS, 160, 401

\bibitem[\protect\citeauthoryear{Prosser, Randich, Stauffer, Schmitt \&
  Simon}{Prosser et~al.}{1996}]{prosser96}
Prosser C.~F.,  Randich S.,  Stauffer J.~R.,  Schmitt J. H. M.~M.,    Simon T.,
   1996, AJ, 112, 1570

\bibitem[\protect\citeauthoryear{Queloz, Allain, Mermilliod, Bouvier \&
  Mayor}{Queloz et~al.}{1998}]{queloz98}
Queloz D.,  Allain S.,  Mermilliod J.~C.,  Bouvier J.,    Mayor M.,  1998,
  A\&A, 335, 183

\bibitem[\protect\citeauthoryear{Randich}{Randich}{1998}]{randich98supersat}
Randich S.,  1998, in Donahue R. A. amd~Bookbinder J.~A.,  ed., 10th Cambridge
  Workshop on Cool Stars, Stellar Systems and the Sun, ASP Conference Series,
  Vol 154 ASP, San Francisco, p.~501

\bibitem[\protect\citeauthoryear{Randich}{Randich}{2000}]{randichxrayrev00}
Randich S.,  2000, in Pallavicini R.,  Micela G.,   Sciortino S.,  eds, Stellar
  Clusters and Associations: Convection, Rotation, and Dynamos: ASP Conference
  Series Vol. 198 ASP, San Francisco, p.~401

\bibitem[\protect\citeauthoryear{Rosner, Tucker \& Vaiana}{Rosner
  et~al.}{1978}]{rosner78}
Rosner R.,  Tucker W.~H.,    Vaiana G.~S.,  1978, ApJ, 220, 643

\bibitem[\protect\citeauthoryear{Saxton}{Saxton}{2003}]{saxton03}
Saxton R.~D.,  2003, Technical Report XMM-SOC-CAL-TN-0023, A statistical
  evaluation of the EPIC flux calibration, Version 2.0.
XMM-Newton SOC

\bibitem[\protect\citeauthoryear{Scelsi, Maggio, Peres \& Pallavicini}{Scelsi
  et~al.}{2005}]{scelsi05}
Scelsi L.,  Maggio A.,  Peres G.,    Pallavicini R.,  2005, A\&A, 432, 671

\bibitem[\protect\citeauthoryear{Siess, Dufour \& Forestini}{Siess
  et~al.}{2000}]{siess00}
Siess L.,  Dufour E.,    Forestini M.,  2000, A\&A, 358, 593

\bibitem[\protect\citeauthoryear{Sills, Pinsonneault \& Terndrup}{Sills
  et~al.}{2000}]{sills00}
Sills A.,  Pinsonneault M.~H.,    Terndrup D.~M.,  2000, ApJ, 534, 335

\bibitem[\protect\citeauthoryear{Simon \& Patten}{Simon \&
  Patten}{1998}]{simon98}
Simon T.,  Patten B.~M.,  1998, PASP, 110, 283

\bibitem[\protect\citeauthoryear{Skinner \& Walter}{Skinner \&
  Walter}{1998}]{skinner98}
Skinner S.~L.,  Walter F.~M.,  1998, ApJ, 509, 761

\bibitem[\protect\citeauthoryear{Stassun, Ardila, Barsony, Basri \&
  Mathieu}{Stassun et~al.}{2004}]{stassun04}
Stassun K.,  Ardila D.~R.,  Barsony M.,  Basri G.,    Mathieu R.~D.,  2004, AJ,
  127, 3537

\bibitem[\protect\citeauthoryear{Stauffer, Caillault, Gagn\'{e}, Prosser \&
  Hartmann}{Stauffer et~al.}{1994}]{stauffer94}
Stauffer J.~R.,  Caillault J.~P.,  Gagn\'{e} M.,  Prosser C.~F.,    Hartmann
  L.~W.,  1994, ApJS, 91, 625

\bibitem[\protect\citeauthoryear{Stelzer \& Neuhauser}{Stelzer \&
  Neuhauser}{2001}]{stelzer01}
Stelzer B.,  Neuhauser R.,  2001, A\&A, 377, 538

\bibitem[\protect\citeauthoryear{Stelzer, Neuhauser \& Hambaryan}{Stelzer
  et~al.}{2000}]{stelzer00}
Stelzer B.,  Neuhauser R.,    Hambaryan V.,  2000, A\&A, 356, 949

\bibitem[\protect\citeauthoryear{Stern, Schmitt \& Kahabka}{Stern
  et~al.}{1995}]{stern95}
Stern R.~A.,  Schmitt J. H. M.~M.,    Kahabka P.~T.,  1995, ApJ, 448, 683

\bibitem[\protect\citeauthoryear{Struder}{Struder}{2001}]{struder01}
Struder L. e.~a.,  2001, A\&A, 365, L18

\bibitem[\protect\citeauthoryear{Telleschi, Gudel, Briggs, Audard, Ness \&
  Skinner}{Telleschi et~al.}{2005}]{telleschi05}
Telleschi A.,  Gudel M.,  Briggs K.,  Audard M.,  Ness J.~U.,    Skinner S.~L.,
   2005, ApJ, 622, 653

\bibitem[\protect\citeauthoryear{Turner}{Turner}{2001}]{turner01}
Turner M. J. L. e.~a.,  2001, A\&A, 365, L27

\bibitem[\protect\citeauthoryear{Wolk, Flaccomio, Micela, Favata, Glassgold,
  Shang \& Feigelson}{Wolk et~al.}{2005}]{wolk05}
Wolk S.~J.,  Flaccomio E.,  Micela G.,  Favata F.,  Glassgold A.~E.,  Shang H.,
     Feigelson E.,  2005, ApJS, 160, 423

\bibitem[\protect\citeauthoryear{Young E. T. amd~Lada, Teixeira, Muzerolle,
  Muench, Stauffer, Beichman, Rieke, Hines, Su, Engelbracht, Gordon, Misselt,
  Morrison, Stansberry \& Kelly}{Young et~al.}{2004}]{young04}
Young E. T. amd~Lada C.~J.,  Teixeira P.,  Muzerolle J.,  Muench A.,  Stauffer
  J.,  Beichman C.~A.,  Rieke G.~H.,  Hines D.~C.,  Su K. Y.~L.,  Engelbracht
  C.~W.,  Gordon K.~D.,  Misselt K.,  Morrison J.,  Stansberry J.,    Kelly D.,
   2004, ApJS, 154, 428

\end{thebibliography}


\bsp 

\label{lastpage}

\end{document}